\documentstyle[12pt,epsf]{book}

\advance \topmargin by -\headheight

\advance \topmargin by -\headsep

\evensidemargin \oddsidemargin

\def\ie{{\em i.e.}}

\def\ie{\hbox{\it i.e.}}

\def\CC{{\mathchoice
{\rm C\mkern-8mu\vrule height1.45ex depth-.05ex
width.05em\mkern9mu\kern-.05em}
{\rm C\mkern-8mu\vrule height1.45ex depth-.05ex
width.05em\mkern9mu\kern-.05em}
{\rm C\mkern-8mu\vrule height1ex depth-.07ex
width.035em\mkern9mu\kern-.035em}
{\rm C\mkern-8mu\vrule height.65ex depth-.1ex
width.025em\mkern8mu\kern-.025em}}}

\def\RR{{\rm I\kern-1.6pt {\rm R}}}

\def\ZZ{{\rm Z}\kern-3.8pt {\rm Z} \kern2pt}

\def\IB{\relax{\rm I\kern-.18em B}}

\def\ID{\relax{\rm I\kern-.18em D}}

\def\II{\relax{\rm I\kern-.18em I}}

\def\IP{\relax{\rm I\kern-.18em P}}

\def\np{Nucl. Phys.}

\def\pl{Phys. Lett.}

\def\prl{Phys. Rev. Lett.}

\def\pr{Phys. Rev.}

\def\jhep{J. High Energy Phys.}

\newcommand{\beq}{\begin{equation}}

\newcommand{\eeq}{\end{equation}}

\newcommand{\rc}{\nonumber\\}

\newcommand{\bear}{\begin{eqnarray}}

\newcommand{\eear}{\end{eqnarray}}

\def\to{\rightarrow}

\def\to{\rightarrow}

\newfont{\namefont}{cmr10}

\newfont{\addfont}{cmti7 scaled 1440}

\newfont{\boldmathfont}{cmbx10}

\newfont{\headfontb}{cmbx10 scaled 1728}

\begin{document}

\begin{titlepage}

\begin{center} \Large \bf Killing spinors of some supergravity solutions

\end{center}

\vskip 0.3truein
\begin{center}
\ Daniel Are\'an Fraga
\footnote{M.Sc. Thesis, Universidade de Santiago de
Compostela, Spain (May, 2006). \\
Advisor: Alfonso V. Ramallo}

\vspace{0.3in}

Departamento de F\'\i sica de Part\'\i culas, Universidade de
Santiago de Compostela \\and\\
Instituto Galego de F\'\i sica de Altas Enerx\'\i as (IGFAE)\\
E-15782 Santiago de Compostela, Spain\\
\vspace{0.15in}
arean@fpaxp1.usc.es

\vspace{0.3in}

\end{center}

\vskip 1truein

\begin{center}
\bf ABSTRACT
\end{center}

We compute explicitly the Killing spinors of some ten
dimensional supergravity solutions. We begin with a 10d metric of
the form $\RR^{1,3}\times{\cal Y}_6$, where ${\cal Y}_6$ is either
the singular conifold or any of its resolutions. Then, we move on to
the Klebanov-Witten and Klebanov-Tseytlin backgrounds, both
constructed over the singular conifold; and we also study the
Klebanov-Strassler solution, built over the deformed conifold.
Finally, we determine the form of the Killing spinors for the
non-commutative deformation of the Maldacena-N\'u\~nez
geometry.

\smallskip

\vskip2.6truecm
\leftline{hep-th/0605286}
\smallskip

\end{titlepage}
\setcounter{footnote}{0}


\pagestyle{empty}

\setcounter{page}{0}


\tableofcontents

\pagestyle{headings}

\chapter{Introduction}
\label{introduction}
\medskip

\setcounter{equation}{0}
\medskip

In this work we compute explicitly the Killing spinors of some
ten dimensional supergravity solutions. The main interest of these
backgrounds comes up in the context of the $AdS/CFT$ correspondence
established in
\cite{ADSCFT} (see \cite{MAGOO} for a review), for they are dual to
four dimensional supersymmetric field theories \cite{n1duality}.
Let us recall that the SUSY transformations for a background of ten
dimensional type IIB supergravity can be parameterized in terms of a
Majorana spinor
$\epsilon$ made up of 32 real components, which is the number of
charges forming the largest SUSY algebra. For a general background to be
supersymmetric, we must require the vanishing of the SUSY transformations
of the whole set of bosonic and fermionic fields of the theory. In
principle, this will result in a reduction of the number of independent
components of $\epsilon$, and therefore, of the supercharges entering the
SUSY algebra. It is precisely this resulting spinor, subjected, in
general, to some projections relating its components, what we call 
the Killing spinor of the background. Then, by computing the Killing 
spinors one can determine the amount of supersymmetry conserved by a
certain geometry.

Let us point out that the knowledge of the explicit form of the
Killing spinors allows one to apply the kappa symmetry
\cite{swedes} technique when looking for  supersymmetric embeddings
of different D-brane probes.
The addition of D-branes, pioneered by Witten in \cite{Wittenbaryon},
has become very fruitful in the
$AdS/CFT$ field, for it provides a way to uncover different stringy
effects in the Yang-Mills (YM) theories. Indeed, by adding different
D-brane probes to the supergravity backgrounds, one can study several
interesting objects living in the dual field theories. For instance,
in ref. \cite{GK} it has been shown that D3-brane probes wrapped over
three cycles of the internal manifold
$T^{1,1}$ in the so-called Klebanov-Witten model \cite{KW} (whose
geometry, which we will describe in detail in chapter \ref{kwcp}, is
$AdS_5\times T^{1,1}$) describe dibaryon operators in the ${\cal N}=1$
superconformal YM theory living on the boundary of $AdS_5$ (see also
refs. \cite{BHK}-\cite{HMcK} for more results on dibaryons in this model and in some
orbifold theories). Besides describing other exotic objects as domain walls (by
means of D-brane probes of codimension one along the field theory
dimensions, see ref. \cite{GK}), the addition of D-brane probes
permits the introduction of open string degrees of freedom into the
gauge/gravity correspondence. One can try to generalize the $AdS/CFT$
correspondence by adding brane probes and identifying the
fluctuations of the probe, which correspond to degrees of freedom of
open strings connecting the probe and the branes that generated the
background, with fundamental hypermultiplets of dynamical matter
fields of the dual field theory \cite{kk}. Let us mention that,
following this program, in ref.
\cite{SSP} the explicit determination of the Killing spinors of the
Klebanov-Witten model allowed us to systematically apply the
kappa symmetry technique in order to study the possible
supersymmetric embeddings in that background for D3, D5 and D7-brane
probes.

The structure of this work is the following: in the next section of this
chapter we present the SUSY variations of the IIB SUGRA fermionic
fields. In chapter \ref{genconsec} we compute the Killing
spinors for the different resolutions of the conifold by
using a 10d background (constructed in \cite{Twist}) arising from the
uplift of a certain configuration in 8d gauged supergravity consisting of
a D6-brane wrapping an $S^2$. Some results of this chapter, such as
the form of the metrics of the singular and deformed conifold and the
projections satisfied by their Killing spinors, are used in the following chapters where we deal
with 10d SUGRA solutions constructed over the singular conifold or
over its deformation. The aim of chapter \ref{kwcp} is to determine
the Killing  spinors of the Klebanov-Witten solution \cite{KW}; we
solve the SUSY equations in a frame such that the Killing spinors do
not depend on the angular coordinates of the conifold. Chapter
\ref{ktcp} is devoted to the Klebanov-Tseytlin model \cite{KT}: we
briefly introduce it and again we are able to write the Killing
spinors in a frame where they do not depend on the angular
coordinates of the conifold. In chapter \ref{kscp} we deal with the
Klebanov-Strassler solution \cite{KS}, we describe it and by
computing its Killing spinors we show that the requirement of
preserving the same supersymmetries as in the solution corresponding
to a D3-brane at the tip of the deformed conifold fixes the values of
the three-forms to those found in ref \cite{KS}. In chapter
\ref{ncmncp} the Killing spinors of the non-commutative deformation of
the Maldacena-N\'u\~nez solution \cite{NCMN} are explicitly computed. This
calculation follows closely the one performed in \cite{flavoring} for the
commutative case \cite{MN, CV} and, in fact, the Killing spinors of
the non-commutative background can be written in terms of the ones of
the commutative geometry by means of a rotation along the
non-commutative plane. Finally, in chapter \ref{conclcp} we summarize
our results and give some remarks.

The results of the computations performed in chapters \ref{kwcp} and
\ref{kscp} were published in ref. \cite{SSP}; as it was said, the
knowledge of the Killing spinors of the Klebanov-Witten model was
essential for the kappa symmetry analysis carried out there. The
Killing  spinors of the Klebanov-Strassler model were
included in the appendix of \cite{SSP} as an starting  point to extend
the study of supersymmetric embeddings to that more interesting
solution. The form of the Killing spinors of the non-commutative
Maldacena-N\'u\~nez solution was published in the appendix of
\cite{ncflav} where we studied the addition of open string degrees of
freedom to that background.

Last, let us briefly comment on the ten dimensional IIB SUGRA
solutions arising from the whole new class of 5d Sasaki-Einstein
manifolds $Y^{p,q}$, recently constructed in \cite{GMSW1,GMSW2}. It
was shown \cite{ms,sequiver} that the 10d backgrounds $AdS_5\times
Y^{p,q}$ are dual to four dimensional superconformal quiver gauge
theories. The authors of
\cite{ypqflav}, by performing a similar computation to the ones
presented here, determined explicitly the Killing spinors of the 10d
background in order to study the addition of brane probes. 
Recently, in \cite{lpqrflav} that study was extended to more general
backgrounds of the form $AdS_5\times L^{p,q,r}$, where $L^{p,q,r}$ is
the more general family of 5d Sasaki-Einstein manifolds constructed in
\cite{Cvetic:2005ft,Martelli:2005wy}.

\section{SUSY transformations}
\setcounter{equation}{0}

In the backgrounds we will consider the fermionic fields (the
dilatino and the gravitino) have vanishing expectation values, so the
SUSY variations of the bosonic fields are trivially zero. Then, the
Killing spinors are obtained by requiring the vanishing of the
supersymmetry variations of the fermionic fields of the theory.

In the type IIB theory the spinor $\epsilon$ is actually composed
of two Majorana-Weyl spinors $\epsilon_L$ and $\epsilon_R$ of well
defined ten-dimensional chirality, which can be arranged as a
two-component vector:
\beq
\epsilon\,=\,\pmatrix{\epsilon_L\cr\epsilon_R}\,.
\label{sugraspinor}
\eeq
However, one can use complex spinors instead of working with the
real  two-component spinor written in eq. (\ref{sugraspinor}).  In
terms of $\epsilon_R$ and $\epsilon_L$ the complex spinor is simply:
\beq
\epsilon\,=\,\epsilon_L\,+\,i\,\epsilon_R\,\,.
\eeq

For type IIB SUGRA with constant Ramond-Ramond scalar the supersymmetry
variations are
\cite{SUSYIIB}:
\bear
%
%
%
%
&&\delta\lambda\,=\,{i\over2}\,\partial_N\phi\,\Gamma^N\,\epsilon^*-{i\over
24}\,{\cal F}^{(3)}_{N_1N_2N_3}\,
\Gamma^{N_1N_2N_3}\,\epsilon\,\,,\rc\rc
&&\delta\psi_{M}\,=\,D_{M}\,\epsilon\,+\,{i\over 1920}\,
F_{N_1\cdots N_5}^{(5)}\,\Gamma^{N_1\cdots
N_5}\Gamma_M\epsilon\,+\rc\rc
&&\,\,\,\,\,\,\,\,\,\,\,\,\,\,\,\,+\,{1\over 96}\,{\cal
F}^{(3)}_{N_1N_2N_3}\,
\big(\,\Gamma_{M}^{\,\,\,N_1N_2N_3}\,-\,
9\delta_{M}^{N_1}\,\,\Gamma^{N_2 N_3}\,\big)\,\epsilon^{*}\,\,,
\label{sugra}
\eear
where $\Gamma^{N_1\cdots N_n}$ stands for the
antisymmetric product $\Gamma^{[\,N_1}\cdots\Gamma^{N_n\,]}\,$.
$\lambda$($\psi$) is the dilatino (gravitino),
$\phi$ is the dilaton, $F^{(5)}$ is the selfdual Ramond-Ramond (RR)
five-form, and
${\cal F}^{(3)}$ is the following complex combination of the
Neveu-Schwarz-Neveu-Schwarz (NSNS) ($H$) and RR ($F^{(3)}$) three-forms:
\beq
{\cal F}^{(3)}_{N_1N_2N_3}\,=\,g_s^{-{1\over
2}}\,H_{N_1N_2N_3}\,+\,i g_s^{{1\over
2}}\,F_{N_1N_2N_3}^{(3)}\,\,.
\label{comp3form}
\eeq


\chapter{Killing spinors of the resolutions of the conifold}
\label{genconsec}
\setcounter{equation}{0}
\medskip
\section{The resolutions of the conifold in 10d Supergravity}
\label{gcintro}

The conifold is a non-compact Calabi-Yau
threefold with a conical singularity. Its metric can be written as
$ds^2_{6}=dr^2\,+\,r^2\,ds^2_{T^{1,1}}$, where 
$ds^2_{T^{1,1}}$ is the metric of the $T^{1,1}$ coset 
$(SU(2)\times SU(2))/U(1)$, which is the base of the cone. The $T^{1,1}$ space is
an Einstein manifold whose metric can be written \cite{Candelas}
explicitly by using the fact that
it is an $U(1)$ bundle over $S^2\times S^2$. Actually, if $(\theta_1,\phi_1)$ and 
$(\theta_2,\phi_2)$ are the standard coordinates of the $S^2$'s and if
$\psi\in [0,4\pi)$ parameterizes the $U(1)$ fiber, the metric may be
written as
\beq
ds^2_{T^{1,1}}\,=\,{1\over 6}\,\sum_{i=1}^{2}\,
\big(\,d\theta_i^2\,+\,\sin^2\theta_i\,d\phi_i^2\,)\,+\,
{1\over9}\,\left(\,d\psi\,+
\,\sum_{i=1}^{2}\cos\theta_id\phi_i\,\right)^2\,\,.
\label{t11metric}
\eeq

The conical singularity can be resolved in two different ways according
to whether an $S^2$ or an $S^3$ is blown up at the singular
point \cite{Candelas}. The former is known as the resolved conifold,
while the latter is the deformed conifold. Both geometries appear
naturally as supergravity duals of D6-branes wrapping an $S^2$. The
natural framework for this problem is the eight dimensional
Salam-Sezgin gauged supergravity
\cite{SS} where the D6 become domain walls. The eleven dimensional
geometry resulting from uplifting the 8d supersymmetric solution
(remember that this 8d SUGRA comes from compactification of the 11d
SUGRA on an $SU(2)$ manifold) consists of a fibration of the $S^2$
over the
$S^3$ of the compactification due to the twisting that must be performed
to get a supersymmetric solution. The resulting 11d metric
\cite{gaugedsugra} is of the form $R^{1,4}\times{\cal Y}_6$, where
${\cal Y}_6$ is a cone whose base is topologically $S^2\times S^3$
and the radial coordinate of the cone is the distance to the domain 
wall in the 8d geometry. 

It was shown in \cite{Twist} that the singular, deformed and resolved
conifold (and their generalizations with one additional parameter) are
obtained as different solutions of the same system of differential
equations, which follows from the vanishing of the 8d SUGRA
supersymmetry variations ($\delta\chi_i=\delta\psi_\alpha=0\;
;\;i=1,2,3\,;\,\alpha=0,...,7$) for an ansatz of the form:
\bear
&&ds^2_8=e^{2f}dx^2_{1,4}+e^{2h}d\Omega^2_2+dr^2\,, \label{8dmetric}\\
&&A^1=g(r)\,\sigma^1\;,\quad A^2=g(r)\,\sigma^2\;,\quad
A^3=\sigma^3\,,
\label{8dgauge}
\eear
where $d\Omega^2_2=d\theta_1^2+\sin^2\theta_1\,d\phi_1^2$ is the metric
of an $S^2$,
$f=f(r)\; ,\; h=h(r)$, and $A^i$ ($i=1,2,3$) is the gauge field along the
$S^2$, and we have defined $\sigma^i$ ($i=1,2,3$) as the Maurer-Cartan
one-forms, namely:
\beq
\sigma^1=d\theta_1\;,\quad\sigma^2=\sin\theta_1\,d\phi_1\;\;
,\quad\sigma^3=\cos\theta_1\,d\phi_1\, ,
\label{s2forms}
\eeq
which
satisfy $d\sigma^i=-{1\over2}\,\epsilon_{ijk}\,\sigma^j\wedge\sigma^k$.

When uplifting to eleven dimensions we impose that the unwrapped part of
the metric corresponds to flat five dimensional Minkowski spacetime; thus
we get the relation $f={\phi\over3}$. Let  $\omega^i$ for
$i=1,2,3$  be a set of $SU(2)$ left invariant one forms of the external
$S^3$ satisfying
$d\omega^i={1\over2}\epsilon_{ijk}\,\omega^j\wedge\omega^k$. Then, the
eleven dimensional metric is \cite{SS}: 
\bear
ds^2_{11}&=&dx^2_{1,4}+e^{2h-{2\phi\over3}}d\Omega^2_2+e^{-{2\phi\over3}}dr^2+
4e^{{4\phi\over3}+2\lambda}\left(\omega^1+g\,\sigma^1\right)^2+\rc
&+&4e^{{4\phi\over3}+2\lambda}\left(\omega^2+g\,\sigma^2\right)^2+
4e^{{4\phi\over3}-4\lambda}\left(\omega^3+\sigma^3\right)^2\,,
\label{11dconif}
\eear
where $\phi=\phi(r)$ is the dilaton of the 8d solution and
$\lambda=\lambda(r)$ is a scalar in the coset $SL(3,\RR)/SO(3)$ of
the 8d solution \cite{gaugedsugra}.

Therefore, once one imposes the vanishing of the 8d gauged SUGRA
supersymmetry transformations, this uplifted metric is brought into
the form $\RR^{1,4}\times {\cal Y}^6$. ${\cal
Y}^6$ being either the resolved, the deformed, or the singular
conifold, according to the different solutions of the aforementioned
first order system \cite{Twist} resulting from the eight dimensional
SUSY equations.

By performing a Kaluza-Klein reduction along one of the flat spatial
directions of the metric (\ref{11dconif}), we get the following ten
dimensional ansatz:
\bear
ds^2_{10}&=&dx^2_{1,3}+e^{2h-{2\phi\over3}}d\Omega^2_2+e^{-{2\phi\over3}}dr^2+
4e^{{4\phi\over3}+2\lambda}\left(\omega^1+g\,\sigma^1\right)^2+\rc
&+&4e^{{4\phi\over3}+2\lambda}\left(\omega^2+g\,\sigma^2\right)^2+
4e^{{4\phi\over3}-4\lambda}\left(\omega^3+\sigma^3\right)^2\,,
\label{10dconif}
\eear
with no fluxes and constant dilaton. The reduction leading to (\ref{10dconif})
was performed along one flat spatial direction. Therefore, we expect
that by imposing the vanishing of the 10d SUSY transformations for
this 10d  metric, we will arrive at the same first order system as for
the 8d background (\ref{8dmetric}) \cite{Twist}. Thus, the metric will
be of the form
$\RR^{1,3}\times {\cal Y}^6$, where ${\cal Y}^6$ is the resolved, the
deformed or the singular conifold, according to the different
solutions of the system of equations. Moreover, since we are working
directly in the uplifted 10d background, we will get the explicit form
of the Killing spinors for the different ten dimensional metrics
$\RR^{1,3}\times {\cal Y}^6$.

\section{Killing spinors}
\setcounter{equation}{0}
In this section we will compute the Killing spinors of the 10d
background (\ref{10dconif}). By requiring the vanishing of the SUSY
variations written in eq. (\ref{sugra}) we will obtain some projections
to be satisfied by the 10d spinor $\epsilon$, together with some
differential equations for the unknown functions entering the ansatz,
namely $g$, $\phi$, $\lambda$, and $h$. The projections imposed on
$\epsilon$ reduce the number of supersymmetries while the different
solutions of the differential equations give rise to the different
resolutions of the conifold.

The vanishing of the SUSY variations (\ref{sugra}) for the background
(\ref{10dconif}) (which has no fluxes) results in the following
equations:
\beq
D_{\tilde m}\,\epsilon=0\,,
\label{sugrasimp}
\eeq
where $\tilde m$ runs along the basis formed by the differentials of
the coordinates of the geometry and
$\epsilon$ is a 10d spinor.
Henceforth we will use indices with tilde when referring to the basis
formed by the differentials of the coordinates, \ie \ $e^{\tilde
m}=dX^{\tilde m}$.


 Since the geometry
(\ref{10dconif}) comes up in the  framework of 8d gauged
supergravity, the Killing spinors should not depend on the
coordinates of the $SU(2)$ group manifold, and, due to the
aforementioned SUSY twisting, neither should they depend on the
remaining $S^2$. Moreover, the ten dimensional metric can be
expressed as the trivial product
$R^{1,3}\times{\cal Y}_6$, so the Killing spinors should not depend
either on the flat space coordinates. Indeed, let us consider the
natural one-form basis $e^a$ for the ten dimensional
metric (\ref{10dconif}):
\bear
&&e^{x^\alpha}=dx^\alpha\;,\;(\alpha=0,1,2,3)\;,\quad
e^{r}\,=\,e^{-{\phi\over3}}\,dr\,,\rc\rc
&&e^1=e^{h-{\phi\over3}}\,d\theta_1\;,\quad
e^2=e^{h-{\phi\over3}}\,\sin\theta_1\,d\phi_1\,,\rc\rc
&&e^{\hat1}=2e^{{2\phi\over3}+\lambda}\left(\omega^1+g\,\sigma^1\right)
\;,\quad e^{\hat2}=2e^{{2\phi\over3}+
\lambda}\left(\omega^2+g\,\sigma^2\right)\,,\rc\rc
&&e^{\hat3}=2e^{{2\phi\over3}-2\lambda}\left(\omega^3+\sigma^3\right)\,.
\label{conifbasis}
\eear

Let us
point out that the covariant derivative appearing in eq.
(\ref{sugrasimp}) can be written as: 
$D_{\tilde m}=\partial_{\tilde m}+{1\over4}\,\omega^{a\,b}_{\tilde m
}\,\Gamma_{a\,b}$, where
$\partial_{\tilde m}$ denotes the usual partial derivative with
respect to the coordinate $X^{\tilde m}$ and
$\omega^{a\,b}_{\tilde
m}$ stands for the components of the spin connection one-form
$\omega^{a\,b}$, namely:
\beq
\omega^{a\,b}=\omega^{a\,b}_{\tilde m}\,dX^{\tilde m}\,.
\label{gcomegatildef}
\eeq
The indices $a,b$ run along the 
frame (\ref{conifbasis}). So $\Gamma_{a\,b}$ denotes the
antisymmetrized product of two constant Dirac matrices $\Gamma_a$
and $\Gamma_b\,,\;(a,b=x^\alpha,r,1,2,\hat1,\hat2,\hat3)$ associated
to that frame.

The spin connection one-form $\omega^{a\,b}$ is defined by the Cartan
equations:
\beq
de^a+\omega^a_{\,\,b}\,\wedge e^b=0.
\label{cartan}
\eeq
Hence, in order to determine the different components of the spin
connection, we insert the derivatives of the
one-forms (\ref{conifbasis})  and a generic ansatz for
$\omega^{a\,b}$ into eq. (\ref{cartan}). As we will see, it will
become useful to write $\omega^{a\,b}$ in the frame
(\ref{conifbasis}), it takes the form:
\bear
&&\omega^{x^\alpha\,b}=0\;,\;(\alpha=0,1,2,3)\;,\rc\rc
&&\omega^{1\,r}=e^{{\phi\over3}}\left(h'-{\phi'\over3}\right)\,e^1
+e^{4{\phi\over3}+\lambda-h}\,g'\,e^{\hat1}\;,\quad
\omega^{2\,r}=e^{{\phi\over3}}\left(h'-{\phi'\over3}\right)\,e^2
+e^{{4\phi\over3}+\lambda-h}\,g'\,e^{\hat2}\;,\rc\rc
&&\omega^{\hat1\,r}=e^{{\phi\over3}}\left(\lambda'+{2\phi'\over3}\right)\,e^{\hat1}
+e^{4{\phi\over3}+\lambda-h}\,g'\,e^1 \;,\quad
\omega^{\hat2\,r}=e^{{\phi\over3}}\left(\lambda'+{2\phi'\over3}\right)\,e^{\hat2}
+e^{{4\phi\over3}+\lambda-h}\,g'\,e^2 \;,\rc\rc
&&\omega^{\hat3\,r}=e^{{\phi\over3}}\left({2\phi'\over3}-2\lambda'\right)
\,e^{\hat3}\;,\quad
\omega^{\hat1\,1}=-e^{4{\phi\over3}+\lambda-h}\, g'\,e^r\;,\quad
\omega^{\hat2\,2}=-e^{{4\phi\over3}+\lambda-h}\, g'\,e^r\;,\rc\rc
&&\omega^{2\,1}=e^{{\phi\over3}-h}\cot\theta\,e^2+e^{{4\phi\over3}-2\lambda-2h}
\,\left(g^2-1\right)\,e^{\hat3} \;,\rc\rc
&&\omega^{\hat1\,\hat3}=e^{{\phi\over3}-h}\cosh\left(3\lambda\right)\,g\,e^2
-{1\over4}\,e^{-{2\phi\over3}-4\lambda}\,e^{\hat2}
\; ,\rc\rc
&&\omega^{\hat2\,\hat1}=e^{{\phi\over3}-h}\,\cot\theta\,e^2+
{1\over4}\,e^{-{2\phi\over3}}\,\left(e^{-4\lambda}-2e^{2\lambda}\right)\,e^{\hat3}
\;,\rc\rc
&&\omega^{\hat2\,\hat3}={1\over4}\,e^{-{2\phi\over3}-4\lambda}\,e^{\hat1}
-e^{{\phi\over3}-h}\,\cosh\left(3\lambda\right)\,g\,e^1\; , \rc\rc
&&\omega^{\hat1\,2}=\omega^{1\,\hat2}=-e^{{\phi\over3}-h}\,
\sinh\left(3\lambda\right)\,g\,e^{\hat3}\;,
\rc\rc
&&\omega^{\hat3\,2}=-e^{{\phi\over3}-h}\,\sinh\left(3\lambda\right)\,g\,e^{\hat1}
-e^{4{\phi\over3}-2\lambda-2h}\,\left(g^2-1\right)\,e^1\;,\rc\rc
&&\omega^{\hat3\,1}=e^{{\phi\over3}-h}\,\sinh\left(3\lambda\right)\,g\,e^{\hat2}
+e^{4{\phi\over3}-2\lambda-2h}\,\left(g^2-1\right)\,e^2\,.
\label{spincon}
\eear
The prime appearing in these expressions denotes the radial derivative
(for instance $\phi'={d\phi\over dr}$). 
It is not difficult to switch to the basis formed by the
differentials of the coordinates; one can write
$\omega^{a\,b}_{\tilde
m}=E^{\,c}_{\,\tilde m}\,\omega^{a\,b}_c$, where
$E^{\,c}_{\,\tilde m}$ are the coefficients appearing in the
expression of the frame one-forms (\ref{conifbasis}) in terms of the
differentials of the coordinates,
\ie
\ $e^a=E^{\,a}_{\,\tilde n}\,e^{\tilde n}=E^{\,a}_{\,\tilde
n}\,dX^{\tilde n}$.

Once we have determined the spin
connection of the geometry, by writing explicitly eqs. (\ref{sugrasimp})
we will get a system of differential equations for $g$,
$\phi$, $\lambda$, and $h$, together with some algebraic constraints and
some projections imposed on
$\epsilon$. Indeed, we start by subjecting  the spinor to the following
angular projection:
\beq
\Gamma_{12}\,\epsilon=-\Gamma_{\hat1\hat2}\,\epsilon\,\,,
\label{2cproj}
\eeq
which arises naturally \cite{gaugedsugra} in the framework of the 8d
gauged SUGRA when requiring that the D6-brane wraps a two-cycle
inside a $K3$ manifold.

Then, since we are assuming that $\epsilon$ only depends on $r$, it
will become easier to write the equations (\ref{sugrasimp}) directly
in the indices running along the frame (\ref{conifbasis}), \ie \
$D_a\,\epsilon=0$, resulting:
\bear
\omega^{a\,b}_{x^\alpha}\,\Gamma_{a\,b}\,\epsilon=0\;,\label{susyeq1}\\
\rc
\omega^{a\,b}_{\hat1}\,\Gamma_{a\,b}\,\epsilon=
\omega^{a\,b}_{\hat2}\,\Gamma_{a\,b}\,\epsilon=
\omega^{a\,b}_{\hat3}\,\Gamma_{a\,b}\,\epsilon=
\omega^{a\,b}_1\,\Gamma_{a\,b}\,\epsilon=
\omega^{a\,b}_{2}\,\Gamma_{a\,b}\,\epsilon=0\; ,\label{susyeq2}\\ \rc
{\rm and}\quad
e^{\phi\over3}\left(\partial_r+{1\over4}\,\omega^{a\,b}_{\tilde
r}\,\Gamma_{a\,b}
\right)\,\epsilon=0\;,\label{susyeq3}
\eear
where in the last equation we have used that
$D_r\,\epsilon=(E^{\,r}_{\,\tilde r})^{-1}\,D_{\tilde r}\,\epsilon$. 
%
%
Since
$\omega^{a\,b}_{x^\alpha}=0$, eqs. (\ref{susyeq1}) are trivially
satisfied. Inserting the spin connection and using the projection
(\ref{2cproj}) in the first equation of (\ref{susyeq2}), namely
$\omega^{a\,b}_{\hat1}\,\Gamma_{a\,b}\,\epsilon=0$,
one gets:
\beq
\left(\lambda'+{2\over3}\phi'\right)\,\epsilon=\left[e^{\phi+\lambda-h}\,g'\,
\Gamma_{1\hat1}-{1\over4}\,e^{-\phi-4\lambda}\,\Gamma_r\,
\Gamma_{\hat1\hat2\hat3}-e^{-h}\,\sinh\left(3\lambda\right)\,g\,
\Gamma_r\,\Gamma_{1\hat2\hat3}\right]\,
\epsilon\, .
\label{eq1hat}
\eeq
The equation $\omega^{a\,b}_{\hat2}\,\Gamma_{a\,b}\,\epsilon=0$ yields
again eq. (\ref{eq1hat}). While from the third equation in
(\ref{susyeq2}) we get:
\bear
\left({2\over3}\phi'-2\lambda'\right)\,\epsilon&=&\left[\,{1\over4}\,
e^{-\phi}\left(e^{-4\lambda}-2e^{2\lambda}\right)-e^{\phi-2\lambda-2h}
\left(g^2-1\right)\right]\Gamma_r\,\Gamma_{\hat1\hat2\hat3}\,\epsilon+\rc\rc
&+&2\,e^{-h}\,
\sinh\left(3\lambda\right)\,g\,\Gamma_r\,\Gamma_{1\hat2\hat3}\,\epsilon\,.
\label{eq3hat}
\eear
The last two equalities in (\ref{susyeq2}) render the same equation:
\bear
\left(h'-{\phi'\over3}\right)\,\epsilon=\Big[-e^{\phi+\lambda-h}\,g'\,\Gamma_{1\hat1}
+e^{-h}\,\cosh\left(3\lambda\right)\,g\,\Gamma_r\,\Gamma_{1\hat2\hat3}+\rc
\rc+e^{\phi-2\lambda-2h}\,\left(g^2-1\right)\Gamma_r\,\Gamma_{\hat1\hat2\hat3}
\,\Big]\epsilon\,.
\label{eq1}
\eear
One can combine equations (\ref{eq1hat}) and (\ref{eq3hat}) to get
rid of $\lambda'$, resulting:
\beq
\phi'\,\epsilon+e^{\phi+\lambda-h}\,g'\,\Gamma_{\hat11}\,\epsilon+\left[\,
{1\over2}\,e^{\phi-2\lambda-2h}\,\left(g^2-1\right)+{1\over8}\,e^{-\phi}\,
\left(e^{-4\lambda}+2e^{2\lambda}\right)\right]
\Gamma_r\,\Gamma_{\hat1\hat2\hat3}\,\epsilon=0\,.
\label{eq13hat}
\eeq
Then, from this last equation it is clear that the 10d spinor $\epsilon$
must satisfy the following projection \cite{Twist}:
\beq
\Gamma_r\,\Gamma_{\hat1\hat2\hat3}\,\epsilon=-\left(\beta+\tilde{\beta}
\,\Gamma_{\hat11}\right)\epsilon\,,
\label{betaproj}
\eeq
where $\beta$ and $\tilde{\beta}$ are functions of the radial coordinate
given by
\bear
\phi'=\left[\,{1\over2}\,e^{\phi-2\lambda-2h}\left(g^2-1\right)
+{1\over8}\,e^{-\phi}\,\left(e^{-4\lambda}+2e^{2\lambda}\right)\right]\beta\,,
\label{betadef}\\\rc
e^{\phi+\lambda-h}\,g'=\left[\,{1\over2}\,e^{\phi-2\lambda-2h}\left(g^2-1\right)
+{1\over8}\,e^{-\phi}\,\left(e^{-4\lambda}+2e^{2\lambda}\right)\right]
\tilde{\beta}\,.\label{betatildef}
\eear
Since
$\left(\Gamma_r\Gamma_{\hat1\hat2\hat3}\right)^2\,\epsilon=\epsilon$ and
$\left\{\Gamma_r\Gamma_{\hat1\hat2\hat3},\Gamma_{\hat11}\right\}=0$, 
by squaring (\ref{betaproj}) one can check that
$\beta^2+\tilde{\beta}^2=1$ and thus we can represent $\beta$ and
$\tilde{\beta}$ as
\beq
\beta=\cos\alpha \;,\quad\tilde{\beta}=\sin\alpha\,\,.
\label{alphadef}
\eeq
Hence, the projection (\ref{betaproj}) can be written as
\beq
\Gamma_r\,\Gamma_{\hat1\hat2\hat3}\,\epsilon=-e^{\alpha\Gamma_{\hat11}}\,
\epsilon\,\,,\label{betaprexp}
\eeq
and then, solved as
\bear
\epsilon=e^{-{\alpha\over2}\Gamma_{\hat11}}\,\tilde{\epsilon}\; ,\quad
\Gamma_r\,\Gamma_{\hat1\hat2\hat3}\,\tilde{\epsilon}=-\tilde{\epsilon}\,\,.
\label{betaprojsolv}
\eear
Since we are working in type IIB
SUGRA, the 10d spinors have well defined chirality. Then, they
verify the following equality:
$\Gamma_{x^0...x^3}\Gamma_r\Gamma_{12\hat1\hat2\hat3}\,\epsilon=
-\epsilon$. Using this identity together with (\ref{alphadef}) and
the two-cycle  projection (\ref{2cproj}), the projection
(\ref{betaproj}) can be rewritten as
\beq
\Gamma_{x^0...x^3}\left(\cos\alpha\,\Gamma_{12}-\sin\alpha\,\Gamma_{1\hat2}
\right)\epsilon=\epsilon\,,
\eeq
showing that the D6-brane is wrapping a non trivial two-cycle inside the
six dimensional manifold ${\cal Y}_6$. This cycle mixes the $S^2$ of
the eight dimensional geometry (\ref{8dmetric})  with the external $S^3$
(along which, the reduction to 8d SUGRA was done). Thus, the phase
$\alpha$ implements the twisting we mentioned in section \ref{gcintro}
(below (\ref{t11metric})).

Next, by inserting projection
(\ref{betaproj}) and equation (\ref{betadef}) into (\ref{eq3hat}),
one gets:
\bear
&&\left\{-2\lambda'+\left[-{2\over3}\,e^{\phi-2\lambda-2h}\left(g^2-1\right)+
{1\over3}\,e^{-\phi}\left(e^{-4\lambda}-e^{2\lambda}\right)\right]\beta+2e^{-h}
\,\sinh\left(3\lambda\right)\,g\,\tilde{\beta}\right\}\,\epsilon= \rc\rc
&&=\left\{2e^{-h}\,
\sinh\left(3\lambda\right)\,g\,\beta-\left[-e^{\phi-2\lambda-2h}\left(g^2-1\right)
+{1\over4}e^{-\phi}\,\left(e^{-4\lambda}-2e^{2\lambda}\right)\right]
\tilde{\beta}\right\}\Gamma_{\hat11}\,\epsilon\,,
\label{eqlambda}
\eear
which consists of an equation for $\lambda'$ and an algebraic
constraint:
\bear
\lambda'=\left[-{1\over3}\,e^{\phi-2\lambda-2h}\left(g^2-1\right)+{1\over6}
\,e^{-\phi}\left(e^{-4\lambda}-e^{2\lambda}\right)\right]\beta+e^{-h}\,
\sinh\left(3\lambda\right)\,g\,\tilde{\beta}\,,  \label{lambdapr}\\
\rc e^{-h}\,\sinh\left(3\lambda\right)\,g\,\beta+
\left[\,{1\over2}\,e^{\phi-2\lambda-2h}\left(g^2-1\right)-{1\over8}\,
e^{-\phi}\,\left(e^{-4\lambda}-2e^{2\lambda}\right)\right]\tilde{\beta}=0\,.
\label{constr1}
\eear
In order to get an equation for $h'$ we can use equation
(\ref{eq13hat}) to eliminate $\phi'$ from equation (\ref{eq1}), hence
we get:
\bear
h'\,\epsilon=-{2\over3}e^{\phi+\lambda-h}\,g'\,\Gamma_{1\hat1}\,\epsilon+
e^{-h}\,\cosh\left(3\lambda\right)\,g\,\Gamma_r\,\Gamma_{1\hat2\hat3}\,\epsilon
+\rc+{1\over6}\left[5e^{\phi-2\lambda-2h}\,\left(g^2-1\right)-{1\over4}e^{-\phi}
\left(e^{-4\lambda}+2e^{2\lambda}\right)\right]\Gamma_r\,\Gamma_{\hat1\hat2\hat3}
\,\epsilon\,,
\label{hpr}
\eear
which after inserting eq. (\ref{betaproj}) renders a differential equation 
for $h'$ and a new algebraic constraint:
\bear
&&h'=-e^{-h}\,\cosh\left(3\lambda\right)\,g\,\tilde{\beta}+{1\over6}\left[
-5e^{\phi-2\lambda-2h}\,\left(g^2-1\right)+{1\over4}\,e^{-\phi}
\left(e^{-4\lambda}+2e^{2\lambda}\right)\right]\beta\,,\rc
\label{hpreq}\\\rc
&&-e^{-h}\,\cosh\left(3\lambda\right)\,g\,\beta+\left[\,{1\over2}\,
e^{\phi-2\lambda-2h}\,\left(g^2-1\right)-{1\over8}\,e^{-\phi}
\left(e^{-4\lambda}+2e^{2\lambda}\right)\right]\tilde{\beta}=0\,\,,
\label{constr2}
\eear
where we have used eq. (\ref{betatildef}) to get rid of $g'$.

To sum up, from equations (\ref{susyeq2}) we have got a system of
differential equations, namely (\ref{betadef}), (\ref{betatildef}),
(\ref{lambdapr}), and (\ref{hpreq}); two algebraic constraints:
(\ref{constr1}) and  (\ref{constr2}); and the projection (\ref{betaproj}).
This projection is compatible with (\ref{2cproj}) and both leave
unbroken eight supercharges. As it was shown in \cite{Twist},  
the algebraic constraints have two different
solutions resulting in two truncations of the system of differential
equations and therefore, in two different internal manifolds ${\cal
Y}_6$.  One solution leads to the generalized resolved
conifold and the other to the generalized deformed conifold.

It remains to determine the
radial dependence of the 10d spinor; it will be fixed by equation
(\ref{susyeq3}), which for the spin connection (\ref{spincon}) reduces to:
\beq
e^{\phi\over3}\,\epsilon'+{1\over2}\left(\omega^{\hat1\,1}_{
r}\,\Gamma_{\hat11}+
\omega^{\hat2\,2}_{r}\,\Gamma_{\hat22}\right)\epsilon=0\,,
\label{radeq1}
\eeq
where $\epsilon'={d\epsilon\over dr}$, and we have taken into
account that $\omega^{a\,b}_{\tilde
r}=e^{-{\phi\over3}}\,\omega^{a\,b}_r$.  By inserting projection
(\ref{2cproj}) and the corresponding components of the spin
connection into this last equation one arrives at
\beq
\epsilon'+e^{\phi+\lambda-h}\,g'\,\Gamma_{\hat11}\,\epsilon=0\,\,,
\label{radeq2}
\eeq
and after inserting (\ref{betaprojsolv}), it results in the two following
equations:
\bear
\tilde{\epsilon}\,'&=&0\,\,,\label{radep}\\
\alpha'&=&-2e^{\phi+\lambda-h}\,g'\,\,  . \label{alphaeq}
\eear
This last equation determines the radial dependence of the phase $\alpha$,
while (\ref{radep}) implies that the spinor
$\tilde\epsilon$ is independent of
$r$. Therefore, the 10d Killing spinor $\epsilon$ can be written as:
\beq
\epsilon=e^{-{\alpha\over2}\Gamma_{\hat11}}\,\tilde{\epsilon}\;\;,\;\;
\label{genconksp}
\eeq
where $\tilde{\epsilon}$ is a constant 10d spinor satisfying the
projections:
\beq
\Gamma_r\Gamma_{\hat1\hat2\hat3}\,\tilde{\epsilon}=-\tilde{\epsilon}\;,\quad
\Gamma_{12}\,\tilde{\epsilon}=-\Gamma_{\hat1\hat2}\,\tilde{\epsilon}\,.
\eeq
As mentioned above, both projections are compatible since
$\left[\,\Gamma_r\Gamma_{\hat1\hat2\hat3}\,,\Gamma_{12\hat1\hat2}\,\right]
=0$. Thus, the 10d SUGRA solution (\ref{10dconif}) leaves unbroken
eight supersymmetries.

\section{Solving the equations}
\label{genconsolv}
\setcounter{equation}{0}
In this section we will sum up the solutions (obtained in
\cite{Twist}) for the system of  differential equations and algebraic
constraints we got in the last section (eqs. (\ref{betadef}),
(\ref{betatildef}), (\ref{lambdapr}), (\ref{constr1}), (\ref{hpreq})  and 
(\ref{constr2})). This system determines the geometry of the 6d
internal part ${\cal Y}_6$ of the ten dimensional geometry 
$R^{1,3}\times{\cal Y}_6$ (\ref{10dconif}). In this approach we get
the metrics of the generalized resolved and deformed conifold
written in a form which will be very useful in the next chapters when 
computing the Killing spinors of several 10d backgrounds. 

The algebraic constraints (\ref{constr1}) and  (\ref{constr2}) can be combined 
to get:
\beq
\tan\alpha={\tilde{\beta}\over\beta}=-2e^{\phi+\lambda-h}\,g=
{e^{-3\lambda-h}\,g\over e^{\phi-2\lambda-2h}\left(g^2-1\right)
-{1\over4}e^{\phi-4\lambda}}\;.
\eeq
The first part of this equation allows us to write $\alpha$ in terms of
the remaining functions, while the last equality yields the following
constraint:
\beq
g\left[g^2-1+{1\over4}\,e^{-2\phi-2\lambda+2h}\right]=0\;,
\label{constr12}
\eeq 
which clearly has two solutions. One of them is $g=0$, corresponding to
$\tilde{\beta}=0\; , \;\beta=1$ (then $\alpha=0$). In this case the
system of differential equations reduces to the one studied in
\cite{gaugedsugra}, whose integral leads to the generalized resolved
conifold \cite{GRCIII}:
\bear
ds_6^2&=&\left[\kappa(\rho)\right]^{-1}d\rho^2+{\rho^2\over9}\,\kappa(\rho)\left(d\psi+
\sum_{a=1}^2\cos\theta_a\,d\phi_a\right)^2+\rc
&+&{1\over6}\left[\left(\rho^2+6a^2\right)
\left(d\theta_1^2+\sin^2\theta_1\,d\phi_1^2\right)+
\rho^2\left(d\theta_2^2+\sin^2\theta_2\,d\phi_2^2\right)\right]\;,
\label{genres}
\eear
with $\kappa(\rho)$ being:
\beq
\kappa(\rho)={\rho^6+9a^2\rho^4-b^6\over\rho^6+6a^2\rho^4}\;.
\label{kappa}
\eeq
where $a$ and $b$ are constants of integration. In equation
(\ref{genres})
$\rho$ is a new radial variable and
$(\theta_2,\phi_2,\psi)$ are the angular coordinates of the
external $S^3$. We have taken into account that in terms of these
coordinates the left-invariant one-forms of the
three-sphere (referred to above eq. (\ref{11dconif})) can be written as
\bear
w^1&=&\sin\psi\sin\theta_2\,d\phi_2\,+\,\cos\psi\,d\theta_2\,\,,\rc
w^2&=&-\cos\psi\sin\theta_2\,d\phi_2\,+\,\sin\psi\,d\theta_2\,\,,\rc
w^3&=&d\psi\,+\,\cos\theta_2\,d\phi_2\,\,,
\label{omegaforms}
\eear
with $\theta_2\in[0,\pi]$, $\phi_2\in[0,2\pi)$ and $\psi\in[0,4\pi)$. 
So (\ref{genres}) can be equivalently written as
\bear
ds_6^2&=&\left[\kappa(\rho)\right]^{-1}d\rho^2+{\rho^2\over9}\,
\kappa(\rho)\left(\sigma^3+\omega^3\right)^2+\rc\rc
&+&{1\over6}\left[\left(\rho^2+6a^2\right)
\left(\left(\sigma^1\right)^2+\left(\sigma^2\right)^2\right)+
\rho^2\left(\left(\omega^1\right)^2+\left(\omega^2\right)^2\right)\right]\;.
\label{genres2}
\eear
The constants of integration $a$ and $b$ (appearing in (\ref{kappa}))
provide the generalized resolution of the conifold singularity
\cite{GRCI}-\cite{GRCIII}. The case
$b=0$ corresponds to the resolved conifold: it is easy to see that for
$\rho=0$ we get an $S^2$ of finite size $a^2$ instead of a singularity.
For $a=0$,
$b=0$ we get back the metric of the singular conifold written in the
following form:
\beq
ds_6^2=d\rho^2+{\rho^2\over9}\,
\left(\sigma^3+\omega^3\right)^2+{\rho^2\over6}\left[
\left(\sigma^1\right)^2+\left(\sigma^2\right)^2+
\left(\omega^1\right)^2+\left(\omega^2\right)^2\right]\;.
\label{singconif}
\eeq
The second solution of the constraint (\ref{constr12}) leads to a non
trivial relation between $g$ and the remaining functions of the ansatz,
 namely:
\beq
g^2=1-{1\over4}\,e^{-2\phi-2\lambda+2h}\;.
\label{constrsol}
\eeq
The corresponding values of $\beta$ and $\tilde{\beta}$ are:
\beq
\beta={1\over2}\,e^{-\phi-\lambda+h}\quad,\quad
\tilde{\beta}=-g\;.
\label{betasol2}
\eeq
Plugging these results into the differential equations (\ref{betadef}),
(\ref{lambdapr}), and (\ref{hpreq}) one arrives at
the first order system:
\bear
\phi'&=&{1\over8}\,e^{-2\phi+\lambda+h}\,,\rc\rc
\lambda'&=&{1\over24}\,e^{-2\phi+\lambda+h}-{1\over2}\,e^{3\lambda-h}+
{1\over2}\,e^{-3\lambda-h}\,,\rc\rc
h'&=&-{1\over12}\,e^{-2\phi+\lambda+h}+{1\over2}\,e^{3\lambda-h}+
{1\over2}\,e^{-3\lambda-h}\,,
\eear
while from (\ref{betatildef}) one gets:
\beq
g'=-{1\over4}\,e^{-2\phi+\lambda+h}\,g\,.
\label{gendefsys}
\eeq
These equations can be straightforwardly solved, resulting:
\bear
e^\phi&=&\hat\mu\left(\cosh\tau\right)^{1\over2}\,,\rc\rc
e^\lambda&=&\left({3\over2}\right)^{1\over6}\left(\cosh\tau\right)^{1\over6}
K(\tau)^{1\over2}\,,\rc\rc
e^h&=&2^{5\over6}\,3^{1\over6}\,\hat\mu\,{\sinh\tau\over(\cosh\tau)^{1\over3}}\,
K(\tau)^{1\over2}\,,\rc\rc
g&=&{1\over\cosh\tau}\,,
\label{defconsysol}
\eear
with
\beq
\\K(\tau)={\left(\sinh(2\tau)-2\tau+C\right)^{1\over3}\over
2^{1\over3}\sinh\tau}\;,
\label{kdefcon}
\eeq
$\hat\mu$ and $C$ are constants of integration and $\tau$ is a
new radial coordinate defined by means of the  differential equation:
\beq
d\tau={1\over2}\,e^{2\lambda-\phi}\,dr\,.
\label{gctaudef}
\eeq
After inserting the solution (\ref{defconsysol}) the 10d metric
(\ref{10dconif}) becomes:
\beq
ds^2_{10}=dx^2_{1,3}\,+
ds^2_6\,,
\eeq
with
\bear
ds^2_{6}={1\over 2}\,\mu^{{4\over 3}}\,K(\tau)
&\Bigg[&{1\over 3 K(\tau)^3}\,\Big(\,d\tau^2\,+\, (w^3+\sigma^3)^2\,\Big)\,
+\,{\sinh^2\tau\over 2\cosh\tau}\,\Big(\,(\sigma^1)^2+(\sigma^2)^2\,\Big)\,+\rc\rc
&&+\,{\cosh\tau\over 2}\Big[\,\Big(\,w^1+{\sigma^1\over \cosh\tau}\,\Big)^2\,+\,
\Big(\,w^2+{\sigma^2\over \cosh\tau}\,\Big)^2\,\Big]\,\Bigg]\,\,,
\label{10dgendefcon}
\eear
which is the metric of the generalized deformed
conifold
\cite{GRCII}. For $C=0$ it describes the deformed conifold, with
$\mu$ (which is just: $\mu=2^{11\over4}\,3^{1\over4}\,\hat\mu$) being the
deformation parameter. It is not difficult to write this 6d metric in the
standard form of
\cite{KS}:
\bear
ds^2_6={1\over2}\mu^{{4\over3}}K(\tau)\Bigg[{1\over3K(\tau)^3}\left(d\tau^2
+\left(g^5\right)^2\right)+\sinh^2\left({\tau\over2}\right)\left(
\left(g^1\right)^2+\left(g^2\right)^2\right)+\rc\rc
+\cosh^2\left({\tau\over2}\right)
\left(\left(g^3\right)^2+\left(g^4\right)^2\right)\Bigg]\;,
\label{gendef}
\eear
where we have defined the following set of one-forms:
\bear
g^1&=&{1\over\sqrt2}\left(\omega^2-\sigma^2\right)\;,\quad
g^2={1\over\sqrt2}\left(\sigma^1-\omega^1\right)\;,\rc\rc
g^3&=&{-1\over\sqrt2}\left(\sigma^2+\omega^2\right)\;,\quad
g^4={1\over\sqrt2}\left(\sigma^1+\omega^1\right)\;,\rc\rc
g^5&=&\sigma^3+\omega^3\;.
\label{gengforms}
\eear
Furthermore, one can easily see that for $\tau\to0$ the metric of the
deformed conifold degenerates into $d\Omega^2_3={1\over2}\mu^{4\over3}
\left({2\over3}\right)^{1\over3}\left[{1\over2}\left(g^5\right)^2+
\left(g^3\right)^2+\left(g^4\right)^2\right]$, which, as expected, 
is the metric of a round $S^3$.

\subsection{Killing spinors of the deformed conifold}
\label{gcdefconks}

It will become useful to write down explicitly the Killing spinors of the
10d metric $R^{1,3}\times{\cal Y}_6$ when ${\cal Y}_6$ corresponds to
the deformed conifold. One just have to insert the particular solution
(\ref{defconsysol}) corresponding to the deformed conifold into the
general expression for the Killing spinors written in equation
(\ref{genconksp}). Thus, one gets:
\beq
\epsilon=e^{-{\alpha\over2}\Gamma_{\hat11}}\,\eta\;\;,\;\;
\label{gcdcksp}
\eeq
where $\eta$ is a constant 10d spinor satisfying the projections
\beq
\Gamma_\tau\Gamma_{\hat1\hat2\hat3}\,\eta=-\eta\;,\quad
\Gamma_{12}\,\eta=-\Gamma_{\hat1\hat2}\,\eta\,,
\label{gcdckspr}
\eeq
and the angle $\alpha$ is given by:
\beq
\sin\alpha=-{1\over\cosh\tau}\;,\quad\cos\alpha={\sinh\tau\over\cosh\tau}\,.
\label{gcdcalpha}
\eeq
As before, $\Gamma_a\,,\;(a=x^\alpha,\tau,1,2,\hat1,\hat2,\hat3)$ are
constant Dirac matrices associated to the frame  (\ref{conifbasis}),
which for the particular solution (\ref{defconsysol}) becomes:
\bear
&&e^{x^\alpha}=dx^\alpha \;,\;(\alpha=0,1,2,3)\;,\quad
e^{\tau}={\mu^{{2\over 3}}\over \sqrt{6}\,K(\tau)}\,\,d\tau\;,\rc\rc
&&e^{i}={\mu^{{2\over 3}}\,\sqrt{K(\tau)}\over 2}\,\,
{\sinh\tau\over \sqrt{\cosh\tau}}\,\,\sigma^i\;,\; (i=1,2)\;,\rc\rc
&&e^{\hat i}={\mu^{{2\over 3}}\,\sqrt{K(\tau)}\over 2}\,\,
\sqrt{\cosh\tau}\,\,\Big(\,w^i+{\sigma^i\over \cosh\tau}\,\Big)
 \;,\; (i=1,2)\;,\rc\rc
&&e^{\hat 3}={\mu^{{2\over 3}}\,\over \sqrt{6}\,\, K(\tau)}\,\,
(w^3+\sigma^3)\;.
\label{gcdefconframe}
\eear


\chapter{Killing spinors of the Klebanov-Witten model}
\label{kwcp}
\section{Introduction}
\setcounter{equation}{0}
In this chapter we will thoroughly present the computation of the Killing 
spinors of the Klebanov-Witten (KW) model \cite{KW}. 
We will briefly introduce the background and,
using some results of the previous chapter, we will construct a one-form
frame in which we expect that the Killing spinors do not depend 
on the angular coordinates of the conifold. Before solving the equations
resulting from the vanishing of the SUSY variations, we will have to
express the fields of the model in that new frame and also determine the
form of the spin connection. Finally, in order to get the explicit form of
the Killing spinors of the KW model when global coordinates are used for
the $AdS_5$ part of the metric, we will repeat the 
calculations for the corresponding one-form frame. In both cases we will
be able to write the Killing spinors of the theory in terms of a constant
10d spinor satisfying two independent (and compatible) projections, which
reduce the number of independent components of the spinor and thus, the
number of unbroken supercharges, from 32 to 8 real components as it was
expected.

This calculation
was schematically published in ref. \cite{SSP}, since the explicit
expression of the Killing spinors was essential for the kappa symmetry
analysis carried out there.

\subsection{The Klebanov-Witten model}
The so-called Klebanov-Witten background is constructed in ref.
\cite{KW} by placing a stack of $N$ D3-branes at the apex of the singular
conifold.  By adding four Minkowski coordinates to the conifold we
construct a Ricci flat ten dimensional metric. Let us now place a stack
of $N$ coincident D3-branes extended along the Minkowski coordinates and
located at the singular point of the conifold. The resulting IIB
supergravity solution is the KW model. The corresponding
near-horizon metric and Ramond-Ramond selfdual five-form are given by
\bear
ds^2_{10}&=&[h(r)]^{-{1\over 2}}\,dx^2_{1,3}\,+\,[h(r)]^{{1\over 2}}\,
\big(\,dr^2\,+\,r^2\,ds^2_{T^{1,1}}\,\big)\,\,,\rc\rc
h(r)&=&{L^4\over r^4}\,\,,\rc\rc
g_s\,F^{(5)}&=&d^4x\,\wedge dh^{-1}\,+\,{\rm Hodge\,\,\, dual}\,\,,\rc\rc
L^4&=&{27\over 4}\,\pi g_s N\alpha'^2\,\,.
\label{KW}
\eear
By plugging the explicit form of the warp factor into the metric, it can
be written as
\beq
ds^2_{10}\,=\,{r^2\over L^2}\,dx^2_{1,3}\,+\,{L^2\over r^2}\,dr^2\,+\,
L^2\,ds^2_{T^{1,1}}\,\,,
\label{adspoincare}
\eeq
which corresponds to the $AdS_5\times T^{1,1}$ space.

It was shown in ref. \cite{KW} that the gauge theory dual to
this supergravity background is an ${\cal N}=1$ superconformal field
theory with some matter multiplets.

\section{Killing spinors}
\setcounter{equation}{0}

To obtain the explicit form of the Killing spinors, one has to look at the
supersymmetry variations of the dilatino and gravitino (see eq.
(\ref{sugra})). Since the dilaton is constant and there is no
three-form flux, the variation of the dilatino vanishes trivially
($\delta\lambda=0$). We are left with the
equations:
\beq
\delta\psi_{M}\,=\,D_{M}\,\epsilon\,+\,{i\over 1920}\,
F_{N_1\cdots N_5}^{(5)}\,\Gamma^{N_1\cdots N_5}
\Gamma_{M}\,\epsilon=0\,.
\label{sugrakw}
\eeq
The final result of the calculation is greatly simplified if we choose
the basis of the frame one-forms that arises naturally when the $T^{1,1}$
metric is written as in eq. (\ref{singconif}) of the previous chapter,
namely: 
\beq
ds^2_{T^{1,1}}\,=\,{1\over 6}\,\big(\,(\sigma^1)^2\,+\,
(\sigma^2)^2\,+\,(w^1)^2\,+\,(w^2)^2\,\big)\,+\,
{1\over 9}\,\big(\,w^3+\sigma^3\,\big)^2\,\;,
\eeq
with the one-forms $\sigma^i$ and $\omega^i$ being given by equations
(\ref{s2forms}) and (\ref{omegaforms}). 
Let us recall that this form of writing the $T^{1,1}$ metric comes up in
the framework of the eight dimensional gauged supergravity obtained from
a Scherk-Schwarz reduction of eleven dimensional supergravity on an
$SU(2)$ group manifold \cite{SS}. Indeed, it was obtained
as the gravity dual of D6-branes wrapping an
$S^2$ inside a K3 manifold \cite{gaugedsugra}. Then, from the
consistency of the reduction leading to the gauged supergravity, the
Killing spinors should not depend on the coordinates of the $SU(2)$
external manifold and, actually, in the one-form basis we will use
they do not depend on any angular coordinate of the
$T^{1,1}$ space. Accordingly, let us consider the following frame for the
ten dimensional metric (\ref{KW}):
\bear
&&e^{x^{\alpha}}={r\over L}\,dx^{\alpha}\;,\;
(\alpha=0,1,2,3)\;,\quad
e^{r}={L\over r}\,dr\,\,,\rc\rc
&&e^{i}={L\over \sqrt{6}}\,\,\sigma^i\;,\; (i=1,2)\;,\rc\rc
&&e^{\hat i}={L\over \sqrt{6}}\,\,w^i\;,\;(i=1,2)\;,\rc\rc
&&e^{\hat 3}={L\over 3}\,\,(\,w^3+\sigma^3\,)\,\,.
\label{kwframe}
\eear
In this frame, the selfdual RR five-form reads:
\beq
g_s\,F^{(5)}={4\over L}\left(e^{x^0}\wedge e^{x^1}\wedge e^{x^2}\wedge
e^{x^3}\wedge e^{r}+e^1\wedge e^2\wedge e^{\hat1}\wedge e^{\hat2}\wedge
e^{\hat3}\right)\,.
\label{f5}
\eeq
\subsection{Spin connection}
\label{sconn}
Let us recall that, as we have mentioned in the previous chapter, the
covariant derivative appearing in the SUSY equations (\ref{sugrakw})
can be written as
\beq
D_{\tilde m}=\partial_{\tilde m}+{1\over4}\,\omega^{a\,b}_{\tilde m
}\,\Gamma_{a\,b}\,,
\label{kwcodvtdef}
\eeq
in the frame formed by the differentials of the coordinates. So
$\partial_{\tilde m}$ denotes the derivative with respect to
$X^{\tilde m}$ and, as before,
$\omega^{a\,b}_{\tilde m}$ stands for the components of the spin
connection one-form in that basis, namely: 
\beq
\omega^{a\,b}=\omega^{a\,b}_{\tilde m}\,dX^{\tilde m}\,.
\label{kwspinconndef}
\eeq
Then, in order to solve equations
(\ref{sugrakw}) we need the spin connection one-form $\omega^{a\,b}$
of the background (where
$a$ and
$b$ are indices running along the one-form basis  (\ref{kwframe})).
We will compute the spin connection for a metric of the form
(\ref{KW}) but with a generic warp factor $\tilde{h}(r)$ instead of
$h(r)={L^4\over r^4}$. Thus, the
corresponding frame is:
\bear
&&\tilde{e}^{{\,x^\alpha}}=\tilde{h}^{-{1\over4}}\,dx^{\alpha}
\;,\;(\alpha=0,1,2,3)\;,\quad
\tilde{e}^{r}\;=\;\tilde{h}^{1\over4}\,dr\;,\rc\rc
&&\tilde{e}^{\,i}=\tilde{h}^{1\over4}\,{r\over\sqrt6}\,\sigma^i
\;,\; (i=1,2)\;,\rc\rc
&&\tilde{e}^{\,\hat i}=\tilde{h}^{1\over4}\,{r\over \sqrt{6}}\,w^i\;,\;
(i=1,2)\;,\rc\rc
&&\tilde{e}^{\,\hat 3}=\tilde{h}^{1\over4}\,{r\over
3}\left(w^3+\sigma^3\right)\;,
\label{kwgenframe}
\eear
which does not only correspond to the current background (for
$\tilde{h}(r)={L^4\over r^4}$), but it also describes the
Klebanov-Tseytlin metric, where $\tilde{h}(r)$ is a more involved
function of the radial coordinate as one will see in the next chapter.
Let us call $\tilde\omega^{a\,b}$ to the spin connection
corresponding to the generic frame (\ref{kwgenframe}); substituting the
derivatives of the one-forms of the frame (\ref{kwgenframe}) together
with a generic ansatz for
$\tilde\omega^{a\,b}$ into  the Cartan equations (\ref{cartan}) we get:
\bear
&&\tilde\omega^{x^\alpha\,r}=\left(\tilde{h}^{-{1\over4}}\right)'
\,\tilde e^{x^\alpha}\;,\;
(\alpha=0,1,2,3)\;,\rc\rc
&&\tilde\omega^{s\,r}=\tilde h ^{-{1\over4}}\left({1\over
r}+{1\over4}\,\tilde h'\,\tilde h^{-1}\right)\,\tilde e^s\;,\;
(s=1,2,\hat1,\hat2,\hat3)\;,\rc\rc
&&\tilde\omega^{1\,2}={1\over
r}\,\tilde{h}^{-{1\over4}}\,\tilde e^{\hat3}-{\sqrt6\over
r}\,\cot\theta_1\,\tilde{h}^{-{1\over4}}\,\tilde e^2\;,\rc\rc
&&\tilde\omega^{\hat1\,\hat2}={2\over
r}\,\tilde{h}^{-{1\over4}}\,\tilde e^{\hat3}-{\sqrt6\over
r}\,\cot\theta_1\,\tilde{h}^{-{1\over4}}\,\tilde e^2\;,\rc\rc
&&\tilde\omega^{\hat1\,\hat3}=-{1\over
r}\,\tilde{h}^{-{1\over4}}\,\tilde e^{\hat2}\;,\quad
\tilde\omega^{\hat2\,\hat3}={1\over
r}\,\tilde{h}^{-{1\over4}}\,\tilde e^{\hat1}\,\,,\rc\rc
&&\tilde\omega^{\hat3\,2}={1\over
r}\,\tilde{h}^{-{1\over4}}\,\tilde e^1\;,\quad
\tilde\omega^{\hat3\,1}=-{1\over
r}\,\tilde{h}^{-{1\over4}}\,\tilde e^2\,\,.
\label{gencon}
\eear
We have expressed the resulting one-form in the frame
(\ref{kwgenframe}), for, as one will see below, it will be more useful
to work directly in that frame. Applying this result to the present
background, i.e.  plugging
$\tilde{h}(r)={L^4\over r^4}$ into (\ref{gencon}), the spin connection of
the Klebanov-Witten background, written directly in the 
frame (\ref{kwframe}), reads:
\bear
&&\omega^{x^\alpha\,r}={1\over L}
\,e^{x^\alpha}\;,\; (\alpha=0,1,2,3)\;,\rc\rc
&&\omega^{1\,2}={1\over L}\,e^{\hat3}-{\sqrt6\over
 L}\,\cot\theta_1\,e^2\,\,,\rc\rc
&&\omega^{\hat1\,\hat2}={2\over L}\,e^{\hat3}-{\sqrt6\over
 L}\,\cot\theta_1\,e^2\,\,,\rc\rc
&&\omega^{\hat2\,\hat3}={1\over L}\,e^{\hat1} \; ,\quad
\omega^{\hat1\,\hat3}=-{1\over L}\,e^{\hat2}\,\,,\rc\rc
&&\omega^{\hat3\,1}=-{1\over L}\,e^2\; ,\quad
\omega^{\hat3\,2}={1\over L}\,e^1\,\,.
\label{conkw}
\eear
One should keep in mind that the components in the coordinate basis,
\ie \ (\ref{kwspinconndef}), can be easily computed in terms of the
ones in (\ref{conkw}): $\omega^{a\,b}_{\tilde
m}=E^{\,c}_{\,\tilde m}\,\omega^{a\,b}_c$.
$E^{\,c}_{\,\tilde m}$ are the coefficients appearing in the
expression of the frame one-forms (\ref{kwframe}) in terms of the
differentials of the coordinates: $e^a=E^{\,a}_{\,\tilde n}\,
dX^{\tilde n}$.

\subsection{Determining the Killing spinors}
Once we have computed the form of the spin connection, we can go back to equations (\ref{sugrakw}). After
substituting the  selfdual five-form (\ref{f5}) they become:
\beq
D_M\,\epsilon+{i\over
4L}\left(\Gamma^{\,x^0\,x^1\,x^2\,x^3\,r}+
\Gamma^{\,1\,2\,\hat1\,\hat2\,\hat3}\right)\Gamma_{M}\,\epsilon=0\,.
\label{sugrakw2}
\eeq
$\Gamma^{\,a}\,,\;(a=x^\alpha,r,1,2,\hat1,\hat2,\hat3)$ are
constant Dirac matrices associated to the frame (\ref{kwframe}). Using the
identity 
satisfied by the chiral 10d spinors:
$\Gamma_{x^0...x^3}\Gamma_r\Gamma_{12\hat1\hat2\hat3}\,\epsilon=-\epsilon$,
these last equations can be written as
\bear
D_{\mu}\,\epsilon&+&{i\over
2L}\,\Gamma^{\,x^0\,x^1\,x^2\,x^3\,r}\,\Gamma_{\mu}\,\epsilon=0\;,\;
(\mu=x^0,x^1,x^2,x^3,r)\,\,,  \label{kwads}  \\\rc
D_{s}\,\epsilon&+&{i\over
2L}\,\Gamma^{\,1\,2\,\hat1\,\hat2\,\hat3}\,\Gamma_{s}\,\epsilon=0\;,\;
(s=1,2,\hat1,\hat2,\hat3)\,\,,
\label{kwt11}
\eear
working directly in the frame (\ref{kwframe}).

Since the spin connection does not mix $AdS_5$ with $T^{1,1}$ components,
we can solve these two sets of equations separately, though the
projections we get from both sets must be compatible.
Let us begin with the $AdS_5$ equations (\ref{kwads}); we expect the
spinor to depend on the $AdS_5$ coordinates so 
we will need to use the equality:
\beq
D_\mu\,\epsilon=(E^{\,\mu}_{\,\tilde \nu})^{-1}\,D_{\tilde
\nu}\,\epsilon\,.
\label{kwcdvtrns}
\eeq
Then, by inserting the spin connection (\ref{conkw}) and applying
this last expression, one can bring equations (\ref{kwads}) into the
form:
\bear
&&\partial_{x^{\alpha}}\,\epsilon=-{r\over
2L^2}\,\Gamma_{x^{\alpha}}\,\Gamma_r\left(1-\Gamma_*\right)\,\epsilon
\;,\;(\alpha=0,1,2,3)\,\,,\label{kwadsx} \\ \rc
&&\partial_r\,\epsilon={1\over2r}\,\Gamma_*\,\epsilon\,\, ,
\label{kwadsr}
\eear
with $\Gamma_*$ being defined as
\beq
\Gamma_*\equiv i\,\Gamma_{\,x^0\,x^1\,x^2\,x^3}\,\,.\label{gammastar}
\eeq
These equations have two solutions:
\bear
&&\epsilon_1=\sqrt{r}\,\epsilon_+\;,\quad
\Gamma_*\,\epsilon_+=\epsilon_+\,\,,
\label{epsplus}\\\rc
&&\epsilon_2=i\left({1\over \sqrt r}\,\Gamma_r\,\Gamma_*+{\sqrt{r}\over
L^2}\,x^{\alpha}\Gamma_{x^{\alpha}}\right)\epsilon_- \;,\quad
\Gamma_*\,\epsilon_-=-\epsilon_-\,\,, \label{epsminus}
\eear
where $\epsilon_{\pm}$ are 10d spinors independent of the $AdS_5$
coordinates. The parameterization of the dependence of $\epsilon_2$ on
the $AdS_5$ coordinates is the same as in ref. \cite{LPT}. Each solution
is obviously
${1\over2}$ SUSY.  After defining
$\eta_-\equiv-i\,\Gamma_r\,\epsilon_-$ and
$\eta_+\equiv
\epsilon_+$, $\epsilon_1$ and $\epsilon_2$ become:
\bear
&&\epsilon_1=\sqrt{r}\,\eta_+\,\,,\label{etaplus}  \\\rc
&&\epsilon_2=\left({1\over \sqrt r}+{\sqrt{r}\over
L^2}\,x^{\alpha}\,\Gamma_r
\,\Gamma_{x^{\alpha}}\right)\eta_-\,\,,
\label{etaminus}
\eear
with $\eta_{\pm}$ being 10d spinors independent of the $AdS_5$
coordinates and satisfying:
\beq
\Gamma_*\,\eta_{\pm}=\pm\,\eta_{\pm}\,\,.
\label{etaproj}
\eeq
Thus, we have got two independent solutions of the supersymmetry
equations for the
$AdS_5$ part (\ref{kwads}), each one being ${1\over2}$
SUSY. Notice that whereas for the first solution, the
spinor
$\epsilon_1$ is independent of the coordinates $x^{\alpha}$ and satisfies
$\Gamma_*\,\epsilon_1=\epsilon_1\,$; for the second solution, $\epsilon_2$
does depend on $x^{\alpha}$ and it is not an eigenvector of $\Gamma_*$.
Both solutions can be unified in the following expression:
\beq
\epsilon=r^{\Gamma_*\over2}\left[1+{1\over2L^2}\,x^{\alpha}\,\Gamma_r\,
\Gamma_{x^{\alpha}}\left(1-\Gamma_*\right)\right] \eta\,\,,
\label{adskilling}
\eeq
where $\eta$ is a 10d spinor constant along $AdS_5$ and the dependence
on the $AdS_5$ coordinates is parameterized as in ref. \cite{LPT} . If we
decompose $\eta$ according to the different eigenvalues of the matrix
$\Gamma_*$: $\Gamma_*\,\eta_{\pm}=\pm\,\eta_{\pm}$, we recover the
independent solutions (\ref{etaplus}) and (\ref{etaminus}).

It remains to solve the second subset of supersymmetry equations
(\ref{kwt11}), the ones depending on the $T^{1,1}$ part of the metric.
Let us insert the solution we have found (\ref{adskilling}) into that
equations. The $\Gamma$-matrices appearing in (\ref{adskilling}) commute
with the even number of $T^{1,1}$ $\Gamma$-matrices in (\ref{kwt11}),
resulting:
\beq
D_{s}\,\eta+{i\over
2L}\Gamma^{\,1\,2\,\hat1\,\hat2\,\hat3}\,\Gamma_{s}\,\eta=0\;,\;
(s=1,2,\hat1,\hat2,\hat3)\,.
\label{kwt11eq2}
\eeq
One can check, using the spin connection given in (\ref{conkw}), that
these five equations are solved by a constant 10d spinor $\eta$
satisfying the usual projections of the $T^{1,1}$ (see \cite{RCN}):
\beq
\Gamma_{12}\,\eta=i\,\eta\;,\quad
\Gamma_{\hat1\hat2}\,\eta=-i\,\eta\,.
\label{t11proj}
\eeq
Therefore, the Killing spinors of the model are given by 
the expression (\ref{adskilling}) in terms of a constant 10d spinor
satisfying the projections (\ref{t11proj}). Furthermore, notice that the
matrix multiplying $\eta$ in eq. (\ref{adskilling}) commutes with
$\Gamma_{12}$ and $\Gamma_{\hat1\hat2}$ so the spinor $\epsilon$ also
satisfies the projections (\ref{t11proj}), namely:
\beq
\Gamma_{12}\,\epsilon=i\,\epsilon\;,\quad
\Gamma_{\hat1\hat2}\,\epsilon=-i\,\epsilon\,.
\label{t11projeps}
\eeq
It is clear from eqs. (\ref{adskilling}) and (\ref{t11proj}) that our
system is $1/4$ supersymmetric, \ie\ it preserves 8 supersymmetries, as
it corresponds to the supergravity dual of an ${\cal N}=1$
superconformal field theory in four dimensions.
\subsection{Killing spinors using global coordinates}

It is also interesting to
write down the form of the Killing spinors when global coordinates are
used for the
$AdS_5$ part of the metric. In these coordinates the ten dimensional
metric takes the form:
\beq
ds^2_{10}\,=\,L^2\,\Big[\,-\cosh^2\rho \,\,dt^2\,+\,d\rho^2\,+\,
\sinh^2\rho\,\,d\Omega_3^2\,\Big]\,+\, L^2\,ds^2_{T^{1,1}}\,,
\label{globalads}
\eeq
where $d\Omega_3^2$ is the metric of a unit three-sphere parameterized by
three angles $(\alpha^1, \alpha^2,\alpha^3)$:
\beq
d\Omega_3^2\,=\,(d\alpha^1)^2\,+\,\sin^2\alpha^1\Big(\,
(d\alpha^2)^2\,+\,\sin^2\alpha^2\,(d\alpha^3)^2\,\Big)\,,
\eeq
with $0\le\alpha^1,\alpha^2\le \pi$ and $0\le\alpha^3\le 2\pi$. 
In order to write down the Killing spinors in these coordinates, let us
choose the following frame for the $AdS_5$ part of the metric:
\bear
&&e^{t}\,=\,L\cosh\rho\,dt\;,\quad
e^{\rho}\,=\,L\,d\rho\,\,,\rc
&&e^{\alpha^1}\,=\,L\sinh\rho\,d\alpha^1\,\,,\rc
&&e^{\alpha^2}\,=\,L\sinh\rho\,\sin\alpha^1\,d\alpha^2\,\,,\rc
&&e^{\alpha^3}\,=\,L\sinh\rho\,\sin\alpha^1\,\sin\alpha^2\,d\alpha^3\,\,.
\label{adsframeglob}
\eear
We will continue to use the same frame forms as in eq. (\ref{kwframe}) for
the 
$T^{1,1}$ part of the metric. 
The components of the spin connection corresponding to the $AdS_5$
part become:
\bear
&&\omega^{t\,\rho}=\sinh\rho\,dt\;,\quad
\omega^{\alpha_1\,\rho}=\cosh\rho\,d\alpha_1\;,\rc
&&\omega^{\alpha_2\,\rho}=\cosh\rho\,\sin\alpha_1\,d\alpha_2\;,\rc
&&\omega^{\alpha_3\,\rho}=\cosh\rho\,\sin\alpha_1\,\sin\alpha_2\,d\alpha_3
\;,\rc
&&\omega^{\alpha_2\,\alpha_1}=\cos\alpha_1\,d\alpha_2\;,\rc
&&\omega^{\alpha_3\,\alpha_1}=\cos\alpha_1\,\sin\alpha_2\,d\alpha_3\;,\quad
\omega^{\alpha_3\,\alpha_2}=\cos\alpha_2\,d\alpha_3\;.
\label{conadsglob}
\eear
Notice that we have written the spin connection in terms of the
differentials of the $AdS_5$ coordinates. The selfdual five-form reads:
\beq
g_s\,F^{(5)}={4\over L}\left(e^1\wedge e^2\wedge e^{\hat1}\wedge e^{\hat2}\wedge
e^{\hat3}-e^t\wedge e^\rho\wedge e^{\alpha_1}\wedge e^{\alpha_2}\wedge
e^{\alpha_3}\right)\,\,.
\label{f5glob}
\eeq
Now we can solve the supersymmetry equations corresponding
to the $AdS_5$ part (namely eqs. (\ref{kwads})) using global
coordinates. Written in the frame (\ref{adsframeglob}), they
read:
\beq
D_{\mu}\,\epsilon-{i\over
2L}\,\gamma\,\Gamma_{\mu}\,\epsilon=0\;,\;
(\mu=t,\rho,\alpha_1,\alpha_2,\alpha_3)\,\,,
\label{kwadsglob}
\eeq
where we have defined:
\beq
\gamma\equiv\Gamma^{\,t\,\rho\,\alpha_1\,\alpha_2\,\alpha_3}\,,
\eeq
and $\Gamma_\mu\;,(\mu=t,\rho,\alpha_1,\alpha_2,\alpha_3)$ are
constant Dirac matrices associated to the frame (\ref{adsframeglob}).
We expect the Killing spinors to depend on the coordinates so we must
proceed as in (\ref{kwcdvtrns})
to write the covariant derivative in terms of the derivatives of the
spinor with respect to the global coordinates. Let us begin with the
equation for
$\mu=\rho$, which yields:
\beq
\partial_\rho\,\epsilon-{i\over2}\,\Gamma^\rho\,\gamma\,\epsilon=0.
\label{rhoeq}
\eeq
This can be easily solved as
\beq
\epsilon=e^{i\,{\rho\over2}\,\Gamma^\rho\,\gamma}\,\tilde{\epsilon}\,,
\label{epsrho}
\eeq
where $\tilde{\epsilon}$ is a ten dimensional spinor independent of
$\rho$. The equation for $\mu=t$ renders:
\beq
\partial_t\,\epsilon=
-{i\over2}\,\Gamma^t\,\gamma\,e^{-i\,\rho\,\gamma\,\Gamma^\rho}\,
\epsilon\,.
\label{teq}
\eeq
Inserting the form of $\epsilon$ written in (\ref{epsrho})
into this last equation, we can solve for $\tilde{\epsilon}$ in terms 
of a spinor $\bar\epsilon$ independent of $\rho$ and
$t$, namely:
\beq
\tilde{\epsilon}=e^{-i\,{t\over2}\,\Gamma^t\,\gamma}\,\bar\epsilon\,,
\label{epsbar}
\eeq 
so we can write $\epsilon$ as
\beq
\epsilon=e^{i\,{\rho\over2}\,\Gamma^\rho\,\gamma}\,
e^{-i\,{t\over2}\,\Gamma^t\,\gamma}\,\bar\epsilon\,.
\label{epst}
\eeq
The equations for the angular components are:
\bear
&&\partial_{\alpha_1}\,\epsilon=-{1\over2}\,\Gamma^{\alpha_1\,\rho}\,
e^{-i\,\rho\,\Gamma^\rho\,\gamma}\,\epsilon\,,
\label{eqalpha1}\\\rc
&&\partial_{\alpha_2}\,\epsilon=-{1\over2}\left(\sin\alpha_1\,\Gamma^{\alpha_2\,\rho}\,
e^{-i\,\rho\,\Gamma^\rho\,\gamma}-\cos\alpha_1\,
\Gamma^{\alpha_2\,\alpha_1}\right)\epsilon\,,
\label{eqalpha2}\\\rc
&&\partial_{\alpha_3}\,\epsilon=-{1\over2}\left(\sin\alpha_1\,\sin\alpha_2\,
\Gamma^{\alpha_3\,\rho}\,e^{-i\,\rho\,\Gamma^\rho\,\gamma}-\cos\alpha_1\,
\sin\alpha_2\,\Gamma^{\alpha_3\,\alpha_1}-\cos\alpha_2\,
\Gamma^{\alpha_3\,\alpha_2}\right)\epsilon\,.\rc
\label{eqalpha3}
\eear
It is easy to solve these three equations in the order we have written
them. After plugging (\ref{epst}) into the first equation we
determine the dependence of $\epsilon$ on $\alpha_1$. Then, the second
equation fixes the
$\alpha_2$-dependence  and finally, from the third equation we get
$\epsilon$ in terms of a constant (along $AdS_5$) 10d spinor
$\epsilon_0$ \cite{Globalads}: 
\beq
\epsilon=e^{i\,{\rho\over2}\,\Gamma^\rho\,\gamma}\,
e^{-i\,{t\over2}\,\Gamma^t\,\gamma}\,e^{-{\alpha_1\over2}\,
\Gamma^{\alpha_1\,\rho}}\,e^{-{\alpha_2\over2}\,
\Gamma^{\alpha_2\,\alpha_1}}\,e^{-{\alpha_3\over2}\,
\Gamma^{\alpha_3\,\alpha_2}}\,\epsilon_0\,.
\label{adskillinglob}
\eeq
As it happened when using cartesian
coordinates, all the matrices in this last expression commute with the
$\Gamma$-matrices appearing in equations (\ref{kwt11}) for the $T^{1,1}$.
Hence, $\epsilon_0$ must satisfy the same projections as the ones in
(\ref{t11proj}), namely:
\beq
\Gamma_{12}\,\epsilon_0=i\,\epsilon_0\;,\quad
\Gamma_{\hat1\hat2}\,\epsilon_0=-i\,\epsilon_0\,.
\label{t11projglob}
\eeq
Then, the Killing spinors of the KW model (when using global
coordinates for the $AdS_5$ part) are given by the expression
(\ref{adskillinglob}) in terms of a 10d constant spinor satisfying the
projections (\ref{t11projglob}). It becomes clear that this solution
leaves unbroken eight supersymmetries, as it was expected.


\chapter{Killing spinors of the Klebanov-Tseytlin model}
\label{ktcp}
\section{Introduction}
\setcounter{equation}{0}
The goal of this chapter is to obtain the explicit expression of the
Killing spinors of the 10d IIB supergravity solution known as the
Klebanov-Tseytlin (KT) model. Proceeding as in the last chapter, we will
solve  the SUSY equations in a frame such that 
the Killing
spinors are not expected to depend on the compact coordinates of the
conifold. We will manage to express them in terms of a
constant spinor subjected to three independent (and compatible)
projections reducing the number of independent real components from 32 to
4, as it should be for a background that leaves unbroken 4 supercharges.

\subsection{The Klebanov-Tseytlin model}
This solution is constructed in \cite{KT} by placing $N$ D3-branes and $M$
fractional D3-branes (wrapped D5-branes) at the singular point of the 
conifold. The D5-branes wrap a 2-cycle inside $T^{1,1}$ and 
serve as sources of the magnetic RR three-form flux through the $S^3$ of
$T^{1,1}$. The dual field theory is
${\cal N}=1$ SYM with gauge group
$SU(N+M)\times SU(N)$ and two chiral multiplets. The non-vanishing
three-form flux in the SUGRA solution is the source of the conformal
symmetry breaking in the dual field theory. Thus, we expect the
corresponding IIB SUGRA solution to have four supersymmetries. The
near-horizon metric and the selfdual RR five-form of the solution are:
\bear
ds^2_{10}&=&[\hat h(r)]^{-{1\over 2}}\,dx^2_{1,3}\,+\,[\hat h(r)]^{{1\over
2}}\,
\big(\,dr^2\,+\,r^2\,ds^2_{T^{1,1}}\,\big)\,\,,\rc\rc
\hat h(r)&=&
{27\pi\left(\alpha'\right)^2\over4r^4}\left[g_sN+a\left(g_sM\right)^2
\ln\left({r\over r_0}\right)+{a\over4}\left(g_sM\right)^2\right]\,,
\label{ktmetric}
\eear
with $a={3\over2\pi}$.
The RR selfdual five-form reads:
\beq
F^{(5)}=27\pi\left(\alpha'\right)^2
N_{eff}\,d\,{\rm Vol}\left(\,T^{1,1}\,\right)+ {\rm Hodge\;dual}\,,
\label{ktf5}
\eeq
where $d{\rm Vol}\left(\,T^{1,1}\,\right)$ is the volume
five-form of the $T^{1,1}$ space and $N_{eff}$ is the following
function of
$r$:
\beq
N_{eff}=N+{3\over2\pi}\,g_s\,M^2\,\ln\left({r\over r_0}\right),
\label{neff}
\eeq
and one can readily check that
\beq
{1\over\left(4\pi^2\alpha'\right)^2}\int_{T^{1,1}}F^{(5)}=N_{eff}\,.
\label{efflux}
\eeq
Hence, the five-form flux acquires a radial dependence and it is
not quantized. It can still be identified with the quantity $N_{eff}$
defining the gauge group $SU(N_{eff}+M)\times SU(M)$ only at special 
radii $r_k=r_0\,{\rm exp}\left({-2\pi\,k\over3\,g_s\,M}\right)$ where
$k$ is an integer, so $N_{eff}=N-k\,M$. In fact, the logarithmic
decreasing of
$N_{eff}(r)$, related to a continuous reduction in the numbers of
degrees of freedom, is known as the RG cascade. This is mapped in the
gauge theory side to a Seiberg duality cascade.

The RR and NSNS three-forms can be written as
\beq
F^{(3)}={M\alpha'\over2}\,\hat\omega_3\;,\quad
H={3g_s\,\alpha'\,M\over2r}\, dr\wedge\hat\omega_2\,,
\label{kt3forms}
\eeq
where $\hat\omega_2$ and
$\hat\omega_3$ are the closed two- and three-forms of the conifold, which
in terms of the left invariant $SU(2)$ one-forms (\ref{omegaforms}) and
the Maurer-Cartan one-forms (\ref{s2forms}) become:
\bear
\hat\omega_2={1\over2}\left(\sigma^1\wedge\sigma^2+\omega^1\wedge\omega^2\right)\;,
\quad\hat\omega_3=\left(\omega^3+\sigma^3\right)\wedge\hat\omega_2\,.
\label{clforms}
\eear

As we have said in the subsection \ref{sconn} of the previous chapter, the
metric of this geometry is described by the one-form basis
(\ref{kwgenframe}) simply by changing the generic  warp factor $\tilde
h(r)$ to $\hat h(r)$ written in (\ref{ktmetric}). Then, let us define the
following one-form basis:
\beq
\hat e^{\,a}=\tilde e^{\,a}\left(\hat
h(r)\right)\;,\; \left(a=x^0,...,x^3,r,1,2,\hat1,\hat2,\hat3\right)\,,
\label{ktframe}
\eeq
where $\tilde e^{\,a}\left(\hat h(r)\right)$ stands for the one-form
frame resulting from the generic one written in eq.
(\ref{kwgenframe}), after making $\tilde h(r)=\hat h(r)$.

\section{Killing spinors}
\setcounter{equation}{0}

In order to determine the Killing spinors of this solution we have to
solve the equations resulting from the vanishing of the SUSY variations
(\ref{sugra}). For this model with constant dilaton and three- and
five-form fluxes they are:
\bear
&&-{i\over
24}\,{\cal
F}^{(3)}_{N_1N_2N_3}\,
\Gamma^{N_1N_2N_3}\,\epsilon=0\,,
\label{ktdilvar}\\\rc
&&D_{M}\,\epsilon\,+\,{i\over 1920}\,
F_{N_1\cdots N_5}^{(5)}\,\Gamma^{N_1\cdots
N_5}\Gamma_{M}\,\epsilon
+{1\over96}\,{\cal F}^{(3)}_{N_1N_2N_3}\,
\big(\,\Gamma_{M}^{\,\,\,N_1N_2N_3}\,-\,
9\,\delta_{M}^{N_1}\,\,\Gamma^{N_2 N_3}\,\big)\,\epsilon^{*}=0\,,\rc
\label{ktgravar}
\eear
where ${\cal F}^{(3)}$ is the complex combination of the RR and NSNS
three-forms defined in (\ref{comp3form}). Let us write
the RR five-form and the complex combination of the RR and NSNS
three-forms in the one-form basis (\ref{ktframe}):
\bear
F^{(5)}&=&-\hat h'\hat h^{-{5\over4}}\left(\hat e^{x^0}\wedge \hat
e^{x^1}\wedge \hat e^{x^2}\wedge \hat e^{x^3}\wedge \hat e^r+\hat
e^1\wedge \hat e^2\wedge \hat e^{\hat1}\wedge \hat e^{\hat2}\wedge
\hat e^{\hat3}\right)\,,\label{ktf5flat}\\
{\cal F}^{(3)}&=&{9M\alpha'\over2r^3}\hat h^{-{3\over4}}\left(\hat
e^1\wedge\hat e^2+\hat e^{\hat1}\wedge\hat
e^{\hat2}\right)\wedge\left(\hat e^r+i\hat e^{\hat3}\right)\,,
\label{ktf3flat}
\eear
where for simplicity we have taken $g_s=1$ and we have used that
\beq
\hat h'\,r^5=-27\pi\left(\alpha'\right)^2\,g_s\,N_{eff}\,,
\label{kthprime}
\eeq
which results from differentiating the expression of $\hat h(r)$ given
in eq. (\ref{ktmetric}).  

Next, in order to write down the spin connection of the
background we plug the warp factor $\hat h(r)$ into the generic spin
connection (\ref{gencon}) we computed in section \ref{sconn}. Then, the
spin connection one-form for the KT model, expressed in the
frame (\ref{ktframe}), reads:
\bear
\omega^{x^\alpha\,r}&=&\left(\hat{h}^{-{1\over4}}\right)'
\,\hat e^{\,x^\alpha}\;,\;
(\alpha=0,1,2,3)\,\,,\rc\rc
\omega^{s\,r}&=&\hat{h}^{-{1\over4}}\left({1\over
r}+{1\over4}\,\hat h'\,\hat{h}^{-1}\right)\,\hat e^{\,s}\;,\;
(s=1,2,\hat1,\hat2,\hat3)\,\,,\rc\rc
\omega^{1\,2}&=&{1\over r}\,\hat{h}^{-{1\over4}}\,
\hat e^{\hat3}-{\sqrt6\over
r}\,\cot\theta_1\,\hat{h}^{-{1\over4}}\,\hat e^2\,\,,\rc\rc
\omega^{\hat1\,\hat2}&=&{2\over
r}\,\hat{h}^{-{1\over4}}\,\hat e^{\hat3}-{\sqrt6\over
r}\,\cot\theta_1\,\hat{h}^{-{1\over4}}\,\hat e^2\,\,,\rc\rc
\omega^{\hat1\,\hat3}&=&-{1\over
r}\,\hat{h}^{-{1\over4}}\,\hat e^{\hat2} \; ,\quad
\omega^{\hat2\,\hat3}={1\over
r}\,\hat{h}^{-{1\over4}}\,\hat e^{\hat1}\,\,,\rc\rc
\omega^{\hat3\,2}&=&{1\over
r}\,\hat{h}^{-{1\over4}}\,\hat e^1\; ,\quad
\omega^{\hat3\,1}=-{1\over
r}\,\hat{h}^{-{1\over4}}\,\hat e^2\,\,.
\label{ktcon}
\eear
We begin by solving the equation $\delta\psi_{x^1}=0$ for a 10d
spinor $\epsilon$ independent of the $x^\alpha$ coordinates. After
inserting the three- and five-forms written in (\ref{ktf3flat}) and
(\ref{ktf5flat}) and the spin connection we have just computed, one gets:
\bear
&&-{1\over8}\,\hat h'\,\hat h^{-{5\over4}}\,
\Gamma_{x^1r}\,\epsilon+{i\over8}\,\hat h'\,\hat h^{-{5\over4}}\,
\Gamma_{x^0 x^1\,x^2\,x^3}\,\Gamma_{r\,x^1}\,\epsilon+\rc\rc
&&+{9\over16}{M\alpha'\over2r^3}\,\hat
h^{-{3\over4}}\,\Gamma_{x^1}\left(\Gamma_{12r}+\Gamma_{\hat1\hat2r}
+i\Gamma_{12\hat3}+i\Gamma_{\hat1\hat2\hat3}\right)\,\epsilon^*=0\,,\rc
\label{ktgravx1}
\eear
where $\Gamma_a\,,\;(a=x^\alpha,r,1,2,\hat1,\hat2,\hat3)$ are
constant Dirac matrices associated to the frame (\ref{ktframe}) and we
have inserted the equality
$\Gamma_{x^0...x^3}\Gamma_r\Gamma_{12\hat1\hat2\hat3}\,\epsilon=
-\epsilon$, following from the well defined chirality of the 10d
spinor $\epsilon$. Let us impose the SUSY cycle projection 
\beq
\Gamma_{12}\,\epsilon=-\Gamma_{\hat1\hat2}\,\epsilon\,\,,
\label{kt2cproj}
\eeq
which is again projection (\ref{2cproj}), arising naturally when
the conifold is obtained from the 8d gauged SUGRA. Thus, the third
term of the supersymmetry equation (\ref{ktgravx1}) vanishes, and
from the remaining ones we get the projection
\beq
\Gamma_{\,x^0\,x^1\,x^2\,x^3}\,\epsilon=-i\,\epsilon\,.
\label{ktd3proj}
\eeq
This is the projection corresponding to a D3-brane extended along the
Minkowski space. It can be straightforwardly checked that the remaining
equations
$\delta\psi_{x^\alpha}=0$ are solved by the same projections
(\ref{kt2cproj}) and (\ref{ktd3proj}), which can be inserted in the 
equality
$\Gamma_{x^0...x^3}\Gamma_r\Gamma_{12\hat1\hat2\hat3}\,\epsilon=-
\epsilon$
to get the following useful relation:
\beq
\Gamma_{r\hat3}\,\epsilon=-i\,\epsilon\,.
\label{ktusproj}
\eeq
Now, we try to solve the equations for the angular
components of the gravitino assuming that $\epsilon$ is also
independent of the coordinates of the conifold. The equation
$\delta\psi_1=0$ becomes:
\bear
&&{1\over2}\hat h^{-{1\over4}}\left[\left({1\over r}+{1\over4}\,\hat h'\,
\hat h^{-1}\right)\Gamma_{1r}+{1\over r}\,\Gamma_{\hat32}\right]\,\epsilon
+{i\over8}\,\hat h'\,\hat h^{-{5\over4}}\,
\Gamma_{x^0 x^1 x^2 x^3}\,\Gamma_{r1}\,\epsilon+\rc\rc&&+
{9\over16}{M\alpha'\over2r^3}\,\hat
h^{-{3\over4}}\,\left(\Gamma_{1\hat1\hat2}-3\Gamma_2\right)\left(\Gamma_r
+i\Gamma_{\hat3}\right)\,\epsilon^*=0\,,
\label{ktgrav1}
\eear
where in the second term we have inserted the total chirality
projection $\Gamma_{x^0...x^3}\Gamma_r\Gamma_{12\hat1\hat2\hat3}\,\epsilon=
-\epsilon$. The last term of this equation vanishes after imposing
the complex conjugate of eq. (\ref{ktusproj}), so we are left with 
\beq
{1\over2r}\left(1-\Gamma_{r12\hat3}\right)\,\epsilon={1\over8}\,
\hat h'\,\hat h^{-1}\left(i\,\Gamma_{\,x^0\,x^1\,x^2\,x^3}-1\right)\,
\epsilon\,,
\eeq
where the right-hand side vanishes when imposing the projection
(\ref{ktd3proj}). Hence, the equation renders the projection
$\Gamma_{r12\hat3}\,\epsilon=\epsilon$, which, after making use of
(\ref{kt2cproj}), can be written as
\beq
\Gamma_{r\hat1\hat2\hat3}\,\epsilon=-\epsilon\,.
\label{ktradproj}
\eeq
The equations for the remaining angular components of the gravitino and
the equation for the dilatino ($\delta\lambda=0$) are easily solved by
imposing the three independent projections we have got, namely
(\ref{kt2cproj}), (\ref{ktd3proj}) and (\ref{ktradproj}). Finally, we
write down the equation for the radial component of the gravitino
assuming that
$\epsilon=\epsilon(r)$. Thus, as we have done in previous chapters, we
should use that $D_r\,\epsilon=(E^{\,r}_{\,\tilde r})^{-1}\,D_{\tilde
r}\,\epsilon\,$ in order to write the covariant derivative in terms of
$\epsilon'\equiv\partial_r\,\epsilon$. This time $E^{\,c}_{\,\tilde m}$
are the coefficients appearing when writing the one-forms
(\ref{ktframe}) in terms of the differentials of the coordinates.
The equation $\delta\psi_r=0$ reads:
\bear
\epsilon'+{i\over8}\,\hat h'\,\hat h^{-1}\,\Gamma_{\,x^0\,x^1\,x^2\,x^3}
\,\epsilon+{9\over16}{M\alpha'\over2r^3}\,\hat
h^{-{1\over2}}\,\left(\Gamma_{12}+\Gamma_{\hat1\hat2}\right)\left(-3+i\,
\Gamma_{r\hat3}\right)\,\epsilon^*=0.
\eear
The third contribution cancels out by virtue of  (\ref{kt2cproj}) and
if we also impose the projection (\ref{ktd3proj}) we arrive at
\beq
\epsilon'+{1\over8}\,\hat h'\,\hat h^{-1}\,\epsilon=0\,.
\eeq
Therefore, the Killing spinor of the Klebanov-Tseytlin model can be
expressed in terms of a 10d constant spinor $\epsilon_0$ as
\beq
\epsilon=\hat h^{-{1\over8}}\,\epsilon_0\,,
\label{ktepsradep}
\eeq
where $\epsilon_0$ satisfies three independent projections, namely:
\beq
\Gamma_{\,x^0\,x^1\,x^2\,x^3}\,\epsilon_0=-i\,\epsilon_0\;,\quad
\Gamma_{12\hat1\hat2}\,\epsilon_0=\epsilon_0\;,\quad
\Gamma_{r\hat1\hat2\hat3}\,\epsilon_0=-\epsilon_0\,.
\label{ktdefproj}
\eeq
Thus, the model has 4 independent spinors as it should be for the
supergravity dual of a 4d ${\cal N}=1$ field theory.

Recalling that
$\Gamma_{x^0...x^3}\Gamma_r\Gamma_{12\hat1\hat2\hat3}\,\epsilon_0=
-\epsilon_0$,
the projections (\ref{ktdefproj}) can be reformulated as
\beq
\Gamma_{\,x^0\,x^1\,x^2\,x^3}\,\epsilon_0=-i\,\epsilon_0\;,\quad
\Gamma_{12}\,\epsilon_0=i\epsilon_0\;,\quad
\Gamma_{\hat1\hat2}\,\epsilon_0=-i\,\epsilon_0\,.
\label{ktdefprojcnf}
\eeq
These projections, which in view of (\ref{ktepsradep}) are also satisfied
by $\epsilon$, can be identified as the projection corresponding to a
D3-brane along the Minkowski directions and the two projections of the
$T^{1,1}$ (see \cite{RCN}). Hence, 
recalling the results of the previous chapter, one readily notices that
these projections are the same as the ones fulfilled
by the Killing spinors $\epsilon_1$ of eq. (\ref{etaplus}) in the last
chapter. Those are the four spinors corresponding to the ordinary
supersymmetries of the Klebanov-Witten background. In fact, the only
difference between
$\epsilon$ written in eq. (\ref{ktepsradep}) and the four Killing spinors
$\epsilon_1$ of the KW solution relies on the different radial dependence.
Therefore, the breaking of conformal invariance due to the addition of the
fractional branes in the Klebanov-Tseytlin model, translates into the
loss of the four Killing spinors $\epsilon_2$ (\ref{etaminus})
realizing the superconformal symmetries.


\chapter{Killing spinors of the Klebanov-Strassler model}
\label{kscp}

\section{Introduction}
\setcounter{equation}{0}
In this chapter we compute explicitly the Killing spinors of the
Klebanov-Strassler (KS) solution \cite{KS}. This background has
attracted a lot of interest during the last years since it is a gravity
dual of  ${\cal N}=1$ SYM with very nice features. It is constructed by
placing fractional D3-branes and D3-branes on the deformed conifold, so
in the UV it approaches the Klebanov-Tseytlin solution described in the
last chapter and therefore, it incorporates the logarithmic flow of
couplings. On the other hand, in the IR, where the KT model was
singular, the deformation of the conifold gives a geometrical realization
of chiral symmetry breaking and confinement.

The structure of this chapter is as follows: in the first section we
characterize the SUGRA background giving some hints into its
construction. In section \ref{kskilling} we solve the SUSY equations in a
frame where the Killing spinors do not depend on the angular coordinates
of the conifold. Finally, in section \ref{ksdiffeqs} we show that the
differential equations for the functions entering the KS ansatz (see
below) that we get from the SUSY equations are equivalent to the first
order system appearing in ref. \cite{KS} plus an extra differential
equation.

The results of this calculation were published in \cite{SSP} as the
initial step of an extension of the kappa symmetry analysis carried
out there to the more interesting KS background. 

\subsection{The Klebanov-Strassler model}
The Klebanov-Tseytlin geometry described in the last chapter becomes
singular at sufficiently small $r$, precisely at the end of the RG
cascade. Then, in order
to construct a SUGRA dual of the IR region of ${\cal N}=1$ SYM, one
can substitute the singular conifold by its deformation. Hence, while
for large $r$ the geometry approaches the KT solution, at $r=0$
the geometry does not collapse but degenerates into a finite $S^3$.
This fact gives a geometric realization of confinement and chiral
symmetry breaking, which are fundamental features of the ${\cal
N}=1$ SYM expected at the end of the cascade.  
The resulting background is the so-called Klebanov-Strassler
\cite{KS} solution. The warped 10d metric is:
\beq
ds^2_{10}=[h(\tau)]^{-{1\over 2}}\,dx^2_{1,3}\,+\,[h(\tau)]^{{1\over
2}}\, ds^2_6\,,
\label{ks10dmetric}
\eeq  
where the six dimensional metric $ds^2_6$ is the one corresponding to the
deformed conifold and $\tau$ is the radial coordinate defined in
(\ref{gctaudef}). The metric of the deformed conifold
is obtained from the one of the generalized deformed conifold written
in eq. (\ref{gendef}) simply by setting 
$C=0$. Thus, it reads:
\bear
ds^2_6={1\over2}\mu^{{4\over3}}K(\tau)\Bigg[{1\over3K(\tau)^3}\left(d\tau^2
+\left(g^5\right)^2\right)+\sinh^2\left({\tau\over2}\right)\left(
\left(g^1\right)^2+\left(g^2\right)^2\right)+\rc\rc
+\cosh^2\left({\tau\over2}\right)
\left(\left(g^3\right)^2+\left(g^4\right)^2\right)\Bigg]\;,
\label{ksdefcon}
\eear
with
\beq
K(\tau)={\left(\sinh(2\tau)-2\tau\right)^{1\over3}\over
2^{1\over3}\sinh\tau}\;,
\label{ksk}
\eeq
and the one-forms $g^i\;(i=1,...,5)$ are defined in equation
(\ref{gengforms}) in terms of the usual angular coordinates.

As we have said in chapter \ref{genconsec}, the metric of the deformed
conifold reduces to the one of an $S^3$ when $\tau\to 0$ while it
coincides with the metric of the singular conifold for $\tau\to\infty$.
Therefore, the RR three-form flux for this model reads:
\beq
F^{(3)}={M\alpha'\over2}\left[(1-F)\,g^5\wedge g^3\wedge g^4
+F\,g^5\wedge g^1\wedge g^2+F'\,d\tau\wedge\left(g^1\wedge g^3+g^2\wedge
g^4\right)\right]\,,
\label{ksf3}
\eeq
with $F=F(\tau)$ satisfying $F(0)=0$ and
$F(\tau\to\infty)={1\over2}$ in order to get an $F^{(3)}$ lying along the
$S^3$ when $\tau\to 0$ while being equal to the one written in
(\ref{kt3forms}) when $\tau\to\infty$, i.e. equal to the RR three-form
flux of the Klebanov-Tseytlin model in the UV. Notice that, as usual, the
prime after any radial function (for instance $F'$) stands for the
radial derivative (${d\over d\tau}$). The NSNS two-form  potential $B$
and its corresponding three-form field strength
$H$ are written as
\bear
B&=&{M\alpha'\over2}\left[f\,g^1\wedge g^2+
k\,g^3\wedge g^4\right]\, ,\label{ksb2form}\\ \rc
H&=&{M\alpha'\over2}\left[d\tau\wedge\left(f'\,g^1\wedge
g^2+ k'\,g^3\wedge g^4\right)+{1\over2}(k-f)\,g^5\wedge\left(g^1\wedge
g^3+ g^2\wedge g^4\right)\right]\,,\rc
\label{ksh3form}
\eear
in terms of two undetermined radial functions $f=f(\tau)$ and
$k=k(\tau)$. We are taking $g_s=1$. Finally, the five-form flux is
constructed by taking:
\bear
F^{(5)}&=&{\cal F}^{(5)}+{\rm Hodge\;dual}\,,\rc\rc
{\cal F}^{(5)}&=&B\wedge F^{(3)}={M^2\,(\alpha')^2\over4}\,l(\tau)\,
g^1\wedge g^2\wedge g^3\wedge g^4\wedge g^5=\rc\rc
&=&27\,M^2\,(\alpha')^2\,l(\tau)\,d{\rm Vol}\left(\,T^{1,1}\,\right)\,,
\label{ksf5}
\eear
where we have defined:
\beq
l(\tau)\equiv f(\tau)\,\left(1-F(\tau)\right)+k(\tau)\,F(\tau)\,.
\label{ksldef}
\eeq
The Hodge dual $*{\cal F}^{(5)}$ becomes:
\beq
*{\cal F}^{(5)}={\alpha\,l(\tau)\over
K^2(\tau)\,h^2\,\sinh^2\tau}\,dx^0\wedge dx^1\wedge dx^2\wedge dx^3\wedge
d\tau\,,
\label{ksf5hd}
\eeq
with $\alpha\equiv4\,M^2\,(\alpha')^2\,\mu^{-{8\over3}}$.

Therefore, $F^{(5)}$ satisfies
by construction the IIB SUGRA equation of motion
$dF^{(5)}=H\wedge F^{(3)}$. However, the ansatz has to verify
the remaining equations of motion for the three-forms:
$d*F^{(3)}=F^{(5)}\wedge H$, and \ $d*H=-F^{(5)}\wedge
F^{(3)}$, and the constant dilaton condition which implies
$\left(F^{(3)}\right)^2=\left(H\right)^2$, together with the
Einstein equation. This renders a system of second order differential
equations determining the unknown functions of the ansatz
($F(\tau),f(\tau),k(\tau)$ and
$h(\tau)$). It is not difficult to find a system of first order
differential equations \cite{KS} that solves those equations. It reads:
\bear
f'&=&(1-F)\,\tanh^2\Big(\,{\tau\over 2}\,\Big)\,,\rc
\,\,\,\,\,\,\,\,\,\,\,\,\,\,\,\,
k'&=&F\coth^2\Big(\,{\tau\over 2}\,\Big)\,,\rc
F'&=&{k-f\over 2}\,\,,
\label{ksystem}
\eear
and,
\beq
h'=-{\alpha\,\,l(\tau)\over K^2(\tau)\,\sinh^2\tau}\,.
\label{ksystemh}
\eeq

In order to arrive at this system let us recall that if $*F^{(3)}$
satisfies the equation $d*F^{(3)}=F^{(5)}\wedge H\,$, it can be written
as $*F^{(3)}=dC^{(6)}+C^{(4)}\wedge H$ in terms of the six-form and 
four-form RR potentials. We will show that the system (\ref{ksystem})
results from requiring the vanishing of the six-form RR potential,
i.e. $C^{(6)}=0$; which, in view of the last expression of $*F^{(3)}$,
is equivalent to:
\beq
*F^{(3)}=C^{(4)}\wedge H\,.
\label{ksc6eq}
\eeq
From eq. (\ref{ksf3}) it
is straightforward to write down
$*F^{(3)}$:
\bear
*F^{(3)}&=&{M\alpha'\over2}\,h^{-1}\,d^4x\wedge\bigg[(1-F)\,
\tanh^2\left(\,{\tau\over2}\,\right)d\tau\wedge g^1\wedge g^2+\rc\rc
&+&F\,\coth^2\left(\,{\tau\over2}\,\right)d\tau\wedge g^3\wedge g^4
+F'g^5\wedge\left(g^1\wedge g^3+g^2\wedge g^4\right)\bigg]\,,
\label{ksf3dual}
\eear
with $d^4x=dx^0\wedge dx^1\wedge dx^2\wedge dx^3$.  

From the equation of motion $dF^{(5)}=H\wedge F^{(3)}$ it is clear
that one can write  $F^{(5)}=dC^{(4)}+B^{(2)}\wedge F^{(3)}$. In
addition, let us write $C^{(4)}$ as $C^{(4)}=\hat C^{(4)}+\tilde
C^{(4)}$, with the four-forms $\hat C^{(4)}$ and $\tilde C^{(4)}$ being
given by:
\beq
d\hat C^{(4)}=*{\cal F}^{(5)}\;,\;\;d\tilde C^{(4)}={\cal
F}^{(5)}-F^{(3)}\wedge B\,,
\label{ksc4sep}
\eeq
so, in view of eq. (\ref{ksf5hd}) it is easy to write down $\hat
C^{(4)}$:
\beq
\hat C^{(4)}=\hat f(\tau)\,dx^0\wedge
dx^1\wedge dx^2\wedge dx^3\,,
\label{kshatc4}
\eeq
where we have defined $\hat f(\tau)$ as a function of the radial
coordinate satisfying:
\beq
\hat f'(\tau)={\alpha\,\,l(\tau)\over
K^2(\tau)\,\left[h(\tau)\right]^2\,\sinh^2\tau}\;.
\label{kshatfprime}
\eeq
Recalling that ${\cal F}^{(5)}\propto d{\rm Vol}(\,T^{1,1}\,)$ and the
expressions for $F^{(3)}$ and $B$ (eqs. (\ref{ksf3}) and
(\ref{ksb2form})) one realizes that all the components of $\tilde
C^{(4)}$ are perpendicular to the
Minkowski space
$(x^0,x^1,x^2,x^3)$. Since $H$ neither has components along any
Minkowski direction (see eq. (\ref{ksh3form})), it becomes clear that
$H\wedge
\tilde C^{(4)}=0$. Hence, one gets:
\bear
H\wedge C^{(4)}=H\wedge \hat C^{(4)}=&&{M\alpha'\over2}\,\hat f(\tau)\,
d^4x\wedge\bigg[d\tau\wedge\left(f'\,g^1\wedge g^2+
k'\,g^3\wedge g^4\right)+\rc\rc
&&+{1\over2}(k-f)\,g^5\wedge\left(g^1\wedge g^3+g^2\wedge
g^4\right)\bigg]\,.
\label{kshwdc4}
\eear
Inserting this result and the expression of $*F^{(3)}$ (eq.
(\ref{ksf3dual})) into the equation (\ref{ksc6eq}) one readily obtains the
first order system (\ref{ksystem}), and the equality $h^{-1}=\hat
f(\tau)$, which after differentiating yields the differential equation
(\ref{ksystemh}).

\section{Killing spinors}
\label{kskilling}
\setcounter{equation}{0}
As we have shown in the last section, the Klebanov-Strassler solution
is formulated in terms of some functions $F(\tau),\;f(\tau),\;k(\tau),$
and $h(\tau)$ defined by means of a system of first order
equations (\ref{ksystem}), (\ref{ksystemh}); which guarantees the
fulfilment of the SUGRA equations of motion.
But this is not the whole story since we should determine if for any
solution of the system we are dealing with a supersymmetric
solution of 10d type IIB supergravity. Indeed, we will show by imposing
the vanishing of the SUSY variations (\ref{sugra}), that the model is
${1\over8}$ SUSY if the functions $F(\tau)\,,\;f(\tau)\,,$ and
$\;k(\tau)$ verify the system of first order differential equations
(\ref{ksystem}) together with an extra algebraic constraint.

As we have seen in the subsection \ref{gcdefconks} of the second
chapter, if we choose the appropriate one-form basis, the Killing
spinors of the 10d solution consisting of adding $\RR^{1,3}$ to the
deformed conifold do not depend on the angular coordinates of the
conifold (see eq.
\ref{gcdcksp}). That basis arises naturally when we write the metric
of the deformed conifold as in eq. (\ref{10dgendefcon}), namely:

\bear
ds^2_{6}={1\over 2}\,\mu^{{4\over 3}}\,K(\tau)
&\Bigg\{&{1\over 3 K(\tau)^3}\,\Big(\,d\tau^2\,+\,
(w^3+\sigma^3)^2\,\Big)\, +\,{\sinh^2\tau\over
2\cosh\tau}\,\Big(\,(\sigma^1)^2+(\sigma^2)^2\,\Big)\,+\rc\rc
&&+\,{\cosh\tau\over 2}\left[\,\Big(\,w^1+{\sigma^1\over
\cosh\tau}\,\Big)^2\,+\,
\Big(\,w^2+{\sigma^2\over \cosh\tau}\,\Big)^2\,\right]\,\Bigg\}\,\,,
\label{ksdefcon2}
\eear
where $K(\tau)$ is defined in eq. (\ref{ksk}). Thus, it is natural
to consider a frame such as (\ref{gcdefconframe}), but now
including the corresponding powers of the warp factor $h(\tau)$ appearing
in the 10d metric (\ref{ks10dmetric}). It reads:
\bear
&&e^{x^\alpha}\,=h^{-{1\over4}}\,dx^\alpha\;,\;(\alpha=0,1,2,3)\;,\quad
e^{\tau}\,={\mu^{{2\over 3}}\,h^{1\over4}\over
\sqrt{6}\,K(\tau)}\,\,d\tau\,,\rc\rc &&e^{i}\,=\,{\mu^{{2\over
3}}\,h^{1\over4}\,\sqrt{K(\tau)}\over 2}\,\, {\sinh\tau\over
\sqrt{\cosh\tau}}\,\,\sigma^i\;,\;(i=1,2)
\,,\rc\rc &&e^{\hat i}\,=\,{\mu^{{2\over
3}}\,h^{1\over4}\,\sqrt{K(\tau)}\over 2}\,\,
\sqrt{\cosh\tau}\,\,\Big(\,w^i+{\sigma^i\over \cosh\tau}\,\Big)\;,\;
(i=1,2)\;,\rc\rc
&&e^{\hat 3}\,=\,{\mu^{{2\over
3}}\,h^{1\over4}\over \sqrt{6}\,K(\tau)}\,\, 
(w^3+\sigma^3)\,.
\label{ksframe}
\eear
Then, the corresponding spin connection one-form will be very similar
to the one written in (\ref{spincon}), when this last one is restricted to
the solution (\ref{defconsysol}). Let us write it schematically as
\bear
&&\omega^{x^\alpha\,\tau}=-{1\over4}h'\,h^{-1}\,C^{-1}\,e^{x^\alpha}
\;,\; (\alpha=0,1,2,3)\,,\rc\rc
&&\omega^{1\,\tau}=A'\,A^{-1}\,C^{-1}\,e^1
+{1\over2}\,g'\,B\,A^{-1}\,C^{-1}\,e^{\hat1}\,,\rc\rc
&&\omega^{2\,\tau}=A'\,A^{-1}\,C^{-1}\,e^2
+{1\over2}\,g'\,B\,A^{-1}\,C^{-1}\,e^{\hat2}\,,\rc\rc
&&\omega^{\hat1\,\tau}=B'\,B^{-1}\,C^{-1}\,e^{\hat1}
+{1\over2}\,g'\,B\,A^{-1}\,C^{-1}\,e^1 \,,\rc\rc
&&\omega^{\hat2\,\tau}=B'\,B^{-1}\,C^{-1}\,e^{\hat2}
+{1\over2}\,g'\,B\,A^{-1}\,C^{-1}\,e^2 \,,  \rc\rc
&&\omega^{\hat3\,\tau}=C'\,C^{-2}\,e^{\hat3}\;,\quad
\omega^{\hat1\,1}=-{1\over2}\,g'\,B\,A^{-1}\,C^{-1}\,e^\tau\;,\quad
\omega^{\hat2\,2}=-{1\over2}\,g'\,B\,A^{-1}\,C^{-1}\,e^\tau\,,\rc\rc
&&\omega^{2\,1}=A^{-1}\,\cot\theta\,e^2+{1\over2}\,C\,A^{-2}
\left(g^2-1\right)e^{\hat3}\,,\rc\rc
&&\omega^{\hat1\,\hat3}={1\over2}\,g\,A^{-1}
\left(B\,C^{-1}+C\,B^{-1}\right)e^2
-{1\over2}\,C\,B^{-2}\,e^{\hat2}\,,\rc\rc
&&\omega^{\hat2\,\hat1}=A^{-1}\,\cot\theta\,e^2+
\left({1\over2}\,C\,B^{-2}-C^{-1}\right)\,e^{\hat3}\,,\rc\rc
&&\omega^{\hat2\,\hat3}={1\over2}\,C\,B^{-2}\,e^{\hat1}
-{1\over2}\,g\,A^{-1}
\left(B\,C^{-1}+C\,B^{-1}\right)\,e^1\,,\rc\rc
&&\omega^{\hat1\,2}=-\omega^{\hat2\,1}={1\over2}\,g\,A^{-1}
\left(C\,B^{-1}-B\,C^{-1}\right)\,e^{\hat3}\,,\rc\rc
&&\omega^{\hat3\,2}={1\over2}\,g\,A^{-1}
\left(C\,B^{-1}-B\,C^{-1}\right)\,e^{\hat1}
-{1\over2}\,C\,A^{-2}\left(g^2-1\right)e^1\,,\rc\rc
&&\omega^{\hat3\,1}={1\over2}\,g\,A^{-1}
\left(B\,C^{-1}-C\,B^{-1}\right)\,e^{\hat2}
+{1\over2}\,C\,A^{-2}\left(g^2-1\right)e^2\,,
\label{kspincon}
\eear
where $A,\;B,\;C$ and $g$ are the following functions of the radial
coordinate:
\bear
&&A={\mu^{2\over3}\,h^{1\over4}\,\sqrt{K(\tau)}\over2}\,{\sinh\tau\over
\sqrt{\cosh\tau}}\;,\quad
B={\mu^{2\over3}\,h^{1\over4}\,\sqrt{K(\tau)}\over2}\,\sqrt{\cosh\tau}\;,
\rc\rc
&&C={\mu^{{2\over
3}}\,h^{1\over4}\over \sqrt{6}\,K(\tau)}\;,
\quad g={1\over\cosh\tau}\;,
\label{ksABC}
\eear
which allow us to write the one-form basis (\ref{ksframe}) in the
following neat form:
\bear
&&e^{x^\alpha}\,=h^{-{1\over4}}\,dx^\alpha\;,\;(\alpha=0,1,2,3)\;,\quad
e^{\tau}\,=C\,d\tau\,,\rc\rc
&&e^{i}\,=A\,\sigma^i\;,\; (i=1,2)\,,\rc\rc
&
&
e^{\hat i}\,=B\,\Big(\,w^i+g\,\sigma^i\,\Big)\;,\; (i=1,2)\,,\rc\rc
&&e^{\hat 3}\,=C\, 
(w^3+\sigma^3)\,.
\label{ksframesp}
\eear
One can check by substituting the expressions of $A,\,B$ and $C$
above into the components of the spin connection written in
(\ref{kspincon}) that, except for the terms proportional to $h'(\tau)$,
the resulting one-form is equal term by term (up to $h(\tau)$ factors)
to the spin connection one-form arising from substituting the
particular solution (\ref{defconsysol}) (describing the deformed
conifold) into the generalized spin connection written in
(\ref{spincon}). We have expressed the spin connection
one-form in the frame (\ref{ksframe}). One
should bear in mind that
\beq
\omega^{a\,b}=\omega^{a\,b}_{\tilde m}\,dX^{\tilde m}=
\omega^{a\,b}_c\,e^c\,,
\eeq
where $e^c$ refers to the frame
(\ref{ksframe}). So, when needed, the components
$\omega^{a\,b}_{\tilde m}$ can be easily computed:
\beq
\omega^{a\,b}_{\tilde
m}=E^{\,c}_{\,\tilde m}\,\omega^{a\,b}_c\,,
\eeq
with $E^{\,c}_{\,\tilde m}$ being the coefficients appearing when one
expresses the one-forms (\ref{ksframe}) in terms of the
differentials of the coordinates, namely: $e^c=E^{\,c}_{\,\tilde
m}\,dX^{\tilde m}\,$.

Let us now write the three- and five-form fluxes in the frame
(\ref{ksframe}). The selfdual RR five-form takes the form:
\beq
F^{(5)}=-{\sqrt{6}\over\mu^{2\over3}}\,K(\tau)\,h^{-{5\over4}}\,h'\left(
e^{x^0}\wedge e^{x^1}\wedge e^{x^2}\wedge e^{x^3}\wedge e^\tau+ e^1\wedge
e^2\wedge e^{\hat1}\wedge e^{\hat2} \wedge e^{\hat3}\right)\,,
\label{ks5formflat}
\eeq 
the RR three-form (\ref{ksf3}) becomes:
\bear
F^{(3)}=&-&{4\sqrt{6}\,h^{-{3\over4}}\over \mu^2}{M\alpha'\over2}\Bigg\{
{1\over2\cosh\tau}\,e^{\hat3}\wedge e^{\hat2}\wedge
e^{\hat1}+{1-g-2F\over2\sinh\tau}\left(e^{\hat3}\wedge e^{\hat2}\wedge
e^1 +e^{\hat3}\wedge e^2 \wedge e^{\hat1}\right)+\rc\rc
&+&{\cosh\tau\over2\sinh^2\tau}\left[1+g\,(g-2+4F)\right]e^{\hat3}\wedge
e^2 \wedge e^1 +{F'\over\sinh\tau}\left(e^\tau \wedge e^{\hat2}\wedge
e^2 +e^\tau \wedge e^{\hat1}\wedge e^1\right)\Bigg\}\,.\rc\rc
\label{ksf3flat}
\eear
The NSNS three-form flux (\ref{ksh3form}) can be written as
\bear
H=&-&{4\sqrt{6}\,h^{-{3\over4}}\over\mu^2}{M\alpha'\over4}\Bigg\{
{f'+k'\over\cosh\tau}\,e^\tau \wedge e^{\hat2}\wedge
e^{\hat1} + {1\over\sinh\tau}\left[(1-g)\,k'-(1+g)\,f'\right]\Big(
e^\tau \wedge e^{\hat2}\wedge e^1+\rc\rc &+&e^\tau \wedge e^2 \wedge
e^{\hat1}\Big)+{\cosh\tau\over\sinh^2\tau}\left[(1-g)^2 \,k'+(1+g)^2\,f'
\right]e^\tau \wedge e^2 \wedge e^1+\rc\rc
&+&{k-f\over\sinh\tau}\left(e^{\hat3}\wedge e^{\hat2}\wedge
e^2+e^{\hat3}\wedge e^{\hat1}\wedge e^1\right)\Bigg\}\,.
\label{kshflat}
\eear

Now we are ready to write the equations resulting from the vanishing of
the SUSY variations (\ref{sugra}). For this background with constant
dilaton and three and five-form fluxes they reduce to the expressions
written in the equations (\ref{ktdilvar}) and (\ref{ktgravar}) of the
previous chapter. Since we expect that the Killing
spinors will only depend on the radial variable $\tau$, we will write
the SUSY equations directly in the frame (\ref{ksframe}).
\subsection{Dilatino SUSY equation}
Using the expressions for the
three-forms written above, and recalling the definition of ${\cal F}^3$
given in (\ref{comp3form}), the equation (\ref{ktdilvar}) resulting
from the vanishing of the variation of the dilatino takes the form:
\bear
&&\Bigg\{{f'+k'\over2\cosh\tau}\Gamma_{\tau\hat2\hat1}+{1\over2\sinh\tau}
\left[(1-g)\,k'-(1+g)\,f'\right]\left(\Gamma_{\tau
\hat2 1}+
\Gamma_{\tau
2\hat1}\right)+\rc\rc&&+{\cosh\tau\over2\sinh^2\tau}\bigg[(1+g)^2\,f'+
(1-g)^2\,k'\bigg]\Gamma_{\tau 21}+{k-f\over2\sinh\tau}\left(
\Gamma_{\hat3\hat{2}2}+\Gamma_{\hat3\hat{1}1}\right)+\rc\rc&&+{i\over2\cosh\tau}
\Gamma_{\hat3\hat{2}\hat1}+{i\over2\sinh\tau}\left(1-g-2F\right)\left(
\Gamma_{\hat{3} 2\hat1}+\Gamma_{\hat3\hat{2}1}\right)+\rc\rc
&&+{i\cosh\tau\over2\sinh^2\tau}\left[1+g\,(4F-2+g)\right]\Gamma_{\hat{3}21}
+{iF'\over\sinh\tau}\left(\Gamma_{\tau\hat{2}2}+\Gamma_{\tau\hat{1}1}\right)
\Bigg\}\,\epsilon=0\,,
\label{ktdileq}
\eear
where $\Gamma_a\,,\;(a=x^\alpha,\tau,1,2,\hat1,\hat2,\hat3)$ are
constant Dirac matrices associated to the frame (\ref{ksframe}) and we
have neglected the common factor
$-{4\sqrt{6}\,h^{-{3\over4}}\over\mu^2}{M\,\alpha'\over2}$.

As we did in chapter \ref{genconsec} (see eq.
(\ref{2cproj})) we will impose the angular projection
\beq
\Gamma_{12}\,\epsilon=-\Gamma_{\hat1\hat2}\,\epsilon\,\,.
\label{ks2cproj}
\eeq
Furthermore, the Killing spinors of the resolutions of the
conifold are subjected to the projection (\ref{betaproj}), which for the
particular case of the deformed conifold (then, taking into account eq.
(\ref{gcdcalpha})) becomes:
\beq
\Gamma_\tau\Gamma_{\hat1\hat2\hat3}\,\epsilon=\left(-{\sinh\tau\over\cosh\tau}
+{1\over\cosh\tau}
\,\Gamma_{\hat11}\right)\epsilon\,.
\label{ksdcproj}
\eeq
We also impose the projection corresponding to a D3-brane extended
along the Minkowski directions:
\beq
\Gamma_{\,x^0\,x^1\,x^2\,x^3}\,\epsilon=-i\,\epsilon\,.
\label{ksd3proj}
\eeq
We will show below that this projection follows from the
vanishing of the gravitino SUSY variation as it happened for the
Klebanov-Tseytlin model (see eq. (\ref{ktd3proj})). Since we are
working in type IIB SUGRA, a 10d spinor $\epsilon$ satisfies the
equality:
\beq
\Gamma_{x^0x^1x^2x^3}\Gamma_\tau
\Gamma_{12\hat1\hat2\hat3}\,\epsilon=-\epsilon\,,
\label{kschirspin}
\eeq
which combined with projections (\ref{ks2cproj}) and (\ref{ksd3proj})
gives rise to
\beq
\Gamma_{\tau\hat3}\,\epsilon=-i\,\epsilon\,.
\label{kscombproj}
\eeq
Using this last projection, the one written in (\ref{ksdcproj}), and
some suitable combinations of both ones, equation (\ref{ktdileq})
becomes:
\beq
\left(\,-i\,P_1\,\Gamma_{\hat3}+P_2\,\Gamma_{\hat3 1\hat1}\,\right)\,
\epsilon=0\,,
\label{ksdileqsch}
\eeq
where
\bear
P_1&=&-i{\sinh\tau\over2\cosh^2\tau}\,(f'+k'+1)+{i\over2\sinh\tau}\,\Big[(1+g)^2\,f'
+(1-g)^2\,k'+1+g\,(4F-2+g)\Big]+\rc\rc&+&{i\over\sinh\tau\,\cosh\tau}\,\left[(1-g)\,k'-
(1+g)\,f'+1-g-2F\,\right]\,,\rc
\label{ktdileqp1}
\eear
and
\bear
P_2&=&{-1\over2\cosh^2\tau}\,(f'+k'+1)+{1\over2\sinh^2\tau}\left[(1+g)^2\,f'+
(1-g)^2\,k'+1+g\,(4F-2+g)\right]+\rc\rc&-&{1\over\cosh\tau}\,\left[(1-g)\,k'
-(1+g)\,f'+1-g-2F\,\right]+{1\over\sinh\tau}\,(2F'-k+f)\;.\rc
\label{ktdileqp2}
\eear
Using that $g={1\over\cosh\tau}$ it is not difficult to see that
$P_1$ automatically vanishes. Hence, we are left with the equation
$P_2=0$, which by substituting the value of $g$ is brought into the
form:
\beq
2F'+\coth\left({\tau\over2}\right)f'-\tanh\left({\tau\over2}\right)k'
+2\coth\tau \,F+f-k=\tanh\left({\tau\over2}\right)\,.
\label{ksdildiffeq}
\eeq
So the vanishing of the dilatino SUSY variation results in this
differential equation for the functions of the ansatz.

\subsection{Gravitino Minkowski components}

Let us now study the SUSY variation of the gravitino, \ie \ equations
(\ref{ktgravar}). All the components along the Minkowski
directions yield the same equation, so for illustrative purposes
we will write the equation corresponding to the  $x^1$ component, namely
$\delta\psi_{x^1}=0$. Looking back at eq. (\ref{ktgravar}) one can write:
\beq
D_{x^1}\,\epsilon\,+\,{i\over 1920}\,
F_{N_1\cdots N_5}^{(5)}\,\Gamma^{N_1\cdots
N_5}\Gamma_{x^1}\,\epsilon
+{1\over96}\,{\cal F}^{(3)}_{N_1N_2N_3}\,
\big(\,\Gamma_{x^1}^{\,\,\,N_1N_2N_3}\,-\,
9\,\delta_{x^1}^{N_1}\,\,\Gamma^{N_2 N_3}\,\big)\,\epsilon^{*}=0\,,
\label{ksgravx1}
\eeq
We will analyze the different pieces of this
equation separately. Recalling the expression of the covariant derivative
and assuming that
$\epsilon$ does not depend on the Minkowski coordinates, the first piece
can be written as
${1\over4}\,\omega^{a\,b}_{x^1}\,\Gamma_{a\,b}\,\epsilon$. Using
the spin connection written in (\ref{kspincon}), one gets that
\beq
D_{x^1}\,\epsilon={1\over4}\,\omega^{a\,b}_{x^1}\,\Gamma_{a\,b}\,\epsilon=-{1\over8}
{\sqrt{6}\over\mu^{2\over3}}\,K(\tau)\,h^{-{5\over4}}\,h'\,\Gamma_{x^1\tau}\,\epsilon\,.
\label{ksgravx1s1}
\eeq
Inserting the five-form written in (\ref{ks5formflat}) into the second
piece of equation (\ref{ksgravx1}) it becomes:
\beq
{i\over 1920}\,
F_{N_1\cdots N_5}^{(5)}\,\Gamma^{N_1\cdots N_5}\Gamma_{x^1}\,\epsilon=
{i\over8}{\sqrt{6}\over\mu^{2\over3}}\,K(\tau)\,h^{-{5\over4}}\,h'\,
\Gamma_{x^0 x^1 x^2 x^3}\Gamma_{\tau x^1}\,\epsilon\,,
\label{ksgravx1s2}
\eeq
where we have used eq. (\ref{kschirspin}). So in view of these
last two equalities (\ie \ (\ref{ksgravx1s1}) and (\ref{ksgravx1s2})),
the first two pieces of eq. (\ref{ksgravx1}) can be written as
\beq
D_{x^1}\,\epsilon\,+\,{i\over 1920}\,
F_{N_1\cdots N_5}^{(5)}\,\Gamma^{N_1\cdots
N_5}\Gamma_{x^1}\,\epsilon={1\over8}
{\sqrt{6}\over\mu^{2\over3}}\,K(\tau)\,h^{-{5\over4}}\,h'\,\Gamma_{\tau
x^1}\,\left(1-i\Gamma_{x^0 x^1 x^2 x^3}\right)\,\epsilon\,.
\label{ksgravs12}
\eeq
In order to make this expression vanish we should impose the projection
(\ref{ksd3proj}). Since, as we will see below, the remaining terms in
equation (\ref{ksgravx1}) do not mix up with these ones; it becomes
clear that we must impose that projection to satisfy the equation
(\ref{ksgravx1}). 

Let us now look at the third term of equation (\ref{ksgravx1}),
the one depending on the three-form ${\cal F}^{(3)}$. Taking into
account the vanishing of the first two pieces and that the three-form
has no components along $x^1$, equation (\ref{ksgravx1}) reduces to
\beq
{\cal F}^{(3)}_{N_1N_2N_3}\,\Gamma^{\,N_1N_2N_3}\,\epsilon^*=0\,,
\eeq
which is very similar to the equation resulting from the vanishing of the
dilatino SUSY variation. In fact, this equation is equal to equation
(\ref{ktdileq}) but with $\epsilon^*$ instead of $\epsilon$. Therefore,
proceeding as we did there, but using the conjugated projections (for
instance $\Gamma_{\tau\hat3}\,\epsilon^*=i\,\epsilon^*$ instead of
(\ref{kscombproj})), we arrive at the following equation:
\beq
\left(\,i\,\hat P_1\,\Gamma_{\hat3}+\hat P_2\,\Gamma_{\hat3
1\hat1}\,\right)\,
\epsilon^*=0\,,
\label{ksgravx1eqsch}
\eeq
where
\bear
\hat P_1&=&i{\sinh\tau\over2\cosh^2\tau}\,(f'+k'-1)-{i\over2\sinh\tau}\,
\Big[(1+g)^2\,f'+(1-g)^2\,k'-1-g\,(4F-2+g)\Big]-\rc\rc&-&{i\over\sinh\tau\,\cosh\tau}\,\left[(1-g)\,k'-
(1+g)\,f'-1+g+2F\,\right]\,,\rc
\label{ksgravx1p1}
\eear
and
\bear
\hat P_2&=&{-1\over2\cosh^2\tau}\,(f'+k'-1)+{1\over2\sinh^2\tau}
\left[(1+g)^2\,f'+ (1-g)^2\,k'-1-g\,(4F-2+g)\right]-
\rc\rc&-&{1\over\cosh\tau}\,\left[(1-g)\,k'-(1+g)\,
f'-1+g+2F\right]-{1\over\sinh\tau}\,(2F'+k-f)\;.\rc
\label{ksgravx1p2}
\eear
Using again that $g={1\over\cosh\tau}$ it is not difficult to see
that $\hat P_1$ is identically zero. In fact, if in the expression of
$\hat P_1$ one changes $f'\to-f'$ and $k'\to-k'$ one arrives at
(\ref{ktdileqp1}), \ie \ at $P_1$. Since $P_1$ vanishes independently
of the form of $f'$ and
$k'$, then $\hat P_1$ must also vanish. Then, in order to satisfy the
equation $\delta\psi_{x^1}=0$, $\hat P_2$ must vanish. Inserting the value
of
$g$, the equation $\hat P_2=0$ can be written as 
\beq
2F'-\coth\left({\tau\over2}\right)f'+\tanh\left({\tau\over2}\right)k'
+2\coth\tau \,F-f+k=\tanh\left({\tau\over2}\right)\,,
\label{ksgravx1diffeq}
\eeq
which is another differential equation for the unknown functions of the
model.

\subsection{Gravitino angular components}

We still have to solve the equations resulting from the angular
components of the gravitino. We will begin with the equation
$\delta\psi_1=0$, namely:
\beq
D_1\,\epsilon\,+\,{i\over 1920}\,
F_{N_1\cdots N_5}^{(5)}\,\Gamma^{N_1\cdots
N_5}\Gamma_1\,\epsilon
+{1\over96}\,{\cal F}^{(3)}_{N_1N_2N_3}\,
\big(\,\Gamma_1^{\,\,\,N_1N_2N_3}\,-\,
9\,\delta_1^{N_1}\,\,\Gamma^{N_2 N_3}\,\big)\,\epsilon^{*}=0\,.
\label{ksgrava1}
\eeq
As for the $x^1$ equation we will study each contribution to this
equation separately. Reading the spin connection from eq.
(\ref{kspincon}), and assuming that $\epsilon$ does not depend on the
compact coordinates of the $T^{1,1}$, the first term becomes:
\beq
D_1\,\epsilon={1\over4}\,\omega^{a\,b}_1\,\Gamma_{a\,b}\,\epsilon={1\over8}
{\sqrt{6}\over\mu^{2\over3}}\,K(\tau)\,h^{-{5\over4}}\,h'\,\Gamma_{1
\tau}\,\epsilon\,.
\label{ksgrava1s1}
\eeq
We have taken into account that all the terms in
$\omega^{a\,b}_1\,\Gamma_{a\,b}\,\epsilon$, except for the one depending
on $h'$, will cancel each other by virtue of eq. (\ref{susyeq2}) in
chapter \ref{genconsec}. This is so because, as we have already said,
the one-form frame (\ref{ksframe}) only differs from the one in eq.
(\ref{gcdefconframe}) in some  $h$ factors. Indeed, one can write
$\omega^{a\,b}_1\,\Gamma_{a\,b}\,\epsilon=
h^{-{1\over4}}\,\omega^{\;\;\bar
a\,\bar b}_{(o)\,1}\,\Gamma_{\bar a\,\bar b}\,\epsilon+{\rm
terms}\,(h')$; where $\omega^{\;\;\bar
a\,\bar b}_{(o)}$ stands for the spin connection of the deformed
conifold (eq. (\ref{spincon}) restricted to the particular solution
(\ref{defconsysol})) and the indices $\bar a, \bar b$ refer to the
corresponding frame, \ie \ (\ref{gcdefconframe}). Finally, from the
analysis done in chapter \ref{genconsec}, it is clear that
$\omega^{\;\;\bar a\,\bar b}_{(o)\,1}\,\Gamma_{\bar a\,\bar
b}\,\epsilon=0$ follows from the more general equation (\ref{susyeq2}).

Inserting the RR five-form as it is written in eq. (\ref{ks5formflat})
and making use of eq. (\ref{kschirspin}), the second term of
(\ref{ksgrava1}) takes the form:
\beq
{i\over 1920}\,
F_{N_1\cdots N_5}^{(5)}\,\Gamma^{N_1\cdots N_5}\Gamma_1\,\epsilon=
{i\over8}{\sqrt{6}\over\mu^{2\over3}}\,K(\tau)\,h^{-{5\over4}}\,h'\,
\Gamma_{x^0 x^1 x^2 x^3}\Gamma_{\tau 1}\,\epsilon\,,
\label{ksgrava1s2}
\eeq
and by adding it to the first term written in eq. (\ref{ksgrava1s1}),
one gets that
\beq
D_1\,\epsilon\,+\,{i\over 1920}\,
F_{N_1\cdots N_5}^{(5)}\,\Gamma^{N_1\cdots
N_5}\Gamma_1\,\epsilon={1\over8}
{\sqrt{6}\over\mu^{2\over3}}\,K(\tau)\,h^{-{5\over4}}\,h'\,\Gamma_{\tau
1}\,\left(-1+i\Gamma_{x^0 x^1 x^2 x^3}\right)\,\epsilon\,,
\label{ksgrava1s12}
\eeq
 which vanishes for a spinor $\epsilon$ satisfying the
projection (\ref{ksd3proj}). 

It only remains the third term in eq. (\ref{ksgrava1}). Let us
study separately the two terms multiplying the complex three-form
${\cal F}^{(3)}$. Reading the three-form fluxes from (\ref{ksf3flat})
and (\ref{kshflat}), the first term, \ie \  
${1\over96}\,{\cal F}^{(3)}_{N_1N_2N_3}\,\Gamma_1^{\,\,\,N_1N_2N_3}\,
\epsilon^*$, becomes:
\bear
&&-{\sqrt{6}\,h^{-{3\over4}}\over\mu^2}{M\,\alpha'\over8}\,\Gamma_1
\Bigg\{{f'+k'\over2\cosh\tau}\Gamma_{\tau\hat2\hat1}+{k-f\over2\sinh\tau}\,
\Gamma_{\hat3\hat{2}2}+{1\over2\sinh\tau}
\left[(1-g)\,k'-(1+g)\,f'\right]\,\Gamma_{\tau
2\hat1}+\rc\rc&&+
{i\over2\cosh\tau}\,\Gamma_{\hat3\hat{2}\hat1}+
{i\over2\sinh\tau}\left(1-g-2F\right)\,\Gamma_{\hat{3} 2\hat1}
+{iF'\over\sinh\tau}\,\Gamma_{\tau\hat{2}2}\Bigg\}\,\epsilon^*\,.
\label{ksgrava1s31}
\eear
Imposing the projections (\ref{ks2cproj}) and (\ref{kscombproj}), which
after complex conjugation become:
$\Gamma_{12}\,\epsilon^*=-\Gamma_{\hat1\hat2}\,\epsilon^*$ and
$\Gamma_{\tau\hat3}\,\epsilon^*=i\,\epsilon^*$, and neglecting the
common factor $-{\sqrt{6}\,h^{-{3\over4}}\over\mu^2}{M\,\alpha'\over8
}\,\Gamma_1$,
equation (\ref{ksgrava1s31}) takes the form:
\bear
&&\Bigg\{i\left[{f'+k'\over2\cosh\tau}-{1\over2\cosh\tau}\right]
\Gamma_{\hat3\hat1\hat2}-{1\over\sinh\tau}\left(F'+{k-f\over2}
\right)\Gamma_{\hat{3} 1\hat1}+\rc\rc
&&+i\left\{{1\over2\sinh\tau}
\left[(1-g)\,k'-(1+g)\,f'\right]-{1\over2\sinh\tau}\left(1-g-2F\right)
\right\}\Gamma_{\hat{3} 1\hat2}\Bigg\}\epsilon^*\,,
\label{ksgrava1s31ex2}
\eear
which by using suitable combinations of the complex
conjugate of projections (\ref{ksdcproj}) and (\ref{kscombproj}) can be
written as
\beq
\left(\, Q_1\,\Gamma_{\hat3}+ Q_2\,\Gamma_{\hat3
1\hat1}\,\right)\,
\epsilon^*\,,
\label{ksgrava1eqsch}
\eeq
with
\bear
Q_1=-{1\over2}\Bigg\{{\sinh\tau\over\cosh^2\tau}\,(f'+k'-1)
-{1\over\sinh\tau\,\cosh\tau}\,\left[(1-g)\,k'-
(1+g)\,f'-1+g+2F\,\right]\Bigg\}\,,\rc
\label{ksgrava1q1}
\eear
and
\bear
Q_2=&-&{1\over2}\Bigg\{{1\over\cosh^2\tau}\,(f'+k'-1)
+{1\over\cosh\tau}\,\left[(1-g)\,k'-
(1+g)\,f'-1+g+2F\,\right]\Bigg\}-\rc\rc&-&{1\over\sinh\tau}\left(F'+
{k-f\over2}\right)
\,.
\label{ksgrava1q2}
\eear
As we will see, if we do not introduce any extra projection, the
remaining terms in equation (\ref{ksgrava1}) will not mix up with
these ones. Therefore one must require that $Q_1=0$, and $Q_2=0$.
Furthermore, recalling that $g={1\over\cosh\tau}$ one can
straightforwardly check that the equation $Q_1=0$ takes the form:
\beq
\coth\left({\tau\over2}\right)f'+\tanh\left({\tau\over2}\right)k'
-{2F\over\sinh\tau}=\tanh\left({\tau\over2}\right)\,,
\label{ksgrava1q1diffeq}
\eeq
and the equation $Q_2=0$ can be written as
\beq
f'-k'-2F+1-2\coth\tau\left(F'+{k-f\over2}\right)=0\,.
\label{ksgrava1q2diffeq}
\eeq
We will now write down the second term depending on the complex
three-form, namely $- {9\over96}\,{\cal
F}^{(3)}_{N_1N_2N_3}\,\delta_1^{N_1}\Gamma^{N_2 N_3}\,\epsilon^*$, which
after inserting the three-forms (\ref{ksf3flat}) and
(\ref{kshflat}) becomes:
\bear
&&{\sqrt{6}\,h^{-{3\over4}}\over\mu^2}{3M\,\alpha'\over8}
\Bigg\{{i\over2\sinh\tau}\left(1-g-2F\right)\,\Gamma_{\hat{3}
\hat2}+{i\cosh\tau\over2\sinh^2\tau}\left[1+g\,(g-2+4F)\right]\,
\Gamma_{\hat32}+\rc\rc &&+{iF'\over\sinh\tau}\,\Gamma_{\tau\hat1}+
{1\over2\sinh\tau}\left[(1-g)\,k'-(1+g)\,f'\right]\,\Gamma_{\tau
\hat2}+{k-f\over2\sinh\tau}\,\Gamma_{\hat3\hat1}+\rc\rc&&+{\cosh\tau\over2\sinh^2\tau}\left[(1-g)^2\,k'
+(1+g)^2\,f'\right]\,\Gamma_{\tau
2}\Bigg\}\,\epsilon^*\,,
\label{ksgrava1s32}
\eear
and, after making use of the complex conjugate of projections
(\ref{ks2cproj}) and (\ref{kscombproj}), and neglecting the common factor
${\sqrt{6}\,h^{-{3\over4}}\over\mu^2}{3M\,\alpha'\over8}\,$, it reads:
\\
\bear
&&\Bigg\{{i\over2\sinh\tau}\left[1-g-2F-(1-g)\,k'+(1+g)
\,f'\right]\Gamma_{\hat3\hat2}+
{1\over2\sinh\tau}\left(2F'+k-f\right)\,\Gamma_{\hat3\hat1}
+\rc\rc&&+{i\cosh\tau\over2\sinh^2\tau}\left[
1+g\,(g-2+4F)-(1-g)^2\,k'-(1+g)^2\,f'\right]\Gamma_{1\hat3\hat1\hat2}
\Bigg\}\,\epsilon^*\,.
\label{ksgrava1s32ex2}
\eear
By imposing suitable combinations of the complex
conjugate of projections (\ref{ksdcproj}) and (\ref{kscombproj}), this
last expression can be written as
\beq
\left(\,\hat Q_1\,\Gamma_{\hat3 1}+ \hat Q_2\,\Gamma_{\hat3
\hat1}\,\right)\,\epsilon^*\,,
\label{ksgrava1eqsch2}
\eeq
where
\bear
\hat
Q_1&=&{1\over2\sinh\tau\,\cosh\tau}\,\left[1-g-2F-(1-g)\,k'+(1+g)
\,f'\right]+\rc\rc&&+\,{1\over2\sinh\tau}\left[
1+g\,(g-2+4F)-(1-g)^2\,k'-(1+g)^2\,f'\right]\,,
\label{ksgrava1hatq1}
\eear
and
\bear
\hat Q_2&=&{1\over2\cosh\tau}\,\left[-1+g+2F+(1-g)\,k'
-(1+g)\,f'\right]+{1\over\sinh\tau}\left(F'+
{k-f\over2}\right)+\rc\rc&&+\,{1\over2\sinh^2\tau}\left[
1+g\,(g-2+4F)-(1-g)^2\,k'-(1+g)^2\,f'\right]\,.
\label{ksgrava1hatq2}
\eear
In order to satisfy the equation $\delta\psi_1=0$ (\ie \ eq.
(\ref{ksgrava1})) without imposing new projections on $\epsilon$ we must
require that
$\hat Q_1=0$, and $\hat Q_2=0$. Using that $g={1\over\cosh\tau}$ one
can easily check that $\hat Q_1=0$ renders the same
differential equation as the expression $Q_1=0$, namely 
equation (\ref{ksgrava1q1diffeq}). After inserting the value of $g$, the
expression
$\hat Q_2=0$ yields the following differential equation:
\bear
\tanh^2\left({\tau\over2}\right)k'-\coth^2\left({\tau\over2}\right)f'
+2\left(\coth^2\tau+{\rm
csch}^2\tau\right)F+2\coth\tau\left(F'+{k-f\over2}\right)=\tanh^2\left({\tau\over2}
\right)\,.\rc
\label{ksgrava1qhat2diffeq}
\eear
Then, from the equation $\delta\psi_1=0$, we have got three differential
equations for the unknown functions of the Klebanov-Strassler ansatz;
these are the equations (\ref{ksgrava1q1diffeq}),
(\ref{ksgrava1q2diffeq}), and (\ref{ksgrava1qhat2diffeq}).

The vanishing of the SUSY variation of
$\psi_2$, (\ie \ $\delta\psi_2=0$) results in the same differential
equations as the ones we got above from requiring that $\delta\psi_1=0$.
As it happened before, the term coming from the covariant derivative and
the one containing the RR five-form cancel each other after using the
projection (\ref{ksd3proj}). Furthermore, if one imposes suitable
combinations of the complex conjugate of projections (\ref{ksdcproj}) and
(\ref{kscombproj}), the terms containing the complex three-form
${\cal F}^{(3)}$ result to be equal to the ones appearing in
$\delta\psi_1=0$ and therefore, the arising differential equations are the
same ones as before.

Let us now impose the cancellation of the SUSY variation of $\psi_{\hat
3}$. The equation we have to solve is:
\beq
D_{\hat 3}\,\epsilon\,+\,{i\over 1920}\,
F_{N_1\cdots N_5}^{(5)}\,\Gamma^{N_1\cdots
N_5}\Gamma_{\hat 3}\,\epsilon
+{1\over96}\,{\cal F}^{(3)}_{N_1N_2N_3}\,
\big(\,\Gamma_{\hat 3}^{\,\,\,N_1N_2N_3}\,-\,
9\,\delta_{\hat 3}^{N_1}\,\,\Gamma^{N_2 N_3}\,\big)\,\epsilon^{*}=0\,.
\label{ksgravahat3}
\eeq
We will follow the same steps as for the preceding components of the
gravitino. Then, let us write down the form of the first term of this
last equation for a spinor $\epsilon$ independent of the compact
coordinates of the $T^{1,1}$. It reads:
\beq
D_{\hat 3}\,\epsilon={1\over4}\,\omega^{a\,b}_{\hat
3}\,\Gamma_{a\,b}\,\epsilon={1\over8}
{\sqrt{6}\over\mu^{2\over3}}\,K(\tau)\,h^{-{5\over4}}\,h'\,\Gamma_{\hat 3
\tau}\,\epsilon\,,
\label{ksgravahat3s1}
\eeq
where again we have used the fact that all terms in
$\omega^{a\,b}_{\hat 3}\,\Gamma_{a\,b}\,\epsilon$, apart from the ones
depending on $h'$, cancel each other as we have explained below eq.
(\ref{ksgrava1s1}). Reading the RR five-form from eq.
(\ref{ks5formflat}) and using eq. (\ref{kschirspin}), the second term 
of eq. (\ref{ksgravahat3}) becomes:
\beq
{i\over 1920}\,
F_{N_1\cdots N_5}^{(5)}\,\Gamma^{N_1\cdots N_5}\Gamma_{\hat 3}\,\epsilon=
{i\over8}{\sqrt{6}\over\mu^{2\over3}}\,K(\tau)\,h^{-{5\over4}}\,h'\,
\Gamma_{x^0 x^1 x^2 x^3}\Gamma_{\tau {\hat 3}}\,\epsilon\,,
\label{ksgravahat3s2}
\eeq
and one can easily check that by imposing the projection (\ref{ksd3proj})
on $\epsilon$, this last expression cancels the term written in eq.
(\ref{ksgravahat3s1}). Thus, as before, the terms coming from the
covariant derivative and from the five-form term cancel each other.
Then, we are left with the terms containing the complex three-form.
Making use of the expressions for the three-forms written
in eqs. (\ref{ksf3flat}) and (\ref{kshflat}) the first term containing
${\cal F}^{(3)}$ can be written as
\bear
&&{1\over96}\,{\cal F}^{(3)}_{N_1N_2N_3}\,\Gamma_{\hat
3}^{\,\,\,N_1N_2N_3}\,
\epsilon^*=-{\sqrt{6}\,h^{-{3\over4}}\over\mu^2}{M\,\alpha'\over8}
\,\Gamma_{\hat 3}\Bigg\{{iF'\over\sinh\tau}\left(\Gamma_{\tau\hat{2}2}+
\Gamma_{\tau\hat{1}1}\right)+
{f'+k'\over2\cosh\tau}\Gamma_{\tau\hat2\hat1}+\rc\rc&&+
{1\over2\sinh\tau}
\left[(1-g)\,k'-(1+g)\,f'\right]\left(\Gamma_{\tau
\hat 2 1}+\Gamma_{\tau
2\hat1}\right)+{\cosh\tau\over2\sinh^2\tau}\left[
(1-g)^2\,k'+(1+g)^2\,f'\right]\,\Gamma_{\tau 2 1}
\Bigg\}\,\epsilon^*\,,\rc
\label{ksgravahat3s31}
\eear
and imposing the projections
$\Gamma_{12}\,\epsilon^*=-\Gamma_{\hat1\hat2}\,\epsilon^*$, and
$\Gamma_{\tau\hat3}\,\epsilon^*=i\,\epsilon^*$ (complex conjugate of
(\ref{ks2cproj}) and (\ref{kscombproj}) respectively) it becomes:
\bear
&&\Bigg\{i\left[{f'+k'\over2\cosh\tau}-{\cosh\tau\over2\sinh^2\tau}
\left[(1-g)^2\,k'+(1+g)^2\,f'\right]\right]\Gamma_{\hat 3\hat1\hat2}
-{2F'\over\sinh\tau}\Gamma_{\hat 3 1\hat1}+\rc\rc&& +{i\over\sinh\tau}
\left[(1-g)\,k'-(1+g)\,f'\right]\Gamma_{\hat 3 1\hat2}
\Bigg\}\,\epsilon^*\,,
\label{ksgravahat3s31ex2}
\eear
where we have neglected the common factor 
$-{\sqrt{6}\,h^{-{3\over4}}\over\mu^2}{M\,\alpha'\over8}\,\Gamma_{\hat
3}$.

Again, if we insert some combinations of the complex
conjugate of projections (\ref{ksdcproj}) and (\ref{kscombproj}), this
last expression can be written as the sum of two independent terms:
\beq
\left(\, M_1\,\Gamma_{\hat3}+ M_2\,\Gamma_{\hat3
1\hat1}\,\right)\,
\epsilon^*\,,
\label{ksgravahat3eqsch}
\eeq
with
\bear
M_1=&-&{\sinh\tau\over2\cosh^2\tau}(f'+k')+{1\over2\sinh\tau}
\left[(1-g)^2\,k'+(1+g)^2\,f'\right]+\rc\rc&+&{1\over\sinh\tau\cosh\tau}
\left[(1-g)\,k'-(1+g)\,f'\right]\,,
\label{ksgravahat3m1}
\eear
which results to be identically zero after substituting $g$ by its value,
\ie \ ${1\over\cosh\tau}$. On the other hand,
\bear
M_2=&-&{f'+k'\over2\cosh^2\tau}+{1\over2\sinh^2\tau}
\left[(1-g)^2\,k'+(1+g)^2\,f'\right]-\rc\rc&-&{1\over\cosh\tau}
\left[(1-g)\,k'-(1+g)\,f'\right]-{2F'\over\sinh\tau}\,. 
\label{ksgravahat3m2}
\eear
and, since the remaining terms in $\delta\psi_{\hat3}=0$
(eq. (\ref{ksgravahat3})) will not mix up with this last one, one must
have $M_2=0$. Hence, by inserting $g={1\over\cosh\tau}$ we get the
following differential equation:
\beq
\coth\left({\tau\over2}\right)f'-\tanh\left({\tau\over2}\right)k'-2F'=0\,.
\label{ksgravahat3diffeq1}
\eeq
The last term in eq. (\ref{ksgravahat3}) is:
\bear
&&-{9\over96}\,{\cal F}^{(3)}_{N_1N_2N_3}\,\delta_{\hat
3}^{N_1}\Gamma^{N_2 N_3}\,\epsilon^*=
{\sqrt{6}\,h^{-{3\over4}}\over\mu^2}{3M\,\alpha'\over8}
\Bigg\{{k-f\over2\sinh\tau}\left(\Gamma_{\hat 2 2}+\Gamma_
{\hat1 1}\right)+{i\over2\cosh\tau}\,\Gamma_{\hat 2 \hat
1}+\rc\rc&&+{i\over2\sinh\tau}\left(1-g-2F\right)\,\left(\Gamma_{\hat 2
1}+
\Gamma_{2 \hat1}\right)+{i\cosh\tau\over2\sinh^2\tau}\left[1+g\,(g-2+4F)
\right]\,\Gamma_{2 1}\Bigg\}\,\epsilon^*\,,\rc
\label{ksgravahat3s32}
\eear
where we have used the expressions of the three-forms written in eqs.
(\ref{ksf3flat}) and (\ref{kshflat}). Making use of the projection 
$\Gamma_{12}\,\epsilon^*=-\Gamma_{\hat1\hat2}\,\epsilon^*$ (see eq.
(\ref{ks2cproj})) and neglecting the common
factor
${\sqrt{6}\,h^{-{3\over4}}\over\mu^2}{3M\,\alpha'\over8}\,$,
this last expression becomes:
\bear
\Bigg\{{k-f\over\sinh\tau}\Gamma_{\hat1 1}+{i\over\sinh\tau}
\left(1-g-2F\right)\Gamma_{\hat 2 1}
+i\left[{\cosh\tau\over2\sinh^2\tau}
\left[1+g\,(g-2+4F)\right]-{1\over2\cosh\tau}\right]
\Gamma_{\hat1\hat2}\Bigg\}\,\epsilon^*\,,\rc
\label{ksgravahat3s32ex2}
\eear
which by imposing suitable combinations of projections (\ref{ksdcproj})
and (\ref{kscombproj}) can be written as
\beq
\left(\, \hat M_1+ \hat M_2\,\Gamma_{\hat1 1}\,\right)\,
\epsilon^*\,,
\label{ksgravahat3eqsch2}
\eeq
where
\bear
\hat
M_1=-{1\over2\sinh\tau}\left[1+g\,(g-2+4F)\right]+
{\sinh\tau\over2\cosh^2\tau}-{1\over\sinh\tau\,\cosh\tau}
\left(1-g-2F\right),
\label{ksgravahat3hatm1}
\eear
and,
\bear
\hat
M_2={1\over2\sinh^2\tau}\left[1+g\,(g-2+4F)\right]-
{1\over2\cosh^2\tau}-{1\over\cosh\tau}\left(1-g-2F\right)+
{k-f\over\sinh\tau}\,.
\label{ksgravahat3hatm2}
\eear
One can readily check that by inserting $g={1\over\cosh\tau}$, \ $\hat
M_1=0$ automatically. In order to satisfy the equation
$\delta\psi_{\hat 3}=0$ we must also have $\hat M_2=0$, which after
substituting
$g={1\over\cosh\tau}$ \ yields an algebraic relation between the
functions entering the Klebanov-Strassler ansatz, namely:
\beq
2\coth\tau\,F+k-f=\tanh\left({\tau\over2}\right)\,.
\label{ksgravahat3alg}
\eeq
Requiring the vanishing of the SUSY variation of the remaining angular
components of the gravitino (\ie \ $\psi_{\hat1}$ and $\psi_{\hat2}$)
will not give rise to any new equation relating the functions of the
ansatz. Indeed, from the equations
$\delta\psi_{\hat1}=\delta\psi_{\hat2}=0$ one gets the same equations as
from imposing $\delta\psi_1=0$; these are eqs. (\ref{ksgrava1q1diffeq}),
(\ref{ksgrava1q2diffeq}) and (\ref{ksgrava1qhat2diffeq}).

\subsection{Gravitino radial component}

Finally, we shall look at the SUSY variation of the radial component of
the gravitino. Then, we must solve the equation
\beq
D_\tau\,\epsilon\,+\,{i\over 1920}\,
F_{N_1\cdots N_5}^{(5)}\,\Gamma^{N_1\cdots
N_5}\Gamma_\tau\,\epsilon
+{1\over96}\,{\cal F}^{(3)}_{N_1N_2N_3}\,
\big(\,\Gamma_\tau^{\,\,\,N_1N_2N_3}\,-\,
9\,\delta_\tau^{N_1}\,\,\Gamma^{N_2 N_3}\,\big)\,\epsilon^{*}=0\,.
\label{ksgravtau}
\eeq
As it obviously implies the radial projection (\ref{ksdcproj}), the spinor
$\epsilon$ depends on the radial coordinate. Then, the covariant
derivative can be written in terms of
$\epsilon'\equiv{d\epsilon\over d\tau}$ as
$D_\tau\,\epsilon=\left(E^\tau_{\tilde \tau}\right)^{-1}
\left(\epsilon'+{1\over4}\,\omega^{a\,b}_{\tilde \tau}\,\Gamma_{a\,b}\,
\epsilon\right)$.  Thus, reading the spin
connection one-form from eq. (\ref{kspincon}), the covariant
derivative becomes:
\beq
D_\tau\,\epsilon={\sqrt{6}\,K(\tau)\over \mu^{{2\over 3}}\,h^{1\over4}}
\left(\epsilon'+{1\over2\cosh\tau}\Gamma_{\hat1 1}\,\epsilon\right)\,,
\label{ksgravtaus1}
\eeq
where we have already imposed the projection (\ref{ks2cproj}) and we have
taken into account that 
$\omega^{ab}_{\tilde
\tau}=E^\tau_{\tilde
\tau}\,\omega^{ab}_\tau={\mu^{2\over3}\over\sqrt2}\,
{h^{1\over4}\over\sqrt{3}\,K(\tau)}\,\omega^{ab}_\tau$.

We will show below that the terms in eq. (\ref{ksgravtau}) containing the
complex three-form will vanish, so it must happen again that the first
two terms in that equation cancel each other. Thus, after inserting the RR
five-form (\ref{ks5formflat}), eq. (\ref{kschirspin}) and projection 
(\ref{ksd3proj}), we get the following equation:
\beq
\epsilon'+{1\over2\cosh\tau}\Gamma_{\hat1 1}\,\epsilon+{1\over8}h^{-1}\,h'
\,\epsilon=0\,.
\label{ksgravtaus12}
\eeq
At this point let us
go back to the radial projection written in eq. (\ref{ksdcproj}) and
notice that it can be solved as
\beq
\epsilon=e^{-{1\over2}\alpha\Gamma_{\hat1
1}}\,\epsilon_0\;\;,\quad\Gamma_{\tau\hat1\hat2\hat3}\,\epsilon_0=
-\epsilon_0\,,
\label{ksgravtaudcproj}
\eeq
with
\beq
\sin\alpha=-{1\over\cosh\tau}\;\;,\;\;\cos\alpha={\sinh\tau\over\cosh\tau}\,.
\label{ksgravtaudcalpha}
\eeq
Plugging (\ref{ksgravtaudcproj}) into eq. (\ref{ksgravtaus12}), one
arrives at
\beq
e^{-{1\over2}\alpha\Gamma_{\hat1
1}}\left(\epsilon_0'-{1\over2}\,\alpha'\,
\Gamma_{\hat1 1}\,\epsilon_0+{1\over2\cosh\tau}\,\Gamma_{\hat1 1}\,
\epsilon_0+{1\over8}h^{-1}\,h'\,\epsilon_0\right)=0\,,
\eeq
which yields the following two equations:
\bear
\alpha'&=&{1\over\cosh\tau}\, ,\label{gravtaualphadiffeq}\\ \rc
\epsilon_0'&=&-{1\over8}h^{-1}\,h'\,\epsilon_0\,.
\label{gravtauradiffeq}
\eear
The first equation is satisfied for $\alpha$ written in eq.
(\ref{ksgravtaudcalpha}). While eq. (\ref{gravtauradiffeq}) determines
the radial dependence of the 10d spinor $\epsilon_0$. So finally, the
Killing spinors of the Klebanov-Strassler model become:
\beq
\epsilon=e^{-{1\over2}\alpha\Gamma_{\hat1 1}}\,h^{-{1\over8}}\,\eta\,,
\label{kskillingdef1}
\eeq
with $\alpha$ being given by eq. (\ref{ksgravtaudcalpha}) and $\eta$ being
a constant 10d spinor satisfying the following projections:
\beq
\Gamma_{\tau\hat1\hat2\hat3}\,\eta=-\eta\;\;,\quad
\Gamma_{1 2}\,\eta=-\Gamma_{\hat1\hat2}\,\eta\;\;,\quad
\Gamma_{\,x^0\,x^1\,x^2\,x^3}\,\eta=-i\,\eta\,.
\label{ksetaprojs}
\eeq
Therefore, the model has 4 independent spinors as it should be for the
SUGRA dual of a 4d ${\cal N}=1$ field theory.

The projections (\ref{ksetaprojs}) can be rewritten as
\beq
\Gamma_{\,x^0\,x^1\,x^2\,x^3}\,\eta=-i\,\eta\;,\quad
\Gamma_{12}\,\eta=i\eta\;,\quad
\Gamma_{\hat1\hat2}\,\eta=-i\,\eta\,,
\label{ksetaprojs2}
\eeq
after making use of the equality $\Gamma_{x^0...x^3}\Gamma_\tau
\Gamma_{12\hat1\hat2\hat3}\,\eta=-\eta$. However,
the last two  projections, corresponding to the ones of the $T^{1,1}$,
are not satisfied by the Killing spinor
$\epsilon$ written in eq. (\ref{ksgravtaudcproj}) due to the factor
$e^{-{1\over2}\alpha\Gamma_{\hat1 1}}$ which anticommutes with them.
Only when $\tau\to\infty$ the angle $\alpha$ vanishes (see
eq. (\ref{ksgravtaudcalpha})) and the Killing spinor satisfies the
projections corresponding to the $T^{1,1}$, as it happened for the
Klebanov-Tseytlin background in the last chapter. This must be so, since
the KS and the KT solutions are identical in the UV, far away from the
tip of the conifold ($\tau=0$). 

Let us show that, as we have said above, the terms in (\ref{ksgravtau})
containing the three-forms effectively vanish. The first term, \ie \ 
${1\over96}\,{\cal
F}^{(3)}_{N_1N_2N_3}\,\Gamma_\tau^{\,\,\,N_1N_2N_3}\,\epsilon^*$, takes
the form:
\bear
&&-{\sqrt{6}\,h^{-{3\over4}}\over\mu^2}{M\,\alpha'\over8}\,\Gamma_\tau
\Bigg\{{k-f\over2\sinh\tau}\left(\Gamma_{\hat3\hat2 2}+
\Gamma_{\hat3\hat1 1}\right)
+{i\over2\sinh\tau}\left(1-g-2F\right)\,\left(\Gamma_{\hat3\hat2
1}+\Gamma_{\hat3 2 \hat1}\right)+\rc\rc&&
+{i\over2\cosh\tau}\,\Gamma_{\hat3\hat2 \hat1}
+{i\cosh\tau\over2\sinh^2\tau}\left[1+g\,(g-2+4F)\right]\,
\Gamma_{\hat3 2 1}\Bigg\}\,\epsilon^*\,,
\label{gravtaus31}
\eear
and by using the projection
$\Gamma_{12}\,\epsilon^*=-\Gamma_{\hat1\hat2}\,\epsilon^*$ (and
neglecting the common factor
$-{\sqrt{6}\,h^{-{3\over4}}\over\mu^2}{M\,\alpha'\over8}\,\Gamma_\tau$),
one can write it as
\bear
\Bigg\{{k-f\over\sinh\tau}\,\Gamma_{\hat3\hat1 1}
+{i\over\sinh\tau}\left(1-g-2F\right)\,\Gamma_{\hat3\hat2 1}
+\left[{i\cosh\tau\over2\sinh^2\tau}\left[1+g\,(g-2+4F)\right]-
{i\over2\cosh\tau}\right]\Gamma_{\hat3\hat1\hat2}\Bigg\}\,\epsilon^*\,,\rc
\label{gravtaus31ex2}
\eear
which is nothing else but $\Gamma_{\hat3}\times$ eq.
(\ref{ksgravahat3s32ex2}) and then, it vanishes once we impose the
differential equation (\ref{ksgravahat3alg}).

The last term of eq. (\ref{ksgravtau}), \ie \
$-{9\over96}\,{\cal
F}^{(3)}_{N_1N_2N_3}\,\delta_\tau^{N_1}\Gamma^{N_2
N_3}\,\epsilon^*$, can be written as
\bear
&&{\sqrt{6}\,h^{-{3\over4}}\over\mu^2}{3M\,\alpha'\over8}
\Bigg\{{f'+k'\over2\cosh\tau}\Gamma_{\hat2\hat1}+{1\over2\sinh\tau}
\left[(1-g)\,k'-(1+g)\,f'\right]\left(\Gamma_{\hat 2
1}+\Gamma_{2 \hat1}\right)+\rc\rc&&
+{\cosh\tau\over2\sinh^2\tau}\left[(1-g)^2\,k'+(1+g)^2\,f'\right]\,
\Gamma_{2 1}+{iF'\over\sinh\tau}\left(\Gamma_{\hat2 2}+
\Gamma_{\hat1 1}\right)
\Bigg\}\,\epsilon^*\,,
\label{gravtaus32}
\eear
which, after using the projection
$\Gamma_{12}\,\epsilon^*=-\Gamma_{\hat1\hat2}\,\epsilon^*$, and
neglecting the common factor
${\sqrt{6}\,h^{-{3\over4}}\over\mu^2}{3M\,\alpha'\over8}$, becomes:
\bear
&\Bigg\{&\left[{\cosh\tau\over2\sinh^2\tau}\left[(1-g)^2\,k'
+(1+g)^2\,f'\right]-{f'+k'\over2\cosh\tau}
\right]\,
\Gamma_{\hat1\hat2}+
\rc\rc
&&+{1\over\sinh\tau}\left[(1-g)\,k'-(1+g)\,f'\right]\,\Gamma_{\hat 2
1}+{2iF'\over\sinh\tau}\Gamma_{\hat1 1}\Bigg\}\,\epsilon^*\,.
\label{gravtaus32ex2}
\eear
Multiplying this last expression by $-i\Gamma_{\hat3}$ one recovers
eq. (\ref{ksgravahat3s31ex2}), thus, the differential eq.
(\ref{ksgravahat3diffeq1}) implies the vanishing of the
last term of eq. (\ref{ksgravtau}).

\section{Differential equations for the KS ansatz}
\label{ksdiffeqs}
\setcounter{equation}{0}
In the last section we have obtained six differential equations
(\ref{ksdildiffeq}),  (\ref{ksgravx1diffeq}), (\ref{ksgrava1q1diffeq}), 
(\ref{ksgrava1q2diffeq}), (\ref{ksgrava1qhat2diffeq}),
(\ref{ksgravahat3diffeq1}) and an algebraic constraint
(\ref{ksgravahat3alg}) relating the functions $f(\tau)$, $k(\tau)$ and 
$F(\tau)$ entering the ansatz of the model. We will see that these
equations reduce to the system (\ref{ksystem}) appearing in \cite{KS}
together with an extra equation.

Plugging eqs. (\ref{ksgravahat3diffeq1}) and (\ref{ksgravahat3alg}) into
eq. (\ref{ksdildiffeq}), one gets:
\beq
F'={k-f\over2}\,,
\label{kssvdeq1}
\eeq
which is one of the differential equations entering the system
(\ref{ksystem}). Inserting this last equation into the algebraic
constraint (\ref{ksgravahat3alg}) one gets the following differential
equation involving only $F$ and its first derivative:
\beq
F'\,+\,\coth\tau\,F\,=\,{1\over 2}\,\tanh\Big({\tau\over 2}\Big)\,\,.
\label{kssvdeq2}
\eeq
This equation can be easily integrated, yielding the explicit form of $F$
(see below). In addition, by substituting the value of $F'$ given by this
last equation into eq. (\ref{ksgravahat3diffeq1}) one arrives at
\beq
\coth\Big({\tau\over 2}\Big) \,f'\,-\,\tanh\Big({\tau\over 2}\Big)
\,k'\,=\,-2\coth\tau\,F\,+\,\tanh\Big({\tau\over 2}\Big)\,\,.
\label{kssvdeq3}
\eeq
Looking at equations (\ref{kssvdeq1}) and (\ref{kssvdeq2}) one can easily
write:
\beq
F'\,+\,{k-f\over 2}\,=\,2F'\,=\,\tanh\Big({\tau\over2}\Big)\,
-\,2\coth\tau\,F\,\,.
\label{kssvdeq12}
\eeq
Let us substitute this last result into equation
(\ref{ksgrava1q2diffeq}). After some calculation we obtain:
\beq
f'-k'\,=\,-\Big[\,\coth^2\Big({\tau\over 2}\Big)\,+\, \tanh^2\Big({\tau\over 2}\Big)
\,\Big]\,F\,+\,\tanh^2\Big({\tau\over 2}\Big)\,\,.
\label{kssvdeq4}
\eeq
By combining this equation with eq. (\ref{kssvdeq3}), one can solve for
$f'$ and $k'$ as functions of $F$, resulting:
\beq
f'=(\,1-F\,)\,\tanh^2\Big({\tau\over 2}\Big)\;\;,\quad
k'=F\,\coth^2\left({\tau\over 2}\right)\,\,.
\label{kssvdeq5}
\eeq
These equations, together with eq. (\ref{kssvdeq1}) form the first-order
system (\ref{ksystem}). 
For the remaining equations one can easily check that eq
(\ref{ksgrava1q1diffeq}) is trivially satisfied after substituting
(\ref{ksystem}), while (\ref{ksgravx1diffeq}) and
(\ref{ksgrava1qhat2diffeq}) are verified after inserting (\ref{ksystem})
and the new equation for $F'$, \ie \ (\ref{kssvdeq2}).


Summing up our results; from imposing the cancellation of the SUSY
variations of the dilatino and the gravitino we have obtained the
first-order system (\ref{ksystem}) appearing in \cite{KS} and, in
addition, we got a new differential equation, namely (\ref{kssvdeq2}), or,
alternatively, the algebraic relation (\ref{ksgravahat3alg}). The
differential equation (\ref{kssvdeq2}) can be easily integrated by the
method of variation of constants, rendering:
\beq
F={1\over 2}\,\,{\sinh\tau\,-\,\tau\over \sinh\tau}\,+\,
{A\over \sinh\tau}\,,
\label{kssvFgen}
\eeq
where $A$ is a constant, which by requiring regularity of $F$ at $\tau=0$
gets fixed to the value $A=0$. Then, it is immediate to
integrate the first order equations for $f$ and $k$ (eqs.
(\ref{kssvdeq5})). The result is the same as in ref. \cite{KS}, namely:
\bear
&&F={1\over 2}\,{\sinh\tau\,-\,\tau\over \sinh\tau}\,,\rc\rc
&&f={1\over 2}\,{\tau\coth\tau\,-\,1\over \sinh\tau}\,
(\,\cosh\tau-1\,)\,,\rc\rc
&&k={1\over 2}\,{\tau\coth\tau\,-\,1\over \sinh\tau}\,
(\,\cosh\tau+1\,)\,.
\eear
Therefore, the requirement of preserving the same supersymmetries as in
the solution corresponding to a D3-brane at the tip of the deformed
conifold (we are imposing the projection corresponding to a D3-brane
(\ref{ksd3proj}) together with the projections satisfied by the
Killing spinors of the deformed conifold \ie \ (\ref{ks2cproj}) and
(\ref{ksdcproj}))  fixes the values of the three-forms to those found
in ref. \cite{KS} (see refs. \cite{GranaPol,Gubser}).


\chapter{Killing spinors of the non-commutative MN
solution}
\label{ncmncp}

\section{Introduction}
\label{mnscintro}
\setcounter{equation}{0}
In this chapter we present the construction of the Killing spinors of the
non-commutative deformation of the so-called Maldacena-N\'u\~nez (MN) background
\cite{MN,CV}. The commutative background is dual to the large $N$
limit of
${\cal N}=1$ super Yang-Mills theory. This geometry, generated by a
fivebrane wrapping a two-cycle, is smooth and leads to confinement
and chiral symmetry breaking. 

The spatial non-commutative theories are field theories living on a spacetime
where two spatial coordinates do not commute, $\ie$ $[x^i,x^j]=\Theta^{ij}\neq0$.
These theories have been thoroughly studied in recent years after the discovery
that they can be obtained as a low energy limit of string theory in the presence
of a Neveu-Schwarz $B$-field \cite{NCString, SW}. In particular, the
non-commutative deformation of the MN background was obtained in \cite{NCMN} by
means of a chain of string dualities and it corresponds to the decoupling limit of
a (D3,D5) bound state with the D3-brane smeared in the worldvolume of the D5 and
wrapped on the two-cycle. The corresponding ten dimensional metric breaks
four dimensional Lorentz invariance since it distinguishes between 
the coordinates of the
non-commutative plane and the other two Minkowski coordinates.
As expected, this solution has a non-vanishing Neveu-Schwarz
$B$-field directed along the non-commutative directions.

After reviewing the details of the non-commutative solution in the
next subsection, we will compute the Killing spinors of the model in 
section \ref{mnsckilling}. 
This computation is similar to the one carried out in \cite{flavoring} for
the commutative model. As in that case, working in the frame arising
naturally when one obtains the MN model as an uplift from 7d gauged
supergravity, the Killing spinors do not depend on the internal
coordinates of the geometry.

This computation was performed in the context of the work published in
\cite{ncflav}, where we studied the addition of flavor degrees of
freedom to the supergravity dual of the non-commutative deformation of
the maximally supersymmetric gauge theories, see refs. \cite{MR,AOSJ}.
There we have also studied the possibility of adding flavor to
non-commutative duals of less supersymmetric theories as it is the
case of the MN background. So in order to do that, using the kappa
symmetry approach when looking for supersymmetric embeddings of probe
branes, we needed the explicit form of the Killing spinors for that
background.

\subsection{The non-commutative Maldacena-N\'u\~nez solution}
The procedure used in \cite{NCMN} to obtain the non-commutative
deformation of the MN solution leads, as we have said, to a metric where
the four dimensional Lorentz symmetry is broken. This metric singles out
the so-called non-commutative plane along which the NSNS B-field is
directed. In the string frame it is given by
\beq
ds^2\,=\,e^{\phi}\,\,\left[\,
dx^2_{0,1}\,+\,h^{-1}\,dx^2_{2,3}\,+\,
e^{2g}\,\big(\,d\theta_1^2+\sin^2\theta_1 d\phi_1^2\,\big)\,+\,
dr^2\,+\,{1\over 4}\,(w^i-A^i)^2\,\right]\,\,,
\label{mnstringmetric}
\eeq
where $\phi$, $h$ and $g$ are functions of the radial coordinate $r$ 
(see below) which have nothing to do with the functions denoted by the
same letters that appeared in previous chapters. $A^i$ is a one-form
which can be written in terms of the angles $(\theta_1, \phi_1)$ and a
function $a(r)$ as follows:
\beq
A^1\,=\,-a(r) d\theta_1\,,
\,\,\,\,\,\,\,\,\,
A^2\,=\,a(r) \sin\theta_1 d\phi_1\,,
\,\,\,\,\,\,\,\,\,
A^3\,=\,- \cos\theta_1 d\phi_1\,.
\label{mnoneform}
\eeq
The $\omega^i$'s appearing in eq. (\ref{mnstringmetric}) are again the
$SU(2)$ left-invariant one-forms defined in (\ref{omegaforms}).
Moreover, the functions $a(r)$, $g(r)$ and  $\phi(r)$ are:
\bear
a(r)&=&{2r\over \sinh 2r}\,\,,\rc\rc
e^{2g}&=&r\coth 2r\,-\,{r^2\over \sinh^2 2r}\,-\,
{1\over 4}\,\,,\rc
e^{-2\phi}&=&e^{-2\phi_0}{2e^g\over \sinh 2r}\,\,,
\label{mnsol}
\eear
where $\phi_0$ is a constant ($\phi_0=\phi(r=0)$).
The function $h(r)$, which distinguishes in the metric the 
coordinates $x^2x^3$ from
$x^0x^1$, can be written in terms of the function $\phi(r)$  as follows:
\beq
h(r)=1+\Theta^2\,e^{2\phi}\,,
\label{mnh}
\eeq
where $\Theta$ is a constant which parameterizes the non-commutative 
deformation, so when $\Theta\neq0$ this background is dual to a gauge
theory in which the coordinates $x^2$ and $x^3$ do not commute, being
$[x^2,x^3]\sim\Theta^2$.

Let us denote by $\hat\phi$ the dilaton field of type IIB 
supergravity. For the solution
of ref. \cite{NCMN} this field takes the value:
\beq
e^{2\hat\phi}=e^{2\phi}\,h^{-1}\,.
\label{mndilaton}
\eeq
Notice that, when the non-commutative parameter $\Theta$ is 
non-vanishing, the dilaton
$\hat\phi$ does not diverge at the UV boundary $r\to\infty$. Indeed, 
$e^{\hat\phi}$ reaches
its maximum value at infinity, where $e^{\hat\phi}\to \Theta^{-1}$. 
This behaviour is in sharp
contrast with the one corresponding to the commutative MN background, 
for which the dilaton
blows up at infinity.

This solution of the type IIB supergravity also includes a RR 
three-form $F^{(3)}$ given by:
\beq
F^{(3)}=-{1\over 4}\,\big(w^1-A^1\big)\wedge
\big(w^2-A^2\big)\wedge \big(w^3-A^3\big)+{1\over 4}\,
\sum_a F^a\wedge \big(w^a-A^a\big)\,\,,
\label{mnRRthreeform}
\eeq
where $F^a$ is the field strength of the $SU(2)$ gauge field $A^a$ of 
eq. (\ref{mnoneform}), defined
as
\beq
F^a=dA^a+{1\over 2}\,\epsilon_{abc}\,A^b\wedge A^c\,\,.
\label{mnfieldstrenght}
\eeq
The different components of $F^a$ can be obtained by plugging the value of the
$A^a$'s on the right-hand side of eq. (\ref{mnfieldstrenght}). One gets:
\beq
F^1=-a'\,dr\wedge d\theta_1\;,\quad
F^2=a'\sin\theta_1 dr\wedge d\phi_1\;,\quad
F^3=(1-a^2)\,\sin\theta_1 d\theta_1\wedge d\phi_1\,\,,
\label{mnsu2f}
\eeq
where the prime denotes derivative with respect to $r$. The NSNS $B$ field
is:
\beq
B=\Theta\,e^{2\phi}\,h^{-1}\,dx^2\wedge dx^3\,.
\label{mnB}
\eeq
It is proportional to the non-commutative parameter $\Theta$ and
it is directed along the $x^2x^3$ coordinates spanning the
non-commutative plane. Indeed, the introduction of the NSNS magnetic
field is the key ingredient in the construction of the non-commutative
deformation. The corresponding three-form field strength
$H=dB$ reads:
\beq
H=2\Theta\,\phi{\,'}\,e^{2\phi}\,h^{-2}\,dr\wedge dx^2\wedge dx^3\,.
\label{mnH}
\eeq
The solution has also a non-vanishing RR five-form $F^{(5)}$, whose 
expression is:
\beq
F^{(5)}=B\wedge F^{(3)}+{\rm Hodge \,\,\,dual}\,,
\label{mnF5}
\eeq
where $B$ and $F^{(3)}$ are given in eqs. (\ref{mnB}) and 
(\ref{mnRRthreeform}) respectively.
The RR field strengths satisfy the equations:
\bear
&&dF^{(3)}=0\,\,,\rc\rc
&&dF^{(5)}=d{}^*F^{(5)}=H\wedge F^{(3)}\,\,,\rc\rc
&&d{}^*F^{(3)}=-H\wedge F^{(5)}\,\,.
\label{mnrreqs}
\eear

\section{Killing spinors}
\label{mnsckilling}
\setcounter{equation}{0}

Once again, we will impose the vanishing of the SUSY variations of
the IIB SUGRA fermionic fields (\ref{sugra}) in order to arrive at an
explicit expression for the Killing spinors of the model. 

This computation follows closely a similar analysis done in ref.
\cite{flavoring} for the commutative MN background. First of all, it is
more convenient to work in Einstein frame, where the metric
(\ref{mnstringmetric}) becomes:
\beq
ds^2_{E}\,=\,e^{{\phi\over 2}}\,\,h^{{1\over 4}}\,\left[\,
dx^2_{0,1}\,+\,h^{-1}\,dx^2_{2,3}\,+\,
e^{2g}\,\big(\,d\theta_1^2+\sin^2\theta_1 d\phi_1^2\,\big)\,+\,
dr^2\,+\,{1\over 4}\,(w^i-A^i)^2\,\right]\,\,.
\label{mnEinsteinmetric}
\eeq
We shall consider the following basis of frame one-forms:
\bear
&&e^{x^{0,1}}=e^{{\phi\over 4}}\,h^{{1\over 8}}\,dx^{0,1}\;,\quad
e^{x^{2,3}}=e^{{\phi\over 4}}\,h^{-{3\over 8}}\,dx^{2,3}\;,\rc\rc
&&e^{r}=e^{{\phi\over 4}}\,h^{{1\over 8}}\,dr\;,\rc\rc
&&e^1=e^{{\phi\over 4}}\,h^{{1\over 8}}\,e^g\,d\theta_1\;,\quad
e^2=e^{{\phi\over 4}}\,h^{{1\over
8}}\,e^g\,\sin\theta_1d\phi_1\,\,,\rc\rc &&e^{\hat i}={1\over
2}\,e^{{\phi\over 4}}\,h^{{1\over 8}}\,(w^i\,-\,A^i)\;,\;
(i=1,2,3)\;.
\label{mnbasis}
\eear
The corresponding spin connection one-form, which results from solving
the Maurer-Cartan equations (\ref{cartan}), reads:
\bear
&&\omega^{x^d\,r}=h^{-{1\over8}}\,e^{-{\phi\over4}}\left({\phi'\over4}+
{1\over8}\,h'\,h^{-1}\right)\,e^{x^d} \;,\; (d=0,1)\,,\rc\rc
&&\omega^{x^i\,r}=h^{-{1\over8}}\,e^{-{\phi\over4}}\left({\phi'\over4}-
{3\over8}\,h'\,h^{-1}\right)\,e^{x^i} \;,\;(i=2,3)\,,\rc\rc
&&\omega^{1\,r}=h^{-{1\over8}}\,e^{-{\phi\over4}}\left[\left(g'+
{\phi'\over4}+{1\over8}\,h'\,h^{-1}\right)\,e^1+{1\over4}\,e^{-g}\,a'
\,e^{\hat1}\right]\,,\rc\rc
&&\omega^{2\,r}=h^{-{1\over8}}\,e^{-{\phi\over4}}\left[\left(g'+
{\phi'\over4}+{1\over8}\,h'\,h^{-1}\right)\,e^2-{1\over4}\,e^{-g}\,a'
\,e^{\hat2}\right]\,,\rc\rc
&&\omega^{\hat1\,r}=h^{-{1\over8}}\,e^{-{\phi\over4}}\left[\left(
{\phi'\over4}+{1\over8}\,h'\,h^{-1}\right)\,e^{\hat1}+{1\over4}\,e^{-g}\,a'
\,e^1\right]\,,\rc\rc
&&\omega^{\hat2\,r}=h^{-{1\over8}}\,e^{-{\phi\over4}}\left[\left(
{\phi'\over4}+{1\over8}\,h'\,h^{-1}\right)\,e^{\hat2}-{1\over4}\,e^{-g}\,a'
\,e^2\right]\,,\rc\rc
&&\omega^{\hat3\,r}=h^{-{1\over8}}\,e^{-{\phi\over4}}\left(
{\phi'\over4}+{1\over8}\,h'\,h^{-1}\right)\,e^{\hat3}\,,\rc\rc
&&\omega^{1\,\hat1}={1\over4}\,h^{-{1\over8}}\,e^{-{\phi\over4}}\,
e^{-g}\,a'\,e^r\;,\quad
\omega^{2\,\hat2}=-{1\over4}\,h^{-{1\over8}}\,e^{-{\phi\over4}}\,
e^{-g}\,a'\,e^r\,,\rc\rc
&&\omega^{1\,2}=h^{-{1\over8}}\,e^{-{\phi\over4}}\left[{1\over4}\,(1-a^2)
\,e^{-2g}\,e^{\hat3}-e^{-g}\cot\theta_1\,e^2\right]\,,\rc\rc
&&\omega^{\hat2\,\hat1}=h^{-{1\over8}}\,e^{-{\phi\over4}}\left(e^{\hat3}
-e^{-g}\,\cot\theta_1\,e^2\right)\,,\rc\rc
&&\omega^{\hat1\,\hat3}=h^{-{1\over8}}\,e^{-{\phi\over4}}\left(e^{\hat2}
+e^{-g}\,a\,e^2\right)\; ,
\quad
\omega^{\hat2\,\hat3}=h^{-{1\over8}}\,e^{-{\phi\over4}}\left(
e^{-g}\,a\,e^1-e^{\hat1}\right)\,,\rc\rc
&&\omega^{\hat3\,2}=h^{-{1\over8}}\,e^{-{\phi\over4}}\left[{1\over4}\,
(1-a^2)\,e^{-2g}\right]\,e^1\,,\rc\rc
&&\omega^{\hat3\,1}=-h^{-{1\over8}}\,e^{-{\phi\over4}}\left[{1\over4}\,
(1-a^2)\,e^{-2g}\right]\,e^2\,,
\label{mnspincon}
\eear
written directly in the frame $e^a$ defined in eq. (\ref{mnbasis}).
One should keep in mind that
$\omega^{a\,b}=\omega^{a\,b}_{\tilde m}\,dX^{\tilde
m}=\omega^{a\,b}_c\,e^c$, so $\omega^{a\,b}_{\tilde
m}=E^{\,c}_{\,\tilde m}\,\omega^{a\,b}_c$ (and  $e^c=E^{\,c}_{\,\tilde
m}\,dX^{\tilde m}$ from (\ref{mnbasis})). Let us also write the RR
and NSNS forms in the frame (\ref{mnbasis}). The selfdual five-form
becomes:
\bear
F^{(5)}&=&\Theta\,h^{-{5\over8}}\,e^{3\phi\over4}\,\bigg[-2\left(e^{\hat1}\wedge
e^{\hat2}\wedge e^{\hat3}\wedge e^{x^2}\wedge e^{x^3}+e^{x^0}\wedge
e^{x^1}\wedge e^1\wedge e^2\wedge e^r\right)+\rc\rc
&&+{1\over2}\,(1-a^2)\,e^{-2g}\,\left(e^1\wedge
e^2\wedge e^{\hat3}\wedge e^{x^2}\wedge e^{x^3}+e^{x^0}\wedge
e^{x^1}\wedge e^r\wedge e^{\hat1}\wedge e^{\hat2}\right)-\rc\rc
&&-{1\over2}\,e^{-g}\,a'\,\left(e^r\wedge
e^1\wedge e^{\hat1}\wedge e^{x^2}\wedge e^{x^3}+e^{x^0}\wedge
e^{x^1}\wedge e^2\wedge e^{\hat2}\wedge e^{\hat3}\right)+\rc\rc
&&+{1\over2}\,e^{-g}\,a'\,\left(e^r\wedge
e^2\wedge e^{\hat2}\wedge e^{x^2}\wedge e^{x^3}+e^{x^0}\wedge
e^{x^1}\wedge e^r\wedge e^{\hat1}\wedge e^{\hat3}\right)\bigg]\,,
\label{mnf5flat}
\eear
and the complex combination of the RR and NSNS three-forms defined in
(\ref{comp3form}) can be written as
\bear
{\cal F}^{(3)}
&=&2\Theta\,e^{3\phi\over4}\,h^{-{9\over8}}\,\phi'\,e^r\wedge
e^{x^2}\wedge e^{x^3}+ie^{-{\phi\over4}}\,h^{-{5\over8}}\,\bigg[-2\,
e^{\hat1}\wedge e^{\hat2}\wedge e^{\hat3}+\rc\rc&&+{1\over2}\,(1-a^2)\,
e^{-2g}\,e^1\wedge e^2\wedge e^{\hat3}-{1\over2}\,e^{-g}\,a'\,e^r\wedge
e^1\wedge e^{\hat1}+{1\over2}\,e^{-g}\,a'\,e^r\wedge e^2\wedge 
e^{\hat2}\bigg]\,.\rc\rc
\label{mnf3flat}
\eear
Now we are ready to solve the SUSY equations arising from
(\ref{sugra}). Up to now in this work we have worked with complex 
spinors, however, from now on it will become easier to switch to real
two-component spinors. It is straightforward to find the following
rules to pass from complex to real spinors:
\beq
\epsilon^*\,\leftrightarrow\,\tau_3\,\epsilon\;,
\quad
i\epsilon^*\,\leftrightarrow\,\tau_1\,\epsilon\;,
\quad
i\epsilon\,\leftrightarrow\,-i\tau_2\,\epsilon\,\,,
\label{mnrule}
\eeq
where $\tau_i$ $(i=1,2,3)$ are Pauli matrices that act on the
two dimensional vector $\pmatrix{\epsilon_1\cr\epsilon_2}$. 

To begin with, we study the vanishing of the dilatino
SUSY variation, which leads to the following equation:
\bear
&&{1\over2}\,e^{-{\phi\over4}}\,h^{-{9\over8}}\,\phi'\,\Gamma_r\,\tau_1
\,\epsilon-{i\over4}\Bigg\{2\Theta\,e^{3\phi\over4}\,h^{-{9\over8}}\,
\phi'\,\Gamma_{r\,x^2x^3}+i\,e^{-{\phi\over4}}\,h^{-{5\over8}}\bigg[\,
{1\over2}\,e^{-g}\,a'\,\Gamma_{r2\hat2}-{1\over2}\,e^{-g}\,a'\,
\Gamma_{r1\hat1}+\rc\rc &&+\left({1\over2}\,(1-a^2)\,e^{-2g}-2\right)\,
\Gamma_{\hat1\hat2\hat3}\bigg]\Bigg\}\,\epsilon=0\,.
\label{mndilvar}
\eear
$\Gamma_a\,,\;(a=x^\alpha,r,1,2,\hat1,\hat2,\hat3)$ are
constant Dirac matrices associated to the frame (\ref{mnbasis}). We
have used that
$\partial_N\hat\phi\,\Gamma^N\,\epsilon^*=\left(E^{\,r}_{\,\tilde
r}\right)^{-1}\,\hat\phi'\,\Gamma_r\,\epsilon^*=e^{-{\phi\over4}}\,h^{-{1\over8}}\,
\hat\phi'\,\Gamma_r\,\epsilon^*$, and $\hat\phi'=h^{-1}\,\phi'$, which
can be easily checked using eq. (\ref{mndilaton}).

As it was done in  \cite{flavoring} for the commutative MN background,
we shall impose the projection:
\beq
\Gamma_{12}\,\epsilon=\Gamma_{\hat 1 \hat 2}\,\epsilon\,.
\label{mnpr1}
\eeq
Thus, after some calculation, one arrives at the equation:
\bear
&&h^{-{1\over2}}\left[{1\over2}\,(1-a^2)\,e^{-2g}-2\right]\,
\Gamma_{r\hat1\hat2\hat3}\,\tau_1\,\epsilon=\rc\rc&&=\bigg(-2h^{-1}\,
\phi'+2\Theta\,e^\phi\,h^{-1}\,\phi'\,\Gamma_{x^2x^3}\,\tau_3+e^{-g}\,
h^{-{1\over2}}\,a'\,\Gamma_{1\hat1}\,\tau_1\bigg)\,\epsilon\,.
\label{mndilvar2}
\eear
Let us now introduce the angle $\alpha$, which also appears in the
commutative case, namely:
\beq
\cos\alpha={\phi'\over 1+{1\over 4}\,e^{-2g}\,(a^2-1)}\;,
\quad
\sin\alpha={1\over 2}\,
{e^{-g}\,a'\over 1+{1\over 4}\,e^{-2g}\,(a^2-1)}\,\,,
\label{mnalphadef}
\eeq
whose value can be obtained from the explicit form  (\ref{mnsol}) of 
the solution, resulting:
\beq
\cos\alpha={\rm \coth} 2r\,-\,{2r\over \sinh^22r}\,.
\label{alphaexplicit}
\eeq
In addition, we define a new angle $\beta$ given by:
\beq
\cos\beta=h^{-{1\over 2}}\;,
\quad
\sin\beta=-\Theta\,e^{\phi}\,h^{-{1\over 2}}\,\,.
\label{mnbetadef}
\eeq
Notice that $\beta=0$ when $\Theta=0$. Moreover, from the definition of
$h$ one can easily check that $\sin^2\beta+\cos^2\beta=1$. 

In terms of the angles $\alpha$ and $\beta$, the equation
(\ref{mndilvar2}) results in a new projection to be imposed on
$\epsilon$, which reads:
\beq
\Gamma_{r\hat 1\hat 2\hat 3}\,\tau_1\,\epsilon=
\Big[\cos\alpha\,\left(\cos\beta+\sin\beta\,\Gamma_{x^2x^3}\,\tau_3
\right)\,-\,\sin\alpha\Gamma_{1\hat 1}\,\tau_1\Big]\,\epsilon\,\,.
\label{mnpr2}
\eeq

We will now study the SUSY variations of the gravitino. We begin with
the components along the Minkowski space. The equation
$\delta\psi_{x^1}=0$ is:
\beq
D_{x^1}\,\epsilon\,+\,{i\over 1920}\,
F_{N_1\cdots N_5}^{(5)}\,\Gamma^{N_1\cdots
N_5}\Gamma_{x^1}\,\epsilon
+{1\over96}\,{\cal F}^{(3)}_{N_1N_2N_3}\,
\big(\,\Gamma_{x^1}^{\,\,\,N_1N_2N_3}\,-\,
9\,\delta_{x^1}^{N_1}\,\,\Gamma^{N_2 N_3}\,\big)\,\epsilon^{*}=0\,.
\label{mngravx1}
\eeq
Considering a spinor independent of the $x^{\alpha}$ coordinates and
inserting the corresponding terms of the spin connection
(\ref{mnspincon}), the first term of (\ref{mngravx1}) takes the form:
\beq
D_{x^1}\,\epsilon={1\over4}\,\omega^{a\,b}_{x^1}\,\Gamma_{a\,b}\,
\epsilon={1\over16}\,e^{-{\phi\over4}}\,h^{-{1\over8}}\left(2\phi'+
h^{-1}\,h'\right)\,\Gamma_{x^1\,r}\,\epsilon\,.
\label{mngravx1s1}
\eeq
We shall plug the five-form (\ref{mnf5flat}) into the second term of
(\ref{mngravx1}). If we also impose the projection
(\ref{mnpr1}), that term becomes:  
\beq
-{i\over8}\,\Theta\,e^{3\phi\over4}\,h^{-{5\over8}}\,\Bigg\{\left[\,{1\over2}
\,(1-a^2)\,e^{-2g}-2\right]\,\Gamma_{\hat1\hat2\hat3\,x^2x^3}\,
\Gamma_{x^1}-e^{-g}\,a'\,\Gamma_{r 1\hat1\,x^2x^3}\,\Gamma_{x^1}\Bigg\}
\,\tau_2\,\epsilon\,,
\label{mngravx1s2}
\eeq
where we have also inserted the total chirality projection
\beq
\Gamma_{x^0x^1x^2x^3}\Gamma_r
\Gamma_{12\hat1\hat2\hat3}\,\epsilon=-\epsilon\,.
\label{mntotchir}
\eeq

Let us now write down the last term of $\delta\psi_{x^1}=0$; reading
the complex three-form from eq. (\ref{mnf3flat}) one arrives at
\bear
{1\over8}\,\Theta\,e^{3\phi\over4}\,h^{-{9\over8}}\,\phi'\,
\Gamma_{x^1r\,x^2x^3}\,\tau_3\,\epsilon+{1\over16}\,e^{-{\phi\over4}}\,h^{-{5\over8}}\,
\Bigg\{\left[\,{1\over2}
\,(1-a^2)\,e^{-2g}-2\right]\,\Gamma_{x^1\hat1\hat2\hat3}
-e^{-g}\,a'\,\Gamma_{x^1r1\hat1}\Bigg\}\tau_1\,\epsilon\,.\rc
\label{mngravx1s3}
\eear
%
Then, gathering the three terms written in eqs. (\ref{mngravx1s1}),
(\ref{mngravx1s2}) and (\ref{mngravx1s3}), and multiplying the
whole equation by $8e^{\phi\over4}\,h^{1\over8}\,\Gamma_{r\,x^1}$, one
gets:
\bear
&&\Bigg\{{1\over2}\,\left(2\phi'+h^{-1}\,h'\right)+i\Theta\,e^{\phi}\,
h^{-{1\over2}}\,\left[\,{1\over2}\,(1-a^2)\,e^{-2g}-2\right]\,
\Gamma_{r\hat1\hat2\hat3}\,\Gamma_{x^2x^3}\,\tau_2-\rc\rc&&-
i\Theta\,e^{\phi}\,e^{-g}\,h^{-{1\over2}}\,a'\,\Gamma_{1\hat1}\,
\Gamma_{x^2x^3}\,\tau_2+\Theta\,e^{\phi}\,h^{-1}\,\phi'\,
\Gamma_{x^2x^3}\,\tau_3+\rc\rc&&+{1\over2}\,h^{-{1\over2}}
\left[\,{1\over2}\,(1-a^2)\,e^{-2g}-2\right]\,\Gamma_{r\hat1\hat2\hat3}\,
\tau_1-{1\over2}\,e^{-g}\,h^{-{1\over2}}\,a'\,\Gamma_{1\hat1}\,\tau_1\,
\Bigg\}\,\epsilon=0\,,
\label{mngravxex2}
\eear 
which, after multiplying by $\left[1+{1\over
4}\,e^{-2g}\,(a^2-1)\right]^{-1}$ can be written in terms of the
angles $\alpha$ and $\beta$, defined in eqs. (\ref{mnalphadef}) and
(\ref{mnbetadef}) respectively, as
\bear
&\bigg[&(1+\sin^2\beta)\cos\alpha+2i\sin\beta\,\Gamma_{r\hat1\hat2\hat3}
\,\Gamma_{x^2x^3}\,\tau_2+2i\sin\alpha\,\sin\beta\,\Gamma_{1\hat1}\,
\Gamma_{x^2x^3}\,\tau_2-\rc\rc&&-\cos\alpha\,\sin\beta\,\cos\beta\,
\Gamma_{x^2x^3}\,\tau_3-\cos\beta\,\Gamma_{r\hat1\hat2\hat3}\,\tau_1-
\sin\alpha\,\cos\beta\,\Gamma_{1\hat1}\,\tau_1\,\bigg]\,\epsilon=0\,.
\label{mngravxex3}
\eear
We have not yet used the projection (\ref{mnpr2}). Notice that by
multiplying that projection by $-i\Gamma_{x^2x^3}\,\tau_3$, one obtains
the following equivalent expression:
\beq
\Gamma_{r\hat 1\hat 2\hat 3}\,\Gamma_{x^2x^3}\,\tau_2\,\epsilon=
i\Big[\cos\alpha\,\sin\beta-\cos\alpha\,\cos\beta\,\Gamma_{x^2x^3}\,
\tau_3+\sin\alpha\,\Gamma_{x^2x^3}\,\Gamma_{1\hat1}\,\tau_3\,\tau_1
\Big]\,\epsilon\,\,.
\label{mnpr2ex2}
\eeq
So, finally, one can readily check that eq. (\ref{mngravxex2}) is
satisfied after imposing (\ref{mnpr2}) and (\ref{mnpr2ex2}).

The equation arising from $\delta\psi_{x^0}=0$ is equal to the
one resulting from $\delta\psi_{x^1}=0$, while from the other two
Minkowski components, namely $\delta\psi_{x^2}=0$ and
$\delta\psi_{x^3}=0$, we get an equation slightly different, which
also vanishes after imposing the projections (\ref{mnpr1}) and
(\ref{mnpr2}).

We shall now look at the SUSY variations of the angular components of
the gravitino. Let us begin with the equation $\delta\psi_{\hat1}=0$:
\beq
D_{\hat 1}\,\epsilon\,+\,{i\over 1920}\,
F_{N_1\cdots N_5}^{(5)}\,\Gamma^{N_1\cdots
N_5}\Gamma_{\hat 1}\,\epsilon
+{1\over96}\,{\cal F}^{(3)}_{N_1N_2N_3}\,
\big(\,\Gamma_{\hat 1}^{\,\,\,N_1N_2N_3}\,-\,
9\,\delta_{\hat 1}^{N_1}\,\,\Gamma^{N_2 N_3}\,\big)\,\epsilon^{*}=0\,.
\label{mngravahat1}
\eeq
We assume that $\epsilon$ does not depend on the angular coordinates
so, after plugging the corresponding components of the spin connection
(\ref{mnspincon}), the first term of this last equation reads:
\beq
D_{\hat1}\,\epsilon={1\over4}\,\omega^{a\,b}_{\hat1}\,\Gamma_{a\,b}\,
\epsilon={1\over2}\,e^{-{\phi\over4}}\,h^{-{1\over8}}\left[{1\over4}e^{-g}\,a'\,
\Gamma_{1r}+\left({\phi'\over4}+{1\over8}\,h^{-1}\,h'\right)
\Gamma_{\hat1r}-\Gamma_{\hat2\hat3}\right]\,\epsilon\,.
\label{mngravhat1s1}
\eeq
Inserting the RR five-form (\ref{mnf5flat}), using eq.
(\ref{mntotchir}), and imposing the projection (\ref{mnpr1}),  the
second term of (\ref{mngravahat1}) takes the form:
\bear
{i\over8}\,\Theta\,e^{3\phi\over4}\,h^{-{5\over8}}\left[\,{1\over2}\,
(1-a^2)\,e^{-2g}+2\right]\,
\Gamma_{\hat1\hat2\hat3\,x^2x^3}\,\Gamma_{\hat1}\,\tau_2\,\epsilon\,.
\label{mngravhat1s2}
\eear
The first term containing the RR three-form (\ref{mnf3flat}), namely
${1\over96}\,{\cal F}^{(3)}_{N_1N_2N_3}\,\Gamma_{\hat1}^
{\,\,\,N_1N_2N_3}\,\epsilon^*$, can be written as
\bear
{1\over8}\,\Theta\,e^{3\phi\over4}\,h^{-{9\over8}}\,\phi'\,
\Gamma_{\hat1r\,x^2x^3}\,\tau_3\,\epsilon+{1\over16}\,e^{-{\phi\over4}}\,h^{-{5\over8}}\,
\left[\,{1\over2}\,(1-a^2)\,e^{-2g}\,\Gamma_{\hat1\hat3\hat1\hat2}-
{1\over2}\,e^{-g}\,a'\,\Gamma_{\hat1r1\hat1}\right]\tau_1\,\epsilon\,,
\rc\label{mngravhat1s31}
\eear
where we have imposed the projection (\ref{mnpr1}). Using that
projection, the last term of eq. (\ref{mngravahat1}), \ie \
$-{9\over96}\,{\cal F}^{(3)}_{N_1N_2N_3}
\,\delta_{\hat 1}^{N_1}\Gamma^{N_2 N_3}\,\epsilon^*$, becomes:
\beq
{3\over16}\,e^{-{\phi\over4}}\,h^{-{5\over8}}\left(2\,\Gamma_{\hat2\hat3}
+{1\over2}\,e^{-g}\,a'\,\Gamma_{r1}\right)\tau_1\,\epsilon\,.
\label{mngravhat1s32}
\eeq
Eventually, we gather the four pieces of $\delta\psi_{\hat1}=0$:
(\ref{mngravhat1s1}), (\ref{mngravhat1s2}), (\ref{mngravhat1s31})
and (\ref{mngravhat1s32}); and we multiply the whole equation by
$8e^{\phi\over4}\,h^{1\over8}\,\Gamma_{r\hat1}$.  We arrive at
\bear
&&\Bigg\{{1\over2}\,\left(2\phi'+h^{-1}\,h'\right)+h^{-{1\over2}}
\left[{1\over4}\,(1-a^2)\,e^{-2g}+3\right]\Gamma_{r\hat1\hat2\hat3}\,
\tau_1+\rc\rc&&+i\Theta\,e^{\phi}\,
h^{-{1\over2}}\,\left[\,{1\over2}\,(1-a^2)\,e^{-2g}+2\right]\,
\Gamma_{r\hat1\hat2\hat3}\,\Gamma_{x^2x^3}\,\tau_2-4
\Gamma_{r\hat1\hat2\hat3}+\rc\rc&&+\Theta\,e^{\phi}\,h^{-1}\,\phi'\,
\Gamma_{x^2x^3}\,\tau_3-e^{-g}\,a'\,\Gamma_{1\hat1}+{1\over2}\,e^{-g}\,
h^{-{1\over2}}\,a'\,\Gamma_{1\hat1}\,\tau_1
\Bigg\}\,\epsilon=0\,.
\label{mngravhat1ex2}
\eear 
Let us proceed as before and multiply this last equation by
$\left[1+{1\over 4}\,e^{-2g}\,(a^2-1)\right]^{-1}$, which  allows us
to write it in terms of the angles $\alpha$ and $\beta$. Next we
substitute the projection (\ref{mnpr2}) and its equivalent expression
(\ref{mnpr2ex2}). After some calculation one gets:
\bear
&&\Bigg\{2\sin\alpha\,\cos\beta\,\Gamma_{1\hat1}\,\tau_1+2\sin\alpha\,\sin\beta\,
\Gamma_{x^2x^3}\,\Gamma_{1\hat1}\,\tau_1\,\tau_3-2\sin\alpha\,
\Gamma_{1\hat1}+\rc\rc&&+{4\over 1+{1\over4}\,e^{-2g}\,(a^2-1)}
\Big(\cos\alpha-\sin\alpha\,\cos\beta\,\Gamma_{1\hat1}\,\tau_1-
\sin\alpha\,\sin\beta\,\Gamma_{x^2x^3}\,\Gamma_{1\hat1}\,\tau_1\,\tau_3
-\Gamma_{r\hat1\hat2\hat3}\Big)\Bigg\}\,\epsilon=0\,.\rc
\label{mngravhat1ex3}
\eear
It is clear that in order to satisfy this equation, one must impose an
extra projection on $\epsilon$. Indeed, we will see that if we require
the term in parentheses to vanish we get a projection that makes
the first three terms cancel each other. The term in parentheses
in eq. (\ref{mngravhat1ex3}) can be written as
\beq
\Big(\cos\alpha+\sin\alpha\,\Gamma_{1\hat1}\Big)\Big(1-\cos\beta\,
\tau_1-\sin\beta\,\Gamma_{x^2x^3}\,\tau_1\,\tau_3\Big)\,\epsilon\,,
\label{mnpr3ex00}
\eeq
where we have used that
$\Gamma_{r\hat1\hat2\hat3}\,\epsilon=\Big[\cos\alpha\,\left(\cos\beta-\sin\beta\,\Gamma_{x^2x^3}\,\tau_3
\right)\,\tau_1-\,\sin\alpha\Gamma_{1\hat 1}\Big]\,\epsilon$,  which
follows from (\ref{mnpr2}). If we want eq. (\ref{mnpr3ex00}) to vanish,
$\epsilon$ must satisfy the following equation:
\beq
\tau_1\,\epsilon=\Big(\cos\beta+\sin\beta\,\Gamma_{x^2x^3}\,\tau_3\Big)
\,\epsilon\,.
\label{mnpr3ex0}
\eeq
Let us write the first three terms of (\ref{mngravhat1ex3}) as
\beq
-2\sin\alpha\,\Gamma_{1\hat1}\,\epsilon+2\sin\alpha\,\Gamma_{1\hat1}\,
\tau_1\left(\cos\beta+\sin\beta\,\Gamma_{x^2x^3}\,\tau_3\right)\,
\epsilon\,,
\eeq
which clearly vanishes after imposing (\ref{mnpr3ex0}). So the equation 
$\delta\psi_{\hat1}=0$ is fulfilled by an $\epsilon$ satisfying the known
projections (\ref{mnpr1}) and (\ref{mnpr2}) together with the new one
(\ref{mnpr3ex0}).

The vanishing of the SUSY variations of the remaining angular components
of the gravitino is guaranteed by the three projections we have just
mentioned and the first order differential equations satisfied by $g$
and $a$, namely:
\bear
g'&=&{1\over2}\,(1-a^2)\,e^{-2g}\,\cos\alpha-e^{-g}\,a\,\sin\alpha\,\,,
\rc\rc a'&=&-2a\cos\alpha-(1-a^2)\,e^{-g}\,\sin\alpha\,\,.
\label{mngadiffeqs}
\eear
$\delta\psi_{\hat2}=0$ takes the same form, up to a global factor, as
$\delta\psi_{\hat1}=0$, and $\delta\psi_{\hat3}=0$ also holds if
we impose the same projections. The equations 
$\delta\psi_1=0$ and
$\delta\psi_2=0$ are equal up to a global factor and they vanish if
one uses the differential equations (\ref{mngadiffeqs}) and again the
projections (\ref{mnpr1}), (\ref{mnpr2}) and (\ref{mnpr3ex0}). 

Notice that the new projection (\ref{mnpr3ex0}) can be written as
\beq
\tau_1\,\epsilon=e^{\beta\,\Gamma_{x^2x^3}\,\tau_3}\,\epsilon\,,
\label{mnpr3ex1}
\eeq
and, in addition, by using (\ref{mnpr3ex0}) on the right-hand side
of (\ref{mnpr2}), one arrives at
\beq
\Gamma_{r\hat 1\hat 2\hat 3}\,\,\epsilon=
\left(\cos\alpha\,-\, \sin\alpha\,\Gamma_{1\hat 1}\right)\,\epsilon\,=\,
e^{-\alpha\Gamma_{1\hat 1}}\,\epsilon\,\,.
\label{mnpr2ex1}
\eeq
Since $[\,\Gamma_{x^2x^3}\tau_3,\Gamma_{1\hat 1}\,]=
\{\tau_1,\Gamma_{x^2x^3}\tau_3\}=
\{\Gamma_{r\hat 1\hat 2\hat 3},\Gamma_{1\hat 1}\}=0$, we can solve 
(\ref{mnpr1}), (\ref{mnpr3ex1}) and (\ref{mnpr2ex1}) as follows:
\beq
\epsilon=e^{{\alpha\over 2}\,\Gamma_{1\hat 1}}\,
e^{-{\beta\over 2}\,\Gamma_{x^2x^3}\,\tau_3}\,\,\eta\,\,,
\label{mndefpr}
\eeq
where $\eta$ is a spinor that can only depend on the radial
coordinate and satisfies:
\beq
\Gamma_{12}\,\eta=\Gamma_{\hat1\hat2}\,\eta\,\,,
\,\,\,\,\,\,\,\,\,\,\,\,
\tau_1\eta=\eta\,\,,
\,\,\,\,\,\,\,\,\,\,\,\,
\Gamma_{r\hat 1\hat 2\hat 3}\,\,\eta=\eta\,\,.
\label{mnetapr}
\eeq

We still have to write the SUSY variation of the
radial  component of the dilatino. In principle we suppose that
$\epsilon$ depends on the radial coordinate. The equation
$\delta\psi_r=0$ will determine such dependence; it takes the form:
\beq
D_r\,\epsilon\,+\,{i\over 1920}\,
F_{N_1\cdots N_5}^{(5)}\,\Gamma^{N_1\cdots
N_5}\Gamma_r\,\epsilon
+{1\over96}\,{\cal F}^{(3)}_{N_1N_2N_3}\,
\big(\,\Gamma_r^{\,\,\,N_1N_2N_3}\,-\,
9\,\delta_r^{N_1}\,\,\Gamma^{N_2 N_3}\,\big)\,\epsilon^{*}=0\,.
\label{mngravrad}
\eeq
Let us write down each term of this equation separately. 
The covariant derivative can be easily written in terms of
${d\epsilon\over dr}\equiv\epsilon'$ as
$D_r\,\epsilon=\left(E^r_{\tilde r}\right)^{-1}
\left(\epsilon'+{1\over4}\,\omega^{a\,b}_{\tilde r}\,\Gamma_{a\,b}\,
\epsilon\right)$.
Then, reading the spin connection
from eq. (\ref{mnspincon}) (recalling that $\omega^{a\,b}_{\tilde
r}=E^r_{\tilde r}\,\omega^{a\,b}_r$), the covariant derivative becomes:
\beq
D_r\,\epsilon=e^{-{\phi\over4}}\,h^{-{1\over8}}\left(\epsilon'+{1\over4}\,
e^{-g}\,a'\,\Gamma_{1\hat1}\,\epsilon\right)\,,
\label{mngravrads1}
\eeq
where we have already used the projection (\ref{mnpr1}). The second
piece of (\ref{mngravrad}), after inserting the five-form written in
eq. (\ref{mnf5flat}), using eq. (\ref{mntotchir}) and imposing again
(\ref{mnpr1}), takes the form:
\bear
{i\over8}\,\Theta\,e^{3\phi\over4}\,h^{-{5\over8}}\Bigg\{\left[{1\over2}\,
(1-a^2)\,e^{-2g}-2\right]
\Gamma_{\hat1\hat2\hat3\,x^2x^3}\,\Gamma_{r}-e^{-g}\,a'\,
\Gamma_{r1\hat1\,x^2x^3}\,\Gamma_{r}
\Bigg\}\,\tau_2\,\epsilon\,.
\label{mngravrads2}
\eear
Next we substitute the RR three-form (\ref{mnf3flat}) into the last terms of eq.
(\ref{mngravrad}) and, after making use of the projection
(\ref{mnpr1}), we arrive at
\bear
&&{1\over96}\,\left(\,\Gamma_r^{\,\,\,N_1N_2N_3}\,-\,
9\,\delta_r^{N_1}\,\,\Gamma^{N_2
N_3}\,\right)\,\epsilon^{*}={1\over8}\,e^{-{\phi\over4}}\,h^{-{5\over8}}\Bigg\{-\left[1+{1\over4}\,(a^2-1)\,e^{-2g}\,
\right]\Gamma_{r\hat1\hat2\hat3}\,\tau_1+\rc\rc&&
\;+\,3\left[{1\over2}\,a'\,e^{-g}
\,\Gamma_{1\hat1}\,\tau_1-\Theta\,e^\phi\,h^{-{1\over2}}\,\phi'\,\Gamma_{x^2x^3}
\,\tau_3\right]\Bigg\}\epsilon\,.
\label{mngravrads3}
\eear
We shall now gather the three contributions to $\delta\psi_r=0$
(namely (\ref{mngravrads1}), (\ref{mngravrads2}), (\ref{mngravrads3}))
and substitute into the equation the form of $\epsilon$ written in eq.
(\ref{mndefpr}). After some calculation, taking into account the
projections (\ref{mnetapr}) satisfied by $\eta$ and the definitions of
$\alpha$ and $\beta$ given in  (\ref{mnalphadef}) and
(\ref{mnbetadef}), we get the following equation:
\bear
\eta'&+&{1\over2}\Bigg[\alpha'\,\Gamma_{1\hat1}-\beta'\,\Gamma_{x^2x^3}\,\tau_3
+e^{-g}\,a'\,\Gamma_{1\hat1}-\Theta\,e^\phi\,\phi'\,h^{-1}\,
\Gamma_{x^2x^3}\,\tau_3-\rc\rc&&-\,{1\over2}\,\Theta^2\,e^{2\phi}\,\phi'\,h^{-1}
-{1\over4}\,\phi'\,h^{-1}\,\Bigg]\eta=0\,.
\label{mngravrad2}
\eear
The fulfilment of this equation requires that the terms
proportional to $\Gamma_{1\hat1}$, to $\Gamma_{x^2x^3}\,\tau_3$ and to
the identity vanish separately, resulting:
\bear
\alpha'+a'\,e^{-g}=0\,,\quad
\beta'+\Theta\,e^\phi\,\phi'\,h^{-1}=0\,,\label{mnalphabetdfeq} \\ \rc
\eta'=\left[{1\over8}\,\phi'\,h^{-1}+{1\over4}\,\Theta^2\,e^{2\phi}\,\phi'\,
h^{-1}\right]\,.
\label{mngravrdepeq}
\eear
One can check that (\ref{mnalphabetdfeq}) follows automatically from
the definitions of $\alpha$ and $\beta$, while (\ref{mngravrdepeq}) fixes
the radial dependence of $\eta$, which can be written as
\beq
\eta=e^{\phi\over8}\,h^{1\over16}\,\epsilon_0\,,
\label{mngravrdepsol}
\eeq
in terms of a constant spinor $\epsilon_0$ satisfying the same
projections as $\eta$, \ie \ (\ref{mnetapr}).

To sum up, the Killing spinors of the non-commutative MN background can
be written in terms of a ten dimensional constant spinor satisfying three
independent and compatible projections reducing the number of independent
spinors from the maximal 32 to 4 as it should be for the gravity dual of a
4d ${\cal N}=1$ supersymmetric theory.
In fact, by looking at \cite{flavoring}, one can check that the only
difference between the Killing spinors of the commutative and
non-commutative models is a rotation along the non-commutative plane.
This should be expected since those are the directions along which the
deformation is performed and the Killing spinors of the MN solution do
not depend on those coordinates.


\chapter{Conclusions}
\label{conclcp}

In this work we have computed explicitly the Killing spinors of five
different solutions of ten dimensional type IIB supergravity. We have
begun by computing the Killing spinors for backgrounds of the form
$\RR^{1,3}\times {\cal Y}_6$, where ${\cal Y}_6$ is either the
singular conifold or any of its resolutions. We have carried out this
calculation using a generalized ansatz proposed in \cite{Twist}
resulting from the uplift of a domain wall setup in eight dimensional
gauged supergravity corresponding to D6-branes wrapping an $S^2$. As
expected, the vanishing of the 10d SUSY variations (\ref{sugra}),
besides determining the form of the 10d Killing spinors, resulted in
the same system of differential equations for the functions entering
the ansatz as in the eight dimensional case. The solutions of that
system realize the different resolutions of the conifold.
Furthermore, it turns out that when written in the natural frame
arising from the uplifted ten dimensional metric, the 10d Killing
spinors do not depend on the angular coordinates of the conifold. In
fact, we have written them in terms of a constant 10d spinor, by
means of a rotation (whose phase depends only on the radial
coordinate) along two internal directions. The constant spinor must
satisfy two independent and compatible projections reducing its
number of independent components from 32 to 8. Thus, as it should be,
these backgrounds leave unbroken eight supersymmetries. These results,
presented in chapter \ref{genconsec}, were very useful for the
development of the subsequent chapters, as we will recall in the
following paragraphs.

We have studied, in chapter \ref{kwcp}, the Killing spinors of the
Klebanov-Witten (KW) model \cite{KW}. This background arises from
placing a stack of
$N$ D3-branes at the tip of the singular conifold. After taking the
usual decoupling limit, the resulting geometry is $AdS_5\times T^{1,1}$
(recall that $T^{1,1}$ is the base of the conifold), and it turns out that
the computation of the Killing spinors is simplified if we write the
$T^{1,1}$ metric in the form (\ref{singconif}) obtained in chapter
\ref{genconsec}. Indeed, we were able to write the Killing spinors in
such a basis that they are independent of the coordinates of the
$T^{1,1}$; one suspected this would be so, since that form of  the metric
comes out  as an uplift from 8d gauged supergravity. Therefore, from the
consistency of the reduction, the Killing spinors should not depend on
any angular coordinate of the group manifold (the $SU(2)$ along which the
reduction takes place) but, in addition, the topological twist needed to
realize supersymmetry with wrapped branes in the eight dimensional theory
results in a fibration of the $SU(2)$ manifold along the remaining
$S^2$, reinforcing the conjecture that the Killing spinors will not
depend on the $T^{1,1}$ coordinates, when written in the appropriate
frame.

Let us also recall that, contrary to what one could naively expect, the
Killing spinors of the KW background do not satisfy the projection
corresponding to a D3-brane extended along the Minkowski directions. In
fact, we have got two independent solutions, namely (\ref{epsplus}) and
(\ref{epsminus}), each one being $1/8$ supersymmetric, and only the first
one satisfies that projection. These are the Killing spinors
corresponding to the four ordinary supersymmetries, while the second ones
(eq. (\ref{epsminus})) realize the four superconformal symmetries.
The complete solution only satisfies the two independent
projections corresponding to the $T^{1,1}$ space. 

We have already pointed out in the introduction that the explicit
knowledge of the Killing spinors of the KW model was essential for the
program carried out in \cite{SSP}, where we systematically explored the
possibilities of adding different D-brane probes in the Klebanov-Witten
background.

The aforementioned form of writing the metric of the $T^{1,1}$,
resulting from (\ref{singconif}) in chapter \ref{genconsec}, has become
useful again for the computation of the Killing spinors of the
Klebanov-Tseytlin (KT) background \cite{KT} performed in chapter
\ref{ktcp}. This solution results from adding $M$ fractional D3-branes
(wrapped D5-branes) to the setup of the KW model. The three-form flux
created by the D5-branes is the source of the conformal symmetry
breaking, so the background preserves only the four usual
supersymmetries corresponding to a four dimensional ${\cal N}=1$ YM
theory. We have written the Killing spinors in a basis where they do
not depend on the coordinates of the $T^{1,1}$ and this time, in
addition to the two projections of the  $T^{1,1}$, they satisfy the
projection corresponding to a D3-brane extended along the Minkowski
space. Therefore, one has 4 independent spinors standing for the four
conserved supersymmetries.

Our next step (chapter \ref{kscp}) was the computation of the Killing
spinors of the Klebanov-Strassler (KS) background \cite{KS}. This
solution is constructed by placing D3-branes and fractional D3-branes
at the tip of the deformed conifold. Then, while in
the UV this solution approaches the KT model, the deformation of the
conifold gives, in the IR, a geometrical realization of confinement
and chiral symmetry breaking. Thus, this background is conjectured to
provide (in the
$\tau\to 0$ limit, where $\tau$ is the holographic coordinate) a dual to
the IR region of ${\cal N}=1$ SYM.

The KS background is formulated in terms of several functions of the
radial coordinate which are defined by means of a system of first order
differential equations solving the equations of motion of type IIB
supergravity. By writing the metric of the deformed conifold in the form
found in chapter \ref{genconsec} and imposing the projections
obtained there for the Killing spinors of the deformed conifold (see
section \ref{gcdefconks}) plus the projection corresponding to a
D3-brane along
$\RR^{1,3}$ we have shown that, in order to have a background preserving
some supersymmetry (in particular a ${1\over8}$ supersymmetric solution),
those functions defining the KS model must satisfy the mentioned first
order system plus an extra algebraic constraint. Then, the Killing
spinors of the KS background can be written, in a frame where they do
not depend on the angular coordinates of the conifold, in terms of a
ten dimensional constant spinor satisfying three independent
projections leaving unbroken four supersymmetries.

Finally, in chapter \ref{ncmncp} we have studied the Killing spinors of
the non-commutative Maldacena-N\'u\~nez (MN) background \cite{NCMN}.
This solution can be obtained from the commutative geometry
\cite{MN,CV} by means of a chain of string dualities resulting in a
deformed background which singles out the two spatial directions along
which the deformation took place (they form the so-called
non-commutative plane). Thus, it breaks the four dimensional Lorentz
invariance and the corresponding dual theory has spatial
non-commutativity along those directions.

The computation of the Killing spinors goes along the same lines as for
the commutative solution (see \cite{flavoring}). Working in the
appropriate frame (the one arising when one obtains the commutative MN
background as an uplift from 7d gauged SUGRA) it turns out that the
Killing spinors do not depend on  the internal angular coordinates of the
geometry and, in fact, we have written them in terms of a constant
spinor satisfying the same three projections as in the commutative case.
The only effect of the deformation is the presence of a rotation along
the non-commutative plane. This was expected since, when working in the
suitable background, the Killing spinors of the commutative solution do
not depend on the directions along which the deformation is performed.
Let us remark that the solution leaves unbroken four supersymmetries as
it corresponds to a four dimensional ${\cal N}=1$ supersymmetric field
theory.



\begin{thebibliography}{99}


\bibitem{ADSCFT} J.~M.~Maldacena, ``The large $N$ limit of superconformal
field theories and supergravity'', {\it Adv.\ Theor.\ Math.\ Phys.}\ 
{\bf 2} (1998) 231, {\rm hep-th/9711200}.

\bibitem{MAGOO} O. Aharony, S. Gubser, J. Maldacena,
H. Ooguri and Y. Oz, ``Large $N$ field theories, string theory and
gravity",  {\sl Phys.  Rept. } {\bf 323} (2000) 183, {\rm
hep-th/9905111}.\\ J. L. Petersen,
``Introduction to the Maldacena Conjecture on AdS/CFT", 
{\sl Int. J. Mod. Phys.} {\bf A14} (1999) 3597, {\rm
hep-th/9902131}.\\
E. D'Hoker and D. Z. Freedman, ``Supersymmetric gauge theories
and the AdS/CFT correspondence'', {\rm hep-th/0201253} 


\bibitem{n1duality} 
C.~P.~Herzog, I.~R.~Klebanov and P.~Ouyang, ``D-branes on the conifold and
N=1 gauge/gravity dualities'', hep-th/0205100.\\
M.~Bertolini, ``Four lectures on the gauge-gravity correspondence'',
Int.\ J.\ Mod.\ Phys. {\bf A18} (2003) 5647, hep-th/0303160.\\
E.~Imeroni, ``The gauge/string correspondence towards realistic gauge
theories'', hep-th/0312070.\\
P.~Di Vecchia, ``N = 1 super Yang-Mills from D
branes'', hep-th/0403216.\\
J.D.~Edelstein, R.~Portugues, ``Gauge/String Duality in Confining
Theories'', hep-th/0602021.






\bibitem{swedes}
M. Cederwall, A. von Gussich, B.E.W. Nilsson, P. Sundell and A.
Westerberg,
``The Dirichlet super-p-branes in Type IIA and IIB supergravity'',
{\sl \np} {\bf B490 }(1997) 179, {\rm hep-th/9611159}:\\
E. Bergshoeff and P.K. Townsend, ``Super D-branes'', 
{\sl \np} {\bf B490 }(1997) 145, {\rm hep-th/9611173};\\
M. Aganagic, C. Popescu and J.H. Schwarz, ``D-brane actions with
local kappa symmetry'', 
{\sl \pl} {\bf B393 }(1997) 311, {\rm hep-th/9610249};
``Gauge-invariant and gauge-fixed D-brane actions'', 
{\sl \np} {\bf B495 }(1997) 99, {\rm hep-th/9612080}.



\bibitem{Wittenbaryon}E. Witten, ``Baryons and branes in Anti-de Sitter 
space", {\sl \jhep} {\bf 9807 }(1998) 006, {\rm hep-th/9805112}.



\bibitem{GK} S. S. Gubser and I. R. Klebanov, ``Baryons and domain walls in a 
${\cal N}=1$ superconformal gauge theory", {\sl \pr} {\bf D58} (1998) 125025,
{\rm hep-th/9808075}.


\bibitem{KW}I. R. Klebanov and E. Witten, ``Superconformal field theory
on threebranes at a Calabi-Yau singularity", 
{\sl \np} {\bf B536 }(1998) 199, {\rm hep-th/9807080}.


\bibitem{BHK}D. Berenstein, C. P. Herzog and I. R. Klebanov,
``Baryon spectra and AdS/CFT correspondence",  
{\sl \jhep} {\bf 0206 }(2002) 047,
{\rm hep-th/0202150}.


\bibitem{Beasley} C. E. Beasley, 
``BPS branes from baryons", 
{\sl \jhep} {\bf 0211 }(2002) 015, {\rm hep-th/0207125}.


\bibitem{GRW}S. Gukov, M. Rangamani and E. Witten,
``Dybaryons, strings, and branes in AdS orbifold models", 
{\sl \jhep} {\bf 9812 }(1998) 025, {\rm hep-th/9811048}.


\bibitem{HMcK}C. P. Herzog and J. McKernan, ``Dibaryon spectroscopy",
{\sl \jhep} {\bf 0308 }(2003) 054, {\rm hep-th/0305048}.

\bibitem{kk} A. Karch and E. Katz, ``Adding flavor to AdS/CFT'',
{\sl \jhep} {\bf 0206 }(2002) 043, {\rm hep-th/0205236};\\
A. Karch, E. Katz and N. Weiner,
``Hadron masses and screening from AdS Wilson loops'',
{\sl \prl} {\bf 90 }(2003) 091601, {\rm hep-th/0211107}.



\bibitem{SSP}D. Arean, D. Crooks and A. V. Ramallo, 
``Supersymmetric probes on the conifold'', 
{\sl \jhep} {\bf 0411 }(2004) 035, {\rm hep-th/0408210}.



\bibitem{Twist} J.D.~Edelstein, A. Paredes and A. V. Ramallo, 
``Singularity resolution in gauged supergravity and conifold
unification'', {\sl \pl} {\bf B554 }(2003) 197, {\rm hep-th/0212139}.




\bibitem{KT}I. R. Klebanov and A Tseytlin, 
``Gravity duals of supersymmetric $SU(N)\times SU(N+M)$ gauge
theories'',  {\sl \np} {\bf B574} (2000) 123, {\rm hep-th/0002159}.



\bibitem{KS} I. R. Klebanov and M. J. Strassler, ``Supergravity and a confining gauge
theory: duality cascades and a chiSB-resolution of naked
singularities'',  {\sl \jhep} {\bf 0008 }(2000) 052, {\rm
hep-th/0007191}.



\bibitem{NCMN}T. Mateos, J. M. Pons and P. Talavera,
``Supergravity dual of non-commutative ${\cal N}=1$ SYM'',
{\sl \np} {\bf B651 }(2003) 291, {\rm hep-th/0209150}.


\bibitem{flavoring}C. N\'u\~nez, A. Paredes and A. V. Ramallo,
``Flavoring the gravity dual of ${\cal N}=1$ Yang-Mills with probes'',
{\sl \jhep} {\bf 0312 }(2003) 024, {\rm hep-th/0311201}.



\bibitem{MN} J.~M.~Maldacena and C. N\'u\~nez, ``Towards the large N 
limit of pure
${\cal N}=1$ super Yang Mills'',
{\sl \prl} {\bf 86 }(2001) 588, {\rm hep-th/0008001}.



\bibitem{CV} A. H. Chamseddine and M. S. Volkov, ``Non-Abelian BPS 
monopoles in $N=4$
gauged supergravity'',
{\sl \prl} {\bf 79 }(1997) 3343, {\rm hep-th/9707176}; ``Non-Abelian 
solitons in $N=4$
gauged supergravity and leading order string theory",
{\sl \pr} {\bf D57 }(1998) 6242, {\rm hep-th/9711181}.



\bibitem{ncflav}  D.~Arean, A.~Paredes and A.~V.~Ramallo,
  ``Adding flavor to the gravity dual of non-commutative gauge
theories'', {\sl \jhep} {\bf 0508} (2005) 017, {\rm hep-th/0505181}.



\bibitem{GMSW1}
J.~P.~Gauntlett, D.~Martelli, J.~Sparks and D.~Waldram, ``Supersymmetric
AdS(5) solutions of M-theory'', {\sl Class.\ Quant.\ Grav.\ }  {\bf 21}
(2004) 4335, {\rm hep-th/0402153}.



\bibitem{GMSW2}
J.~P.~Gauntlett, D.~Martelli, J.~Sparks and D.~Waldram, ``Sasaki-Einstein
metrics on S(2) x S(3)'', {\sl Adv.\ Theor.\ Math.\ Phys.\ } {\bf 8}
(2004) 711, {\rm hep-th/0403002}.


\bibitem{ms}
D.~Martelli and J.~Sparks, ``Toric geometry, Sasaki-Einstein manifolds and
a new infinite class of AdS/CFT duals'', {\sl Commun. Math. Phys. } 
{\bf 262} (2006) 51, {\rm hep-th/0411238}.


\bibitem{sequiver}
S.~Benvenuti, S.~Franco, A.~Hanany, D.~Martelli and J.~Sparks, ``An
infinite family of superconformal quiver gauge theories with
Sasaki--Einstein duals'', {\sl \jhep} {\bf 0506} (2005) 064, 
{\rm hep-th/0411264}.


\bibitem{ypqflav}  F.~Canoura, J.D.~Edelstein, L.A.~Pando
Zayas, A.~V.~Ramallo and D.~Vaman
  ``Supersymmetric branes on $AdS_5\times Y{p,q}$ and their field
theory duals'', 
{\rm hep-th/0512087}.


\bibitem{lpqrflav}  F.~Canoura, J.D.~Edelstein and A.~V.~Ramallo, 
``D-brane probes on $L^{a,b,c}\,$ superconformal field theories'', 
{\rm hep-th/0605260}.


\bibitem{Cvetic:2005ft}
  M.~Cvetic, H.~Lu, D.~N.~Page and C.~N.~Pope,
  ``New Einstein-Sasaki spaces in five and higher dimensions'',
  {\rm hep-th/0504225}.



\bibitem{Martelli:2005wy}
  D.~Martelli and J.~Sparks,
  ``Toric Sasaki-Einstein metrics on $S^2 \times S^3$'',
  Phys.\ Lett.\ B {\bf 621}, 208 (2005)
  {\rm hep-th/0505027}.



\bibitem{SUSYIIB} J. H. Schwarz, ``Covariant field  equations of
chiral N=2 D=10 supergravity'', {\sl \np} {\bf B226} (1983) 269. 



\bibitem{Candelas} P. Candelas and X. de la Ossa, ``Comments on conifolds", 
{\sl \np} {\bf B342} (1990) 246. 



\bibitem{SS} A.~Salam and E.~Sezgin, ``d=8 Supergravity'', Nucl.\ Phys.\ 
B {\bf 258} (1985) 284.


\bibitem{RCN} L.J. Romans, 
``New compactifications of chiral $N=2$, $d=10$ supergravity",
{\sl \pl} {\bf B153 }(1985) 392.



\bibitem{gaugedsugra} 
J.D.~Edelstein and C.~N\'u\~nez, ``D6 branes and M-theory
geometrical transitions from gauged  supergravity'',
{\sl \jhep} {\bf 0104}, (2001) 028, {\rm hep-th/0103167}.


\bibitem{GRCI} 
L.A.~Pando Zayas and A.A.~Tseytlin, ``3-branes on resolved
conifold'', {\sl \jhep} {\bf 0011}, (2000) 028, {\rm hep-th/0010088}.


\bibitem{GRCII} 
G.~Papadopoulos and A.A. Tseytlin, ``Complex geometry of conifolds and
5-brane wrapped on 2-sphere'', {\sl Class. Quant. Grav.} {\bf 18}, 1333
(2001), {\rm hep-th/0012034}.


\bibitem{GRCIII} 
L. A. Pando Zayas and A. A. Tseytlin, ``3-branes on spaces with
$\RR\times S^2\times S^3 $ topology'', {\sl \pr} {\bf D63}
(2001) 086006, {\rm hep-th/0101043}.



\bibitem{LPT}H. Lu, C. N. Pope and P. K. Townsend, 
``Domain walls from Anti-de-Sitter spacetime", 
{\sl \pl} {\bf B391} (1997) 39, {\rm hep-th/9607164}.



\bibitem{Globalads}M. T. Grisaru, R. C. Myers and O. Tafjord, 
``SUSY and Goliath", {\sl \jhep} {\bf 0008 }(2000) 040, {\rm hep-th/0008015}; see
also: H. Lu, C. N. Pope and J. Rahmfeld, ``A Construction of Killing
Spinors on $S^n$\ '', {\sl J. Math. Phys.} {\bf 40 }(1999) 4518, {\rm
hep-th/9805151}.



\bibitem{GranaPol} M. Gra\~na and J. Polchinski, ``Supersymmetric three-form flux
perturbations on $AdS_5$", 
{\sl \pr} {\bf D63 }(2001) 026001, {\rm hep-th/0009211}.



\bibitem{Gubser} S. Gubser, ``Supersymmetry and F-theory realization of the deformed
conifold with flux", {\rm hep-th/0010010}.



\bibitem{NCString}
A. Connes, M. R. Douglas and A. Schwarz, ``Noncommutative geometry 
and matrix theory:
compactifications on tori", {\sl \jhep} {\bf 9802} (1998) 003,  {\rm 
hep-th/9711162};\\
M. R. Douglas and C. Hull, ``D-branes and the noncommutative torus",
{\sl \jhep} {\bf 9802} (1998) 008,  {\rm hep-th/9711165}.



\bibitem{SW}
N. Seiberg and E. Witten, ``String theory and non-commutative geometry",
{\sl \jhep} {\bf 9909} (1999) 032,  {\rm hep-th/9908142}.



\bibitem{MR}
J. M. Maldacena and J. G. Russo,
``Large N limit of non-commutative gauge theories'',
{\sl \jhep} {\bf 9909} (1999) 025,  {\rm hep-th/9908134};\\
A. Hashimoto and N. Itzhaki, ``Noncommutative Yang-Mills and the 
AdS/CFT correspondence",
{\sl \pl} {\bf B465} (1999) 142, {\rm hep-th/9907166}.


\bibitem{AOSJ}M. Alishahiha, Y. Oz and M. M. Sheikh-Jabbari,
``Supergravity and large N noncommutative field theories", 
{\sl \jhep} {\bf 9911} (1999) 007,  {\rm hep-th/9909215}.




\end{thebibliography}
\end{document}